\documentclass[iop,apj]{emulateapj}
\usepackage{amsmath,amssymb,amstext}

\usepackage[breaklinks,colorlinks,citecolor=blue,linkcolor=magenta]{hyperref} 

\usepackage[all]{hypcap} 

\usepackage{natbib}
\usepackage{graphicx}
\usepackage[caption=false]{subfig}
\usepackage{booktabs}

\shorttitle{Short Title}
\shortauthors{Krief et al.}

\begin{document}
	
	\title{Solar opacity calculations using the super-transition-array method}
	
%
%
%


	
	
	\author{M. Krief, A. Feigel and D. Gazit}
	\email{menahem.krief@mail.huji.ac.il}
	\affil{The Racah Institute of Physics, The Hebrew University, 91904 Jerusalem, Israel}
	
\begin{abstract}
		A new opacity model based on the Super-Transition-Array (STA) method for the calculation of monochromatic opacities of local thermodynamic equilibrium plasmas, was developed. The atomic code, named STAR (STA-Revised), is described and used to calculate spectral opacities for a solar model implementing the recent AGSS09 composition. Calculations are carried throughout the solar radiative zone. The relative contributions of different chemical elements and atomic processes to the total Rosseland mean opacity are analyzed in detail. 
		Monochromatic opacities and charge state distributions were compared with the widely used Opacity-Project (OP) code, for several elements near the radiation-convection interface.
		STAR Rosseland opacities for the solar mixture show a very good agreement with OP
		and the OPAL opacity code, throughout the radiation zone. Finally, an explicit STA calculation of the full AGSS09 photospheric mixture, including all heavy metals was performed. It was shown that due to their extremely low abundance, and despite being very good photon absorbers, the heavy elements do not affect the Rosseland opacity.
\end{abstract}

\keywords{dense matter --- plasmas --- atomic processes ---atomic data --- opacity --- Sun: interior}
\maketitle
	
\section{Introduction}
Over the past decade, a new solar problem has emerged 
as solar photospheric abundances have been improved (\cite{asplund2005solar,asplund2009chemical,caffau2011solar,scott2015,scott2015_2,grevesse2015}) and the indicated solar metallicity, which is mainly due to low-Z metallic elements, has been significantly revised
downward than previously assumed (\cite{grevesse1993cosmic,grevesse1998standard}).
Standard solar models (SSMs), which are the fundamental theoretical tools to investigate the solar interior, do not reproduce helioseismic measurements (\cite{basu2004constraining}), such as the convection zone radius, the surface helium abundance and the sound speed profile (\cite{serenelli2009new}), when using these revised abundances. This gave rise to the \textit{solar composition problem}, motivating a rapid increase of research efforts in the field.
 
Rosseland opacity is a key quantity describing the coupling between radiation and matter in the solar interior plasma. Metallic elements significantly contribute to the opacity, although they are only present as a few percent of the mixture, since bound-bound and bound-free absorption can become a major source of opacity for the partially ionized metals. Thus, the amount of metals directly modifies the opacity profile of the sun and the solar composition problem is strongly related to the role of opacity in solar models. 

Among a variety of proposed
explanations, it has been shown that in order to reproduce the helioseismic measurements, changes to the opacity are required to compensate the lower solar metallicity.
Specifically, it was shown, that a smooth increase of
the opacity from the range of $ 2\%-5\% $ in the central regions to
the range of $ 12\%-15\% $ at the radiation-convection interface
is required (see \cite{christensen2009opacity,serenelli2009new,villante2010constraints,villante2010linear,villante2015quantitative} and references therein).
It is believed that the opacities of metals in the stellar mixture should be revised upward to compensate for the decreased low-Z metallic element abundances. 
In addition, recently, the monochromatic opacity of iron, a major contributer to the solar opacity, was measured at very similar conditions to those expected near the
solar convection zone boundary (\cite{bailey2015higher}).
The measured spectrum appears to be larger than calculations of various state of the art and widely used atomic models, by about a factor of two in several spectral regions near the L-shell photoabsorption lines. No satisfactory explanation to this discrepancy is known.

We have recently developed (\cite{krief2015effect}) an atomic code, named STAR (STA-Revised),  for the calculation of opacities of local thermodynamic equilibrium plasmas, by the STA method. The STA method (\cite{BarShalom1989}) is used to group the large number of transition lines between an enormous number of configurations into large assemblies, called a Super-Transition-Arrays (STAs). The STAs moments are calculated analytically using the partition function algebra, and are split into smaller STAs until convergence is achieved. Unlike detailed line accounting (DLA) methods such as the Opacity-Project (OP) (\cite{seaton1994opacities}) and OPAL (\cite{iglesias1996updated}), the STA method is able to handle situations where the number of relevant transitions is immense. On the other hand, unlike average-atom models (\cite{Rozsnyai1972,shalitin1984level}), the STA method reveals the unresolved-transition-array (UTA) spectrum, otherwise obtained by a very costly, sometimes intractable, detailed configuration accounting (DCA) calculation (see \autoref{ssec:bound_bound}).

In this work we use STAR to calculate and analyze solar opacities. The atomic calculations are based on a fully relativistic quantum mechanical theory via the Dirac equation. We analyze in detail the contribution of different chemical elements in the solar mixture to the total Rosseland mean opacity and examine the role of different atomic processes (bound-bound, bound-free, free-free and scattering). The analysis is carried throughout the radiative zone (that is, from the solar core to the radiation-convection  boundary). 

\section{The Solar Model}
\begin{figure}
	\centering
	\resizebox{0.5\textwidth}{!}{\includegraphics{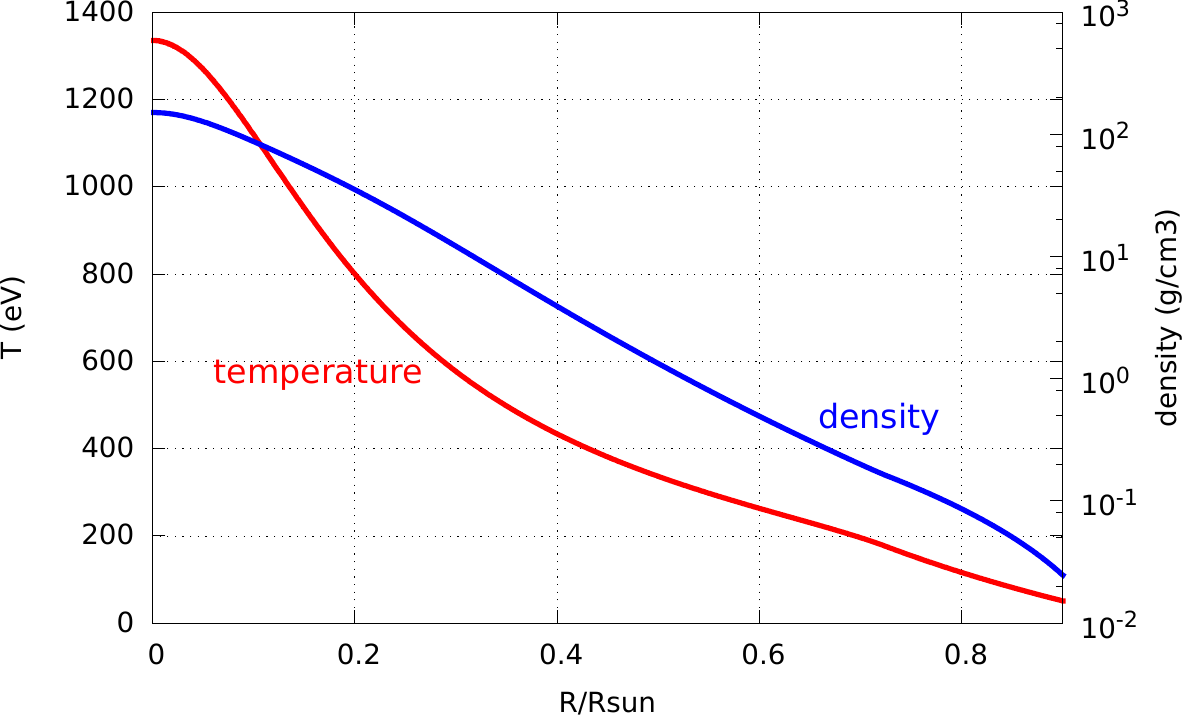}} 
	\caption{The temperature (red) and density (blue) profiles of the SSM.}
	\label{fig:rad_t_rho}
\end{figure}

\begin{figure}
		\centering
		\resizebox{0.5\textwidth}{!}{\includegraphics{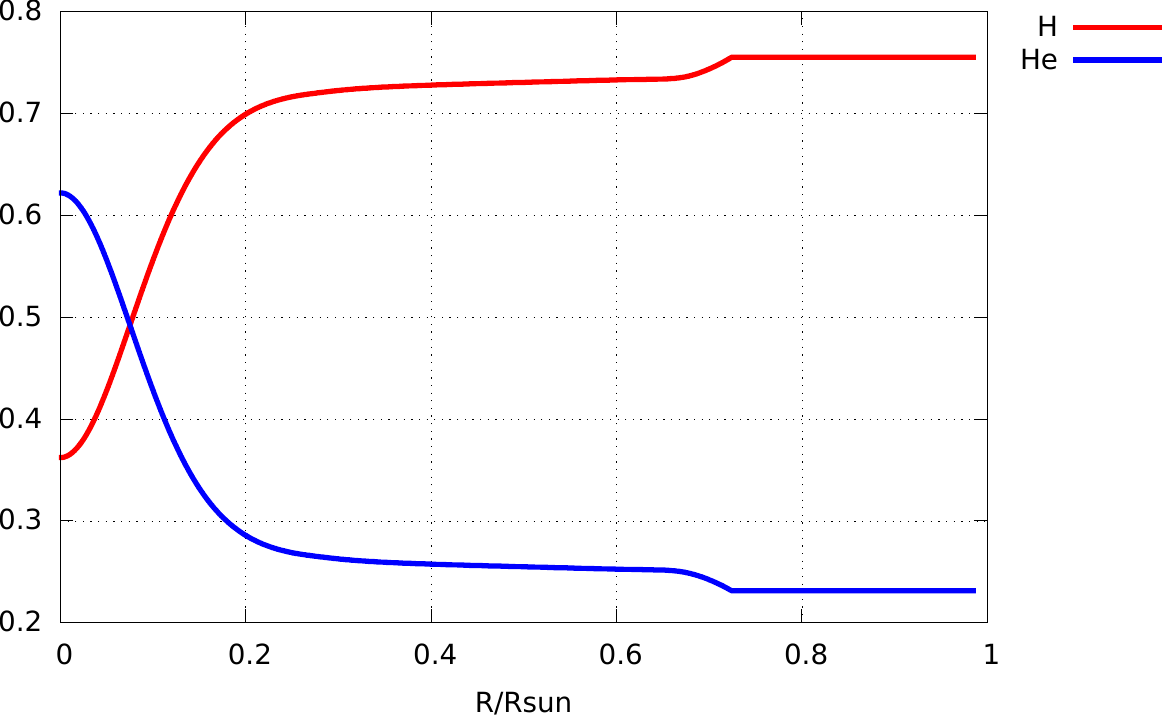}} 		
		\resizebox{0.5\textwidth}{!}{\includegraphics{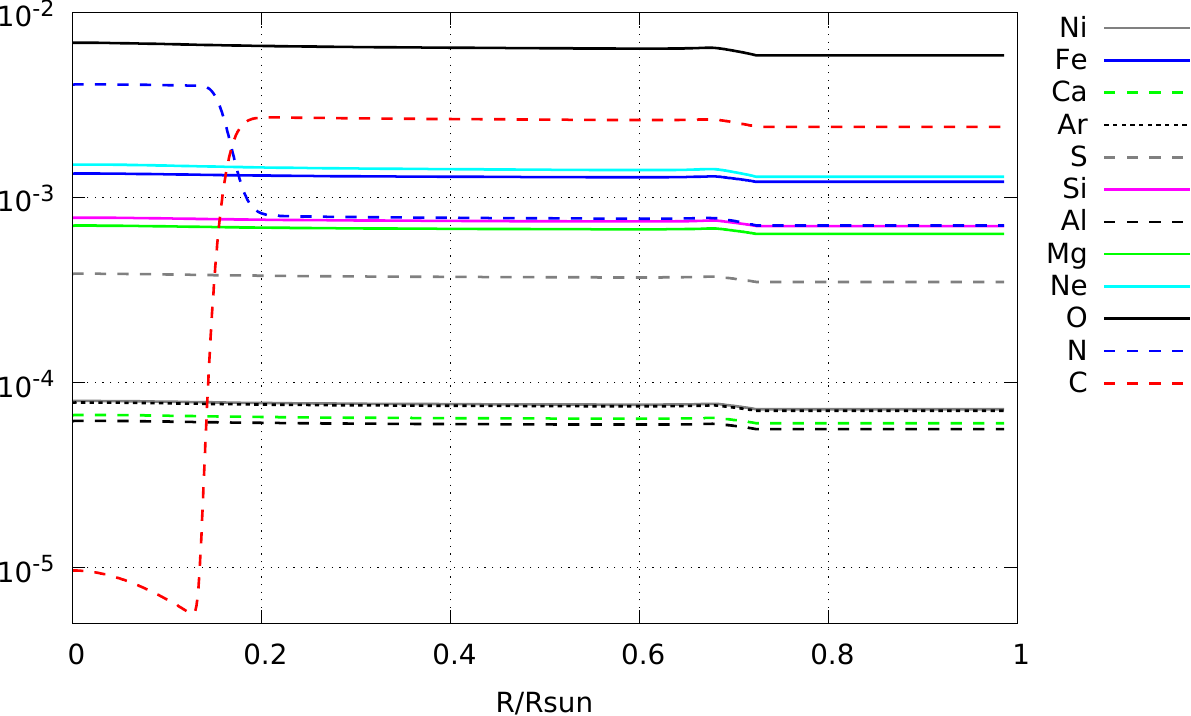}} 
		\resizebox{0.5\textwidth}{!}{\includegraphics{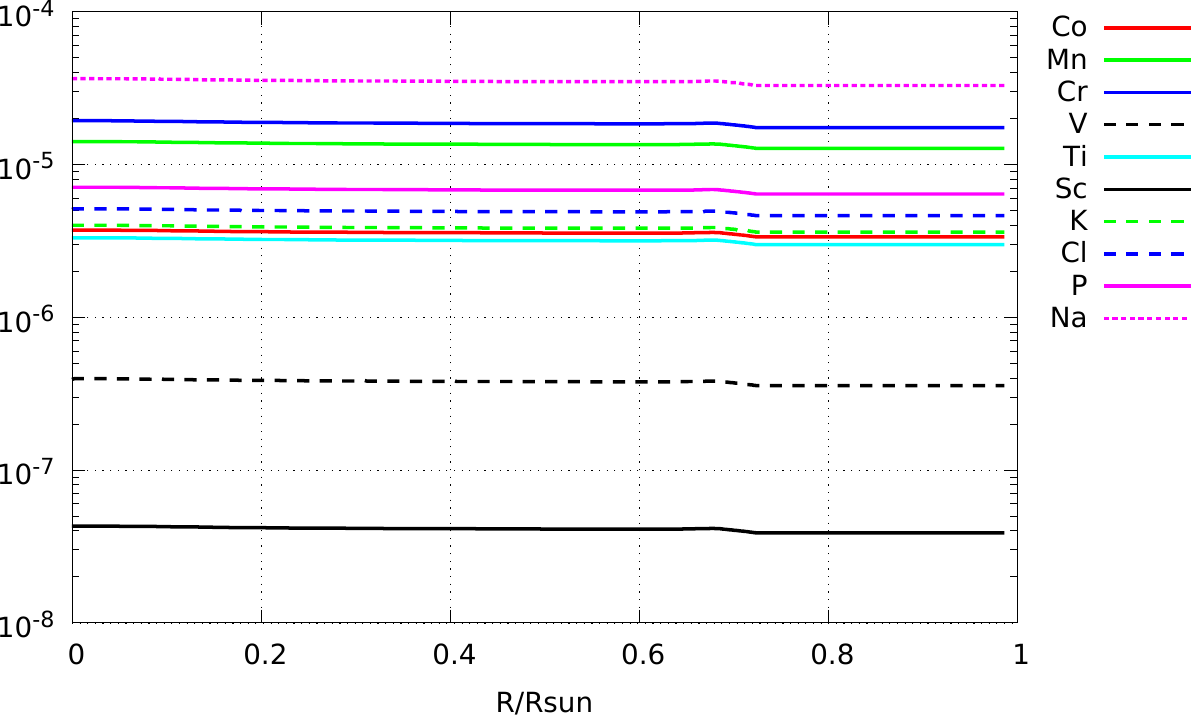}} 
	\caption{Mass fractions profiles for the 24 element in the SSM mixture.}
	\label{fig:abunds}
\end{figure}

The development of advanced stellar atmosphere models resulted in a downward revision of the solar photospheric metal abundances.
In this work we use the recent \cite{asplund2009chemical} (AGSS09)
set of chemical abundances
for volatile elements (C, N, O, and
Ne) together with their recommended meteoritic abundances
for refractory elements (Mg, Si, S, and Fe), where Si is
the anchor of both scales.
This compilation was used as an input for evolutionary solar model
calculations presented in
\cite{villante2014chemical,villante2010constraints,serenelli2011solar,serenelli2009new} (and references therein), which we refer to as "the SSM".
The resulting solar profiles\footnote{Solar data by Serenelli and Villante can be found online \url{ http://www.ice.csic.es/personal/aldos/Solar_Data.html}.} 
of various quantities are used throughout this work.
The  SSM temperature and density profiles are shown in \autoref{fig:rad_t_rho}.
The solar mixture of the SSM consists of the following 24 elements:
H, He, C, N, O, Ne, Na, Mg, Al, Si, P, S, Cl, Ar, K, Ca, Sc, Ti, V, Cr, Mn, Fe, Co and Ni. 
The elemental composition profiles are shown in \autoref{fig:abunds}. These are determined from the initial abundances and from the SSM due to processes such as nuclear burning, diffusion and gravitational settling.

The opacity used in the SSM was calculated by the Opacity-Project (OP) (\cite{seaton1994opacities,badnell2005updated}). 
The effect of intrinsic opacity changes (that is, with a fixed composition) on observable quantities, i.e. the convection-zone radius, surface helium abundance, sound speed profile etc, can be studied 
by the linear solar model (LSM) of \cite{villante2010linear}. This justifies the calculation of different opacity models using the thermodynamic conditions obtained by the SSM. 
\section{The Opacity Model}
In this section we briefly describe various parts of the computational model implemented by the STAR atomic code. A more comprehensive description of the STAR model will be reported elsewhere.
\subsection{The Mixture Model}\label{ssec:mix_densities}
The initial step in the calculation of the opacity of a mixture is the calculation of the effective densities of the different components (\cite{nikiforov1969calculation}). A mixture of temperature $ T $, density $ \rho $ and mass fractions $ m_{i} $ is described as a collection of ion-spheres with different volumes, or equivalently, individual \textit{effective mass densities} $ \rho_{i} $. The individual densities are found by the requirement that the chemical potentials $ \mu_{i} $ of all species are equal:
\begin{eqnarray}
\mu_{i}=\mu,
\end{eqnarray}
 and by the condition that the volume of the mixture is equal to the sum of the volumes occupied by the individual components:
\begin{eqnarray}
\dfrac{1}{\rho}\sum_{i}m_{i}=\sum_{i}\dfrac{m_{i}}{\rho_{i}}.
\end{eqnarray}
The equality of chemical potentials insures the equality of \textit{electronic} pressures only, rather than the total pressure (including the ionic contribution, see 	\cite{carson1968calculation}).
	However, the usual procedure adopted in the present work, is to equalize the chemical potentials
	\cite{nikiforov1969calculation,rose1992calculations,klapisch2013models},
	which, if the free electrons are treated semiclassically, also insures the
	equality of the (free) electronic densities at the ion sphere radius.

\begin{figure}
		\centering
	\resizebox{0.5\textwidth}{!}{\includegraphics{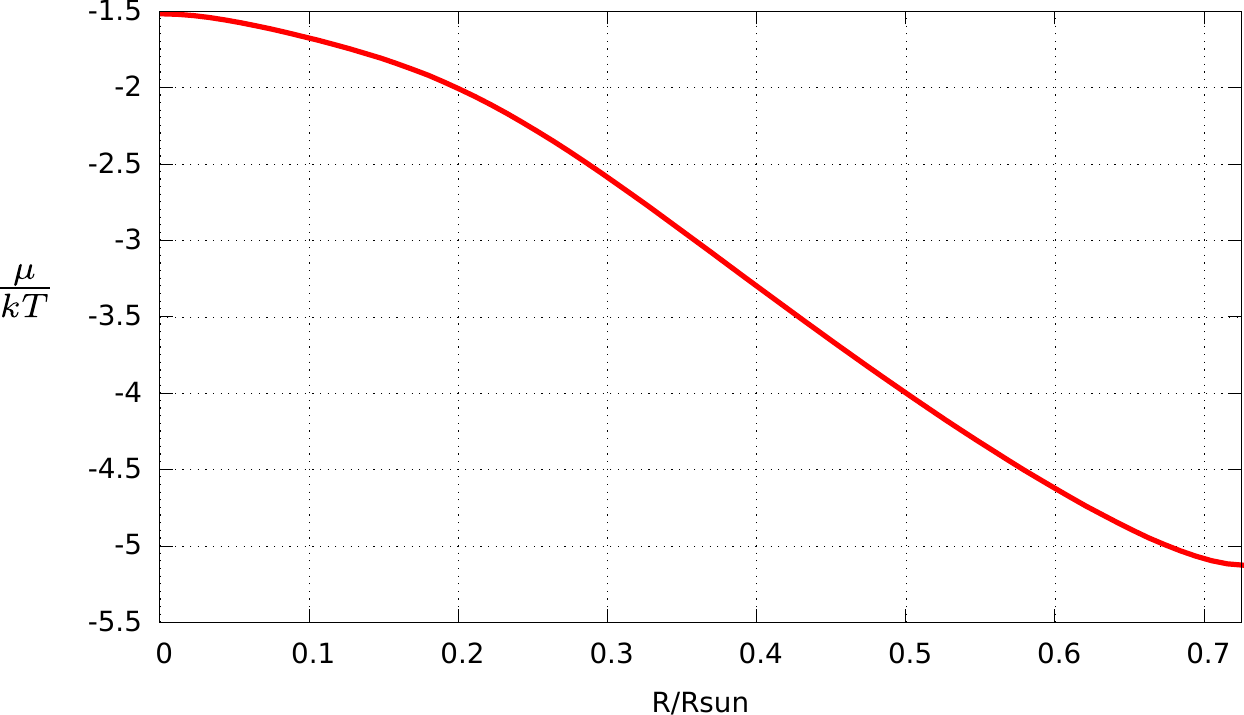}} 
	\caption{The solar mixture degeneracy parameter $ \mu/k_{B}T $ profile.}
	\label{fig:mu}
\end{figure}

\begin{figure}
		\centering
.		\resizebox{0.5\textwidth}{!}{\includegraphics{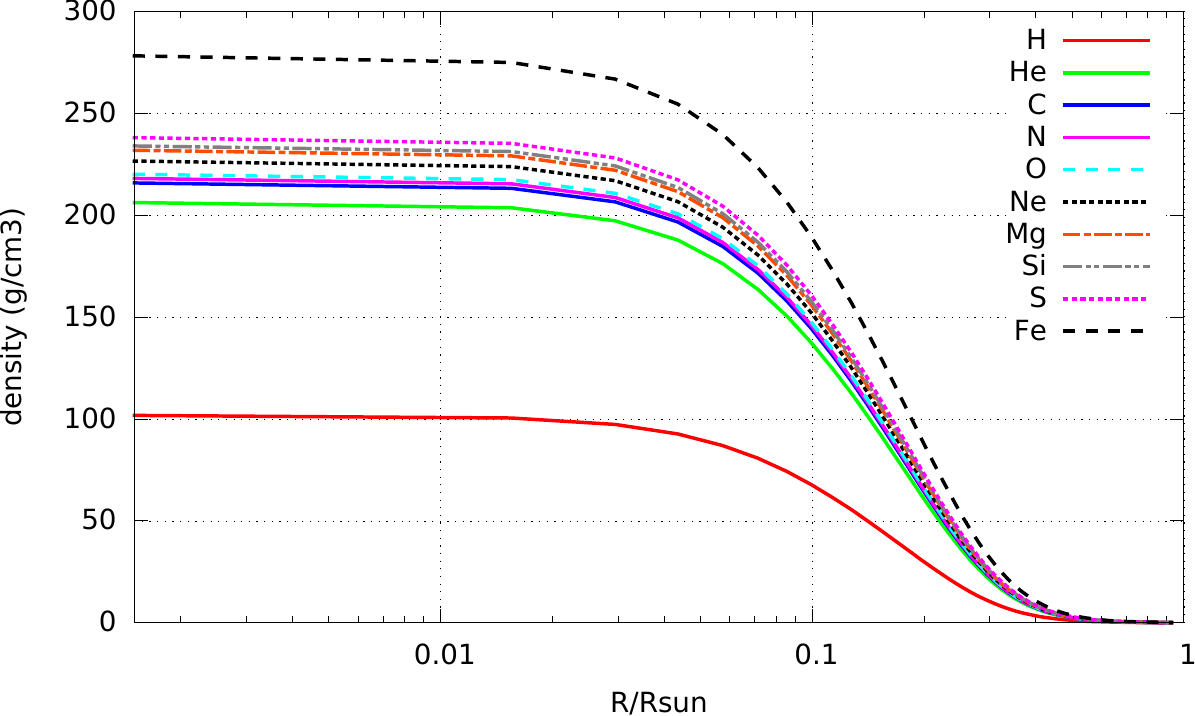}} 
	\caption{Effective densities for the 10 most abundant elements.}\label{fig:eff_densities}
\end{figure}

The chemical potential can be calculated using a finite temperature Thomas-Fermi model (\cite{feynman1949equations}) for which bound and free electrons are treated semiclasically, or alternatively, with more sophisticated ion-sphere models such as the relativistic Hartree-Fock-Slater average atom model for which bound-electrons are treated quantum-mechanically via the Dirac equation (\cite{Rozsnyai1972,iacob2006quantum}), or the Liberman model for which bound and free electrons are treated quantum-mechanically (\cite{liberman1979self,Wilson2006,murillo2013partial,penicaud2009average,ovechkin2014reseos}).
In this work we used the relativistic Hartree-Fock-Slater average atom model for the calculation of effective mass densities for the solar mixture.
 The solar mixture degeneracy parameter $ \eta=\mu/k_{B}T $ (where $ k_{B} $ is the Boltzmann's constant) is shown in \autoref{fig:mu}. It is evident that the degeneracy increases towards the center due to the high increase of matter density, and despite the (lower) increase in temperature. Effective mass densities of several elements are given in \autoref{fig:eff_densities}.

\begin{figure}[]
	\centering
	\resizebox{0.5\textwidth}{!}{\includegraphics{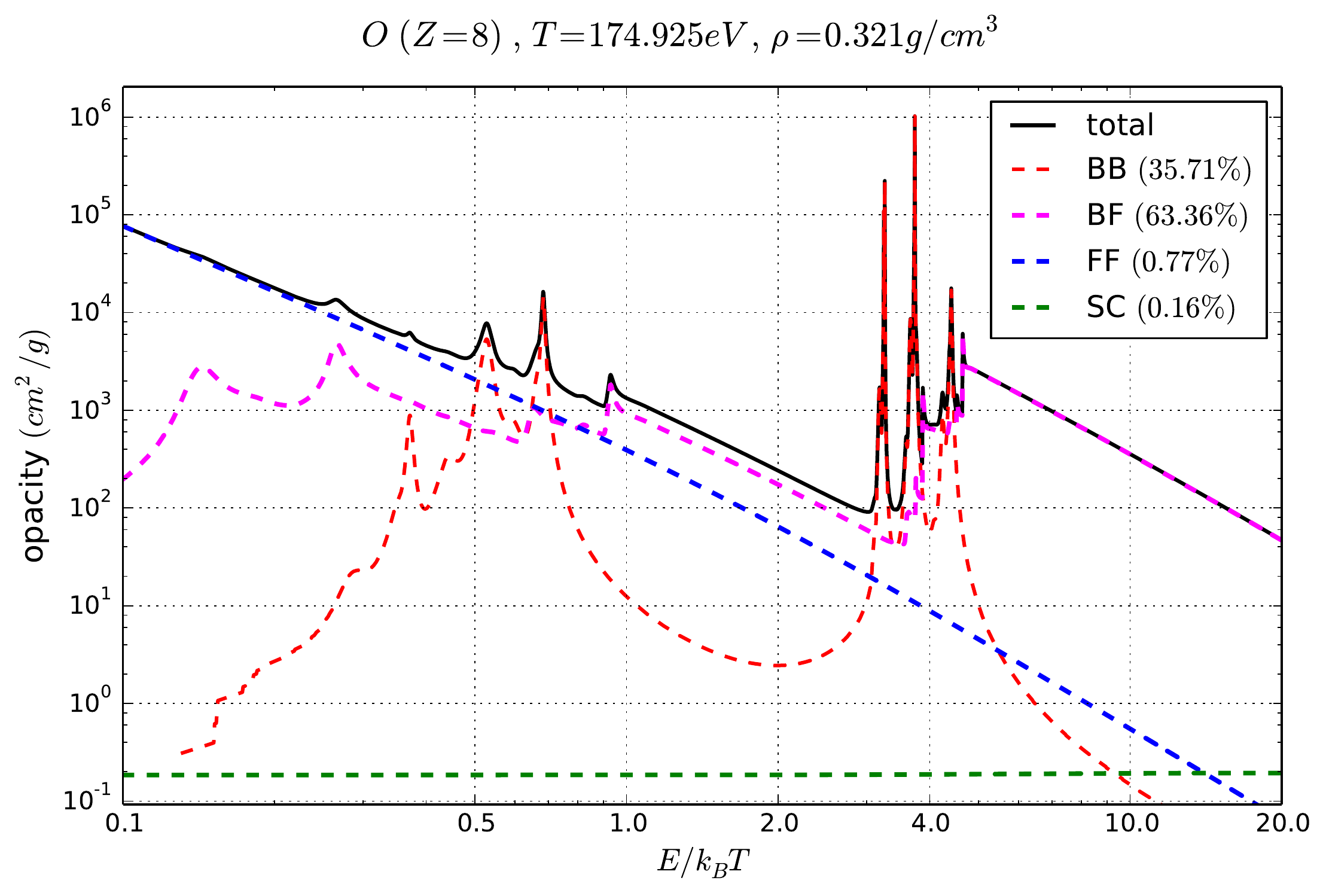}} 
	\resizebox{0.5\textwidth}{!}{\includegraphics{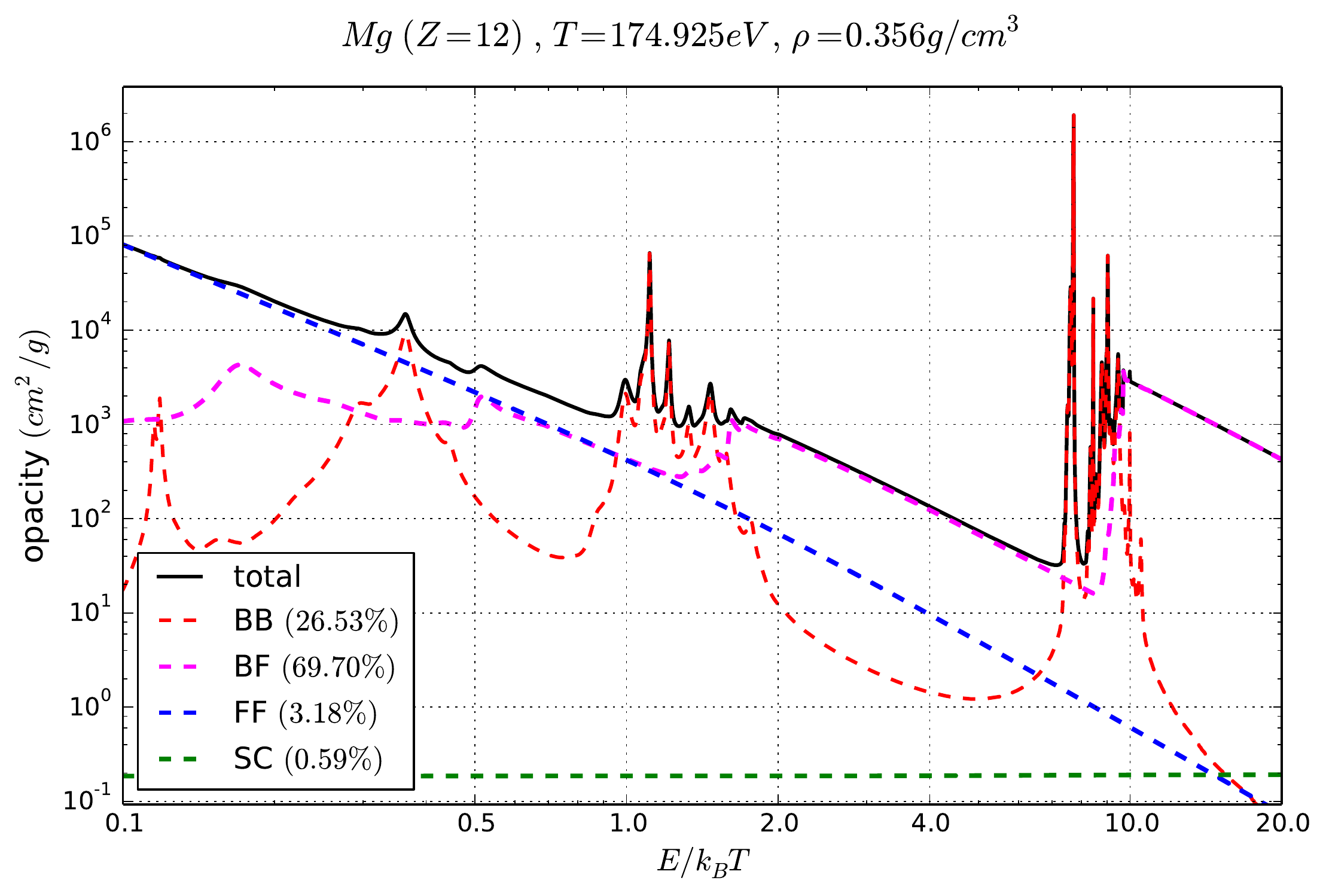}} 
	\resizebox{0.5\textwidth}{!}{\includegraphics{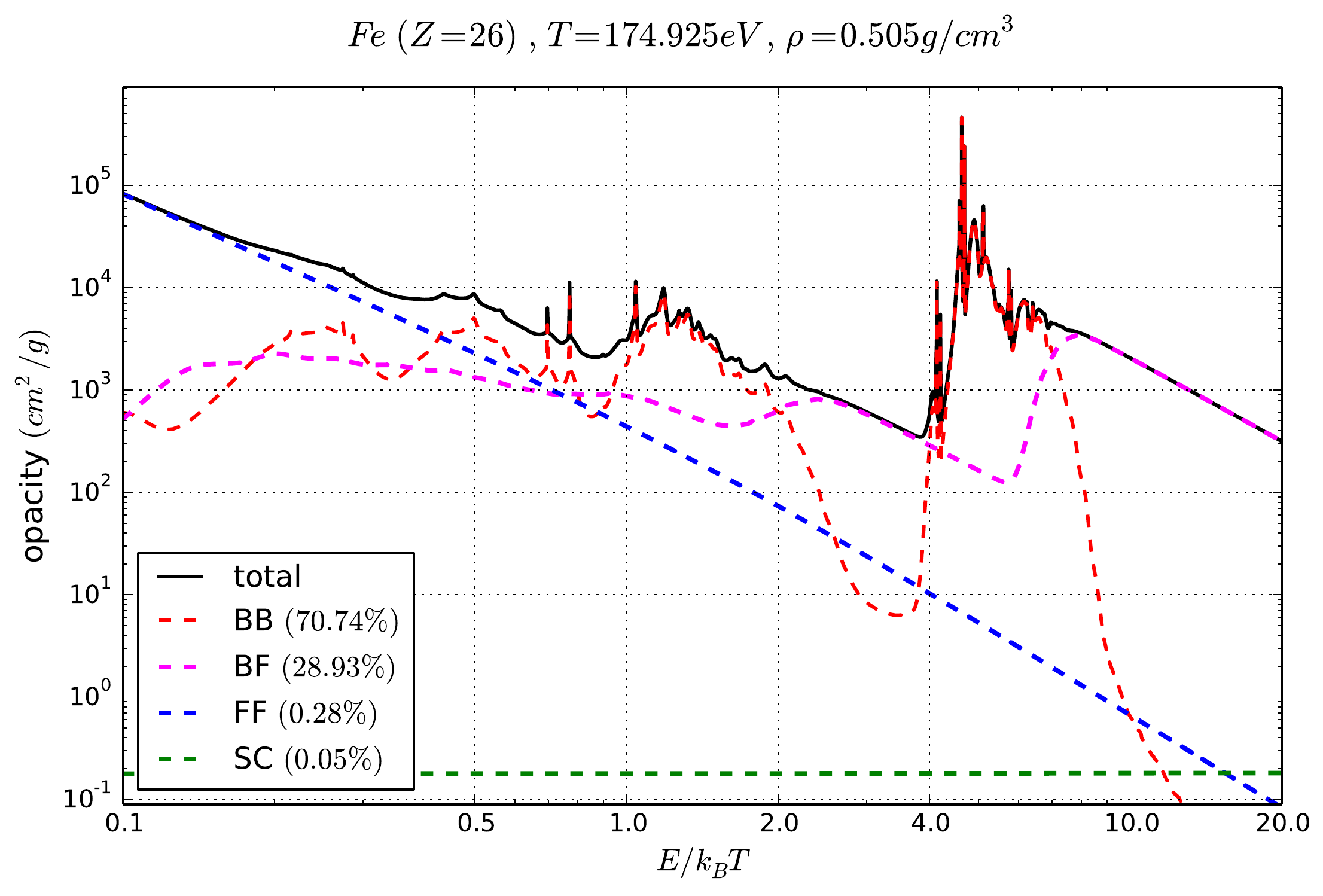}} 
	\caption{
		Total monochromatic opacity (solid line) and the atomic processes individual contributions (dashed lines) - bound-bound (BB), bound-free (BF), free-free (FF) and scattering (SC), at $ R=0.726$ (for which $ T=174.9eV $, $ \rho=0.16 g/cm^{3}$),  for oxygen (upper figure), magnesium (middle figure) and iron (lower figure). The relative contributions of the atomic processes to the Rosseland mean is given in the legends, and the effective mass densities are given in the titles.
		}
	\label{fig:specs_procs}
\end{figure}

The monochromatic opacity is evaluated as a sum of four different contributions involving photon scattering (SC) and the free-free (FF), bound-free (BF) and bound-bound (BB) photoabsorption processes. For each component of the mixture, the photon-absorption (including stimulated emission) cross-section is written as:
\begin{eqnarray}
\begin{gathered}
	\sigma_{i}(E)=
	\sigma_{sc}(E)+
	\\\big[ \sigma_{bb}(E)+\sigma_{bf}(E)+\sigma_{ff}(E)\big] (1-e^{-E/k_{B}T}).
\end{gathered}
\end{eqnarray}
The calculation of the cross-section of each element is carried at the effective density as explained in last paragraph.
The monochromatic opacity is given by the absorption coefficient per unit mass:
\begin{eqnarray}
\kappa_{i}(E)= \dfrac{N_{A}}{A_{i}}\sigma_{i}(E),
\end{eqnarray}
where $ A_{i} $ is the atomic mass of the $ i $ component and $ N_{A} $ is Avogadro's constant. The total monochromatic opacity of the mixture is given by the sum of the individual opacities, weighted by the mass fraction:
\begin{eqnarray}
\kappa(E)=\sum_{i}m_{i}\kappa_{i}(E).
\end{eqnarray}
Total monochromatic opacities, together with the contribution of individual atomic processes are shown in \autoref{fig:specs_procs}, for oxygen, magnesium and iron at the conditions of the radiation-convection boundary.
The Rosseland mean opacity is given by:
\begin{eqnarray}\label{eq:ross_mean_def}
\dfrac{1}{\kappa_{R}}=\int_{0}^{\infty}\dfrac{R(u)du}{\kappa(uk_{B}T)},
\end{eqnarray}
with the Rosseland weight function:
\begin{eqnarray}\label{eq:ross_function}
R(u)=\dfrac{15}{4\pi^{4}}\dfrac{u^{4}e^{u}}{(e^{u}-1)^{2}},
\end{eqnarray}
where $ u=E/k_{B}T $.

We adopt the following simple way to measure fractional contributions (i.e. the contribution of different elements or different atomic processes), to the total Rosseland mean.
Suppose an opacity spectrum is composed of $ N $ individual spectral contributions:
\begin{eqnarray}
\kappa_{tot}(E)=\sum_{i=1}^{N} \kappa_{i}(E).
\end{eqnarray}
We define the fractional Rosseland contributions by:
\begin{eqnarray}
\delta \kappa _{i}=\dfrac{\kappa_{R}^{i}-\kappa_{R}^{i-1}}{\kappa_{R}^{N}},
\end{eqnarray}
where $ \kappa_{R}^{i} $
is the Rosseland mean \eqref{eq:ross_mean_def} of the cumulative spectra $ \sum_{j=1}^{i}\kappa_{j}(E) $, while for the first contributer 
$\delta \kappa_{1}=\kappa_{R}^{1}/\kappa_{R}^{N} $.

\subsection{Photon Scattering}
In this work we calculate the photon scattering with several corrections as summarized in the book by
\cite{huebner2014opacity}.
The photon scattering cross section is written as:
\begin{eqnarray}
\sigma_{sc}(E)=G(u,T^{\prime })R(\eta ,T^{\prime })f(\eta ,\delta )Z_{f}\sigma_{T},
\end{eqnarray}
where $ Z_{f} $ is the average number of free electrons and the Thomson cross-section is 
\begin{eqnarray}
\sigma_{T}=\frac{8\pi }{3}\alpha ^{4}a_{0}^{2},
\end{eqnarray}
with $ \alpha $ the fine structure constant and $ a_{0} $ the Bohr radius.
$G(u,T^{\prime })$ corrects the Klein-Nishina cross-section due to the finite temperature and contains
corrections for inelastic scattering. It was calculated by \cite{sampson1959opacity} using
a relativistic Maxwell-Boltzmann distribution for \textit{nondegenerate} electrons. For nonrelativistic temperatures it is given by the expansion:
\begin{eqnarray}
\begin{gathered}
G(u,T^{\prime }) =1+2T^{\prime }+5T^{\prime 2}+\frac{15}{4}T^{\prime 3}\\
-\frac{1}{5}\left( 16+103T^{\prime }+408T^{\prime 2}\right) uT^{\prime } \\
+\left( \frac{21}{2}+\frac{609}{5}T^{\prime }\right) \left( uT^{\prime
}\right) ^{2}-\frac{2203}{70}\left( uT^{\prime }\right) ^{3},
\end{gathered}
\end{eqnarray}
where $ T^{\prime }=k_{B}T/m_{e}c^{2} $.
The factor $ R(\eta ,T^{\prime }) $ that was calculated by \cite{kilcrease2001plasma}, contains corrections due to the Pauli blocking as a result of partial degeneracy and includes a relativistic electron dispersion relation. For $ T'\ll 1 $ (which holds in the solar interior) it is given by:
\begin{eqnarray}
&& R(\eta ,T^{\prime })=R^{NR}(\eta )+
\frac{15T^{\prime
	}}{8}\left[ 1-R^{NR}(\eta )\frac{%
	I_{3/2}\left( \eta \right) }{I_{1/2}\left( \eta \right) }\right]
 \\&&
-\dfrac{345I_{3/2}(\eta)}{128I_{1/2}(\eta)} \left[1-R^{NR}(\eta)\left( \dfrac{30I_{3/2}(\eta)}{23I_{1/2}(\eta)}-\dfrac{7I_{5/2}(\eta)}{23I_{3/2}(\eta)}\right)  \right]
\notag  T'^{2},
\end{eqnarray}
where $ R^{NR} $ is the analogous non-relativistic correction due to \cite{rose1995effect}:
\begin{eqnarray}
R^{NR}(\eta )=\frac{I_{-1/2}\left( \eta \right) }{2I_{1/2}\left( \eta
	\right) },
\end{eqnarray}
and the Fermi-Dirac integral is defined by:
\begin{eqnarray}
I_{k}(x)=\int_{0}^{\infty}\dfrac{y^{k}dy}{e^{y-x}+1}.
\end{eqnarray}
Finally, 
$ f(\eta ,\delta ) $ is a correction due to plasma collective effects.
We use the formula of \cite{boercker1987collective} 
that is  based on electron correlations at high
densities and includes an exchange-correlation correction:
\begin{eqnarray}
&&f(\eta ,\delta )= 1+\frac{I_{1/2}\left( \eta \right) }{2^{3/2}}
\notag \\ && -R^{NR}(\eta )\left( \frac{\lambda _{D}}{%
	\lambda _{e}}\right) ^{2}\left( 1-\frac{I_{1/2}\left( \eta \right) }{2^{3/2}}%
\right) h(\delta).
\end{eqnarray}
The ion and electron Debye lengths are given by:
\begin{eqnarray}
\lambda _{i}=\left( \frac{\epsilon _{0}k_{B}T}{Z_{f}^{2}n_{i}e^{2}}\right)
^{1/2} \ ; \ \lambda_{e}=\left( \frac{\epsilon _{0}k_{B}T}{n_{e}e^{2}}\right)^{1/2}, 
\end{eqnarray}
where $ \epsilon_{0} $ is the vacuum dielectric constant and $ e $ is the unit electron charge, the total Debye length is:
\begin{eqnarray}
\frac{1}{\lambda _{D}}=\frac{1}{\lambda _{e}}+\frac{1}{\lambda_{i}},
\end{eqnarray}
and
\begin{eqnarray}\label{eq:scatter_h_unstable}
\begin{gathered}
h(\delta )=\frac{3\delta}{8} \Bigg[ \left( \delta ^{3}+2\delta ^{2}+2\delta
\right) \ln \left( \frac{\delta }{2+\delta }\right) \\+2\delta ^{2}+2\delta +%
\frac{8}{3}\Bigg],
\end{gathered}
\end{eqnarray}
with:
\begin{eqnarray}
\delta =\frac{1}{2}\left( \frac{\hbar c}{\lambda _{D}E }\right)^{2}.
\end{eqnarray}
We note that for long photon wavelengths where $\delta \gg 1 $, a direct numerical calculation of Eq. \eqref{eq:scatter_h_unstable} can be unstable. In practice, for $\delta >1000$ it is preferable to use an expansion at $ \delta=\infty $:
\begin{eqnarray}
h(\delta )\approx 1-\frac{7}{5\delta }+\frac{11}{5\delta ^{2}}-\frac{128}{35\delta
	^{3}}+\frac{44}{7\delta ^{4}}-\frac{232}{21\delta ^{5}}.
\end{eqnarray}

\subsection{Free-Free}
\begin{figure}
		\centering
	\resizebox{0.5\textwidth}{!}{\includegraphics{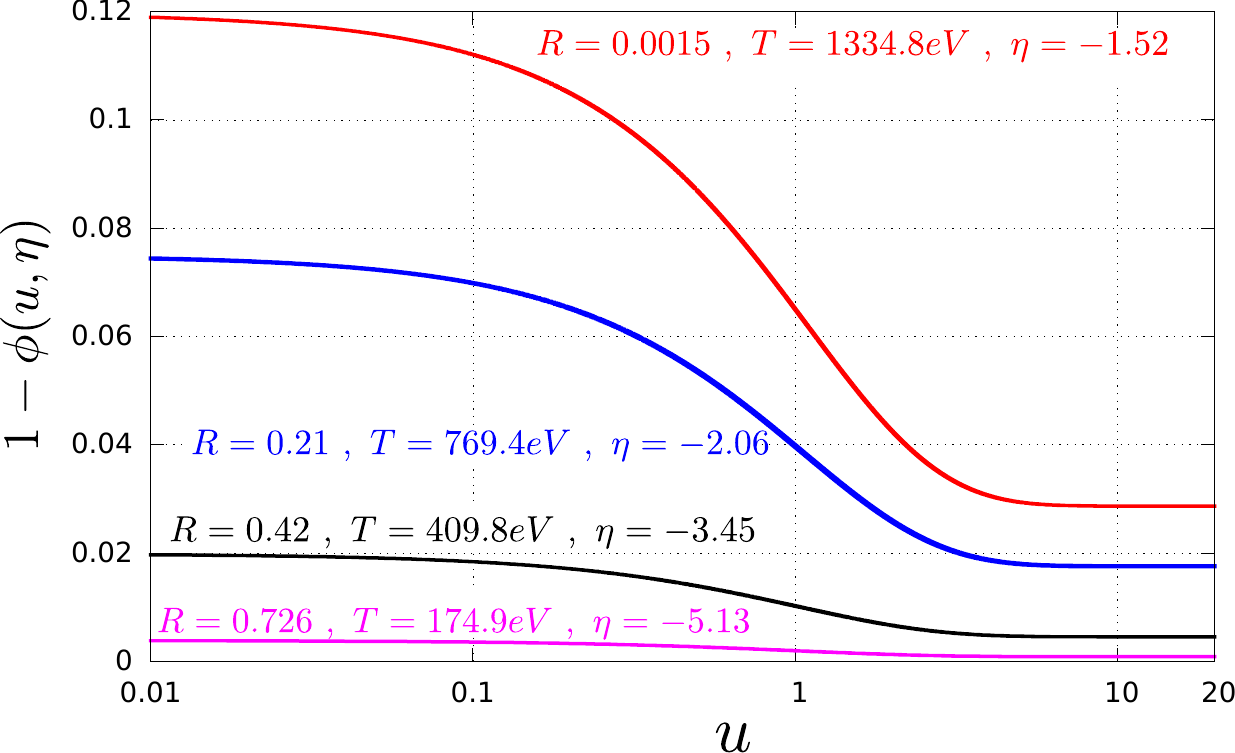}} 
	\caption{Free-free degeneracy correction $ 1-\phi(u,\eta)$  at several solar radii.}
	\label{fig:ff_deg}
\end{figure}

In this work, the free-free photoabsorption is calculated via a screened-hydrogenic approximation with a multiplicative degeneracy correction.
The free-free absorption cross section is given by:
\begin{eqnarray}
\sigma_{ff}(E)=\sigma _{K}(E)\phi(u,\eta)\bar{g}_{ff}, 
\end{eqnarray}
where 
the Kramers cross-section (\cite{kramers1923xciii}) is:
\begin{eqnarray}
\sigma_{K}(E)=
\frac{16\pi ^{2}\hbar^{2}e^{6} }{3m_{e}^{2}c}\left( \frac{2\pi m_{e} }{3k_{B}T}\right) ^{1/2}\dfrac{Z_{f}^{2}n_{e}}{E^{3}},
\end{eqnarray}
with $ n_{e} $ the free electron number density.
To allow a reasonably accurate and fast calculation of the (thermal) averaged free-free Gaunt factor 
$ \bar{g}_{ff} $,
we use a screened-hydrogenic approximation and a Maxwellian distribution via the tables of \cite{van2014accurate}. Degeneracy is incorporated using the correction factor (\cite{green1960fermi,rose1993degeneracy,faussurier2015refractive}):
\begin{eqnarray}
\phi(u,\eta)=\dfrac{\sqrt{\pi}}{2I_{1/2}(\eta)(1-e^{-u})}\ln\left(\dfrac{1+e^{\eta}}{1+e^{\eta-u}} \right).
\end{eqnarray}
We note that since $ \phi $ depends only on the temperature and the chemical potential, the degeneracy correction is the same for all components in the mixture. The effect of degeneracy at different solar radii is presented in \autoref{fig:ff_deg}. It is evident that the correction is vital at the core vicinity, as the plasma is more degenerate (see also \autoref{fig:mu}).

\subsection{Bound-Bound}\label{ssec:bound_bound}
In principle, the bound-bound photoabsorption spectra results from all radiative transitions between all levels from all pairs of electronic configurations. Unfortunately, for a hot dense plasma, the number of lines between each pair of configurations may be enormous (\cite{scott2010advances,gilleron2009efficient}) and statistical methods must be used. The Unresolved-Transition-Array (UTA) method (\cite{moszkowski1962energy,bauche1979variance,bauche1988transition, krief2015variance}) treats all levels between each pair of configurations statistically, using analytic expressions for the energy moments of the transition array. In many cases, the number of configurations may also be extremely large and a detailed-configuration-accounting (DCA) calculation is intractable as well.

\begin{figure}
	\centering
	\resizebox{0.5\textwidth}{!}{\includegraphics{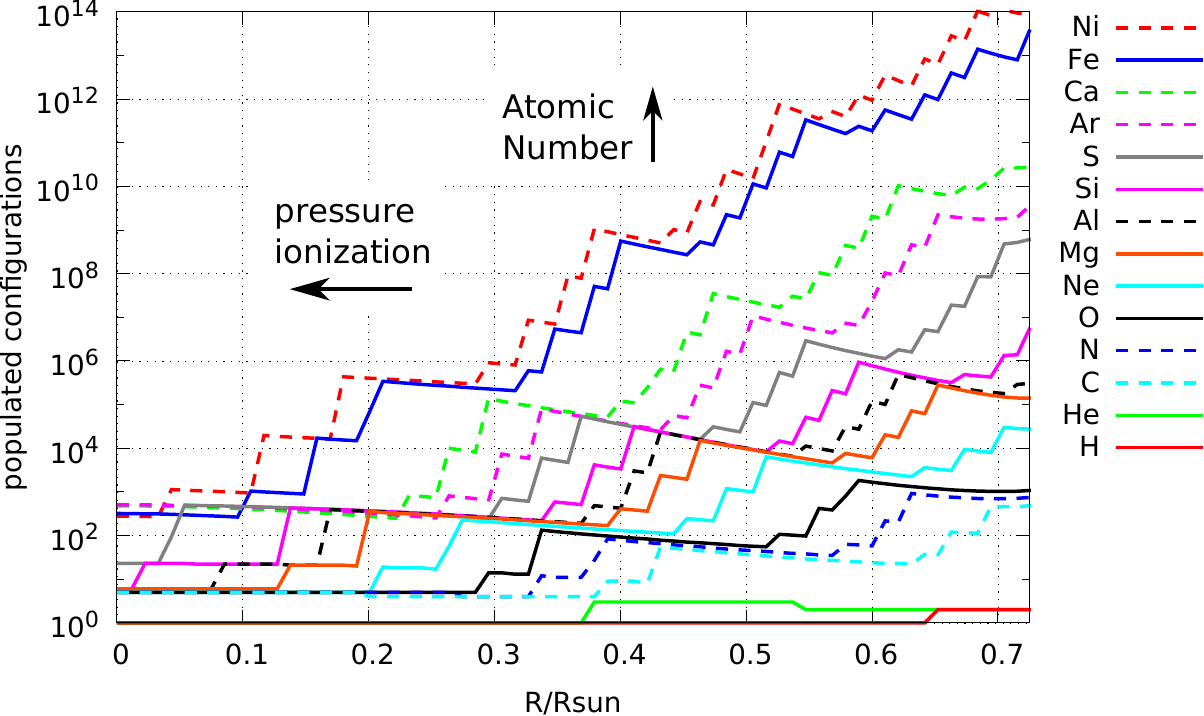}} 
	\caption{Number of populated electronic configurations $ \mathcal{N}_{C} $ for various elements, across the radiation zone.}
	\label{fig:number_configs}
\end{figure}

Let us estimate the number of populated configurations. A (relativistic) configuration is defined by a set of occupation numbers  $ \left\lbrace q_{s} \right\rbrace $ on relativistic $ s=(nlj) $ orbitals, which are full solutions of the Dirac equation. Neglecting  electron correlations, the probability distribution of $ \left\lbrace q_{s} \right\rbrace $ is binomial (\cite{iacob2006quantum}):
\begin{eqnarray}\label{eq:PC_BINOMIAL}
P\left(  \left\lbrace q_{s} \right\rbrace \right) =\prod_{s}\binom{g_{s}}{q_{s}}n_{s}^{q_{s}}(1-n_{s})^{g_{s}-q_{s}},
\end{eqnarray}
where 
$ n_{s}=1/(e^{-(\epsilon_{s}-\mu)/k_{B}T} +1) $ is the Fermi-Dirac distribution, $ g_{s}=2j_{s}+1 $ is the orbital degeneracy and $ \epsilon_{s} $ is the orbital energy.
The variance of the population in each shell is given by:
\begin{eqnarray}
\delta q_{s}=\sqrt{\left\langle (q_{s}-\overline{q_{s}})^{2} \right\rangle}=\sqrt{g_{s}n_{s}(1-n_{s})},
\end{eqnarray}
so that the occupation numbers of shells whose energies are near the Fermi-Dirac step
$ \left| \epsilon_{s}-\mu \right|\approx k_{B}T  $
fluctuate, while the other shells are either filled or empty. The number of populated configurations $ \mathcal{N}_{C}  $ can be estimated by the "width" of the multivariate distribution \eqref{eq:PC_BINOMIAL}:
\begin{eqnarray} \label{eq:num_config_binomial}
\mathcal{N}_{C} \approx \prod_{s} \left( 6\delta q_{s}+1\right). 
\end{eqnarray}
A calculation of $\mathcal{N}_{C} $ for various elements in the solar mixture is shown in \autoref{fig:number_configs}.
 The sharp jumps are a direct result of the sudden decrease in the number of bound orbitals as they dissolve into the continuum. This is the process of pressure ionization which is more dominant towards the solar core as the density increases. It is evident that the number of configurations grows exponentially with the atomic number. Near the convection-radiation boundary, $ \mathcal{N}_{C}>10^{8} $ for atomic numbers $ Z>15 $ and a detailed configuration accounting (DCA) calculation is very costly, while a full detailed-line-accounting (DLA) calculation is probably impossible.

\begin{figure}
	\centering
	\resizebox{0.5\textwidth}{!}{\includegraphics{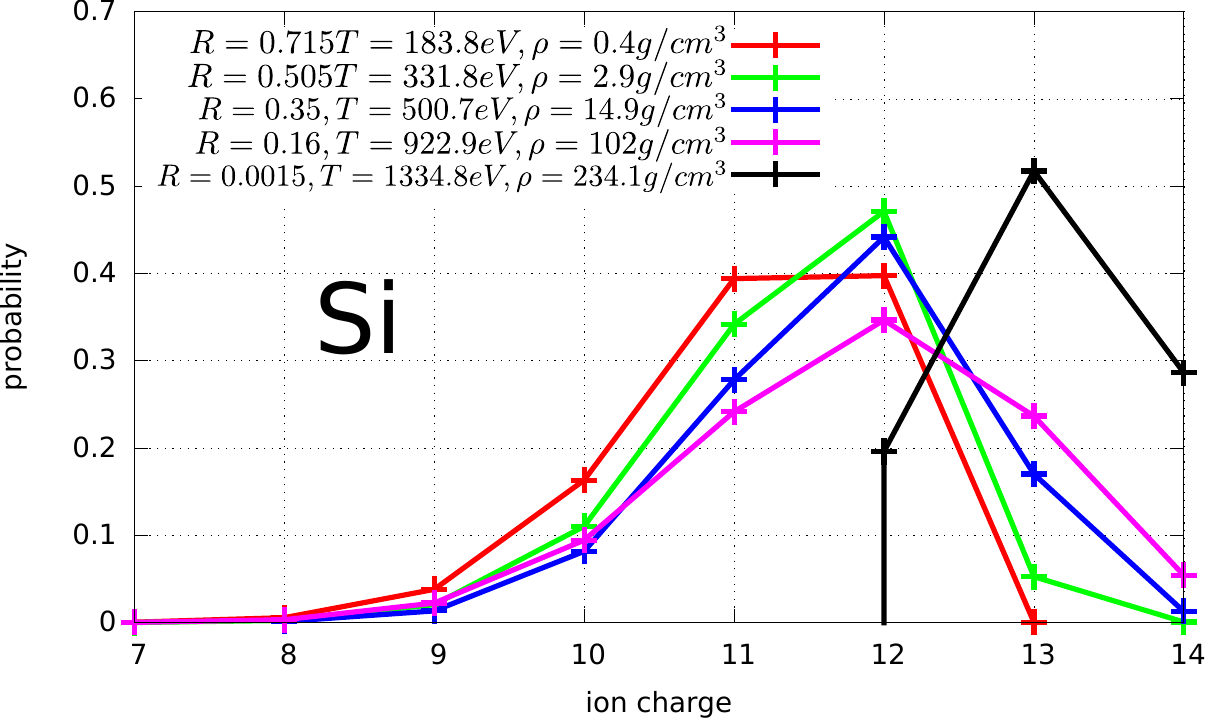}} 
	\resizebox{0.5\textwidth}{!}{\includegraphics{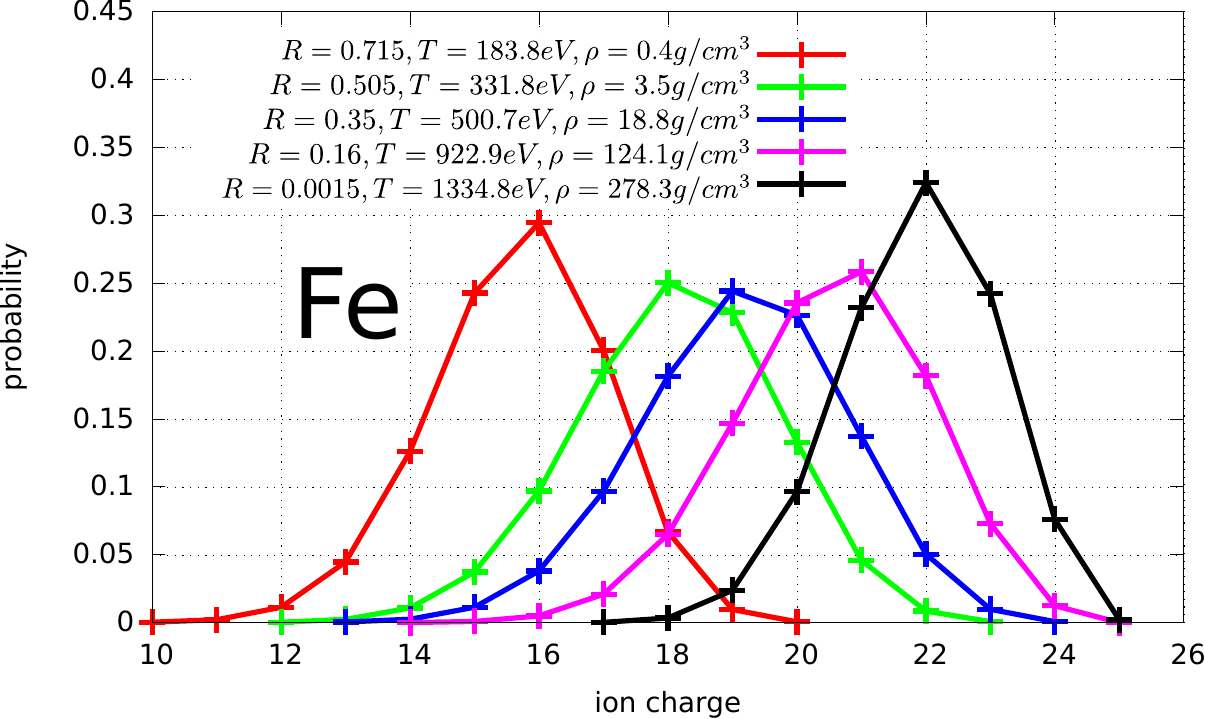}} 
	\caption{Ion charge distributions for iron (lower figure) and silicon (upper figure) at several solar radii.}
	\label{fig:pq}
\end{figure}

\begin{figure}
	\centering
	\resizebox{0.5\textwidth}{!}{\includegraphics{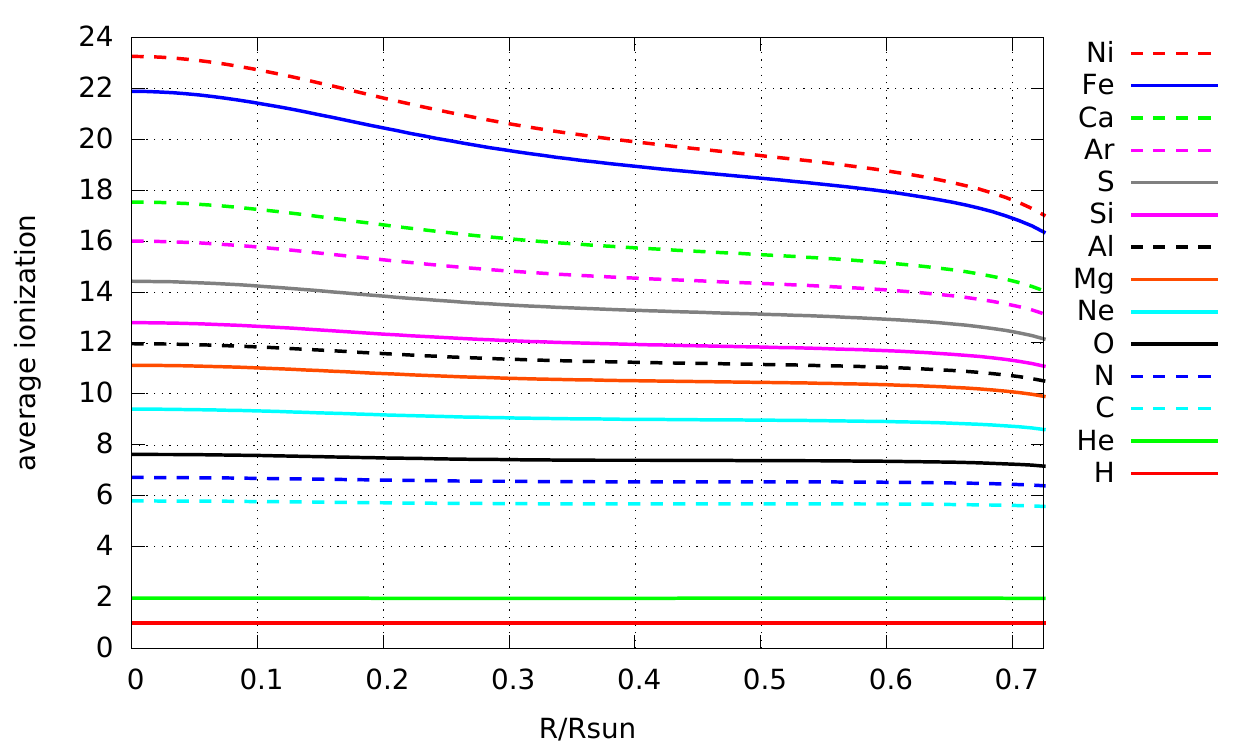}} 
	\caption{Average ionization for various elements, throughout the radiation zone.}
	\label{fig:elements_zf}
\end{figure}

The bound-bound photoabsorption cross-section is calculated by the STA method (\cite{BarShalom1989,blenski1997hartree,hazak2012configurationally}). We follow the notation of \cite{BarShalom1989} for which relativistic  $ s=(nlj) $ orbitals are grouped into \textit{supershells} $ \sigma=\prod_{s\in\sigma}(s) $. Configurations are collected into large groups called \textit{superconfigurations} (SC), defined by a set of occupation numbers $ \left\lbrace Q_{\sigma} \right\rbrace $ over the supershells and denoted by $ \Xi=\prod_{\sigma}(\sigma)^{Q_{\sigma}} $. 
For each orbital jump $ \alpha\rightarrow\beta $ and SC, all transitions between levels in $ \Xi $ to levels in $ \Xi' $ (which is obtained from all configurations in $ \Xi $ via the electron jump), are grouped into a single Super-Transition-Array (STA) whose first three moments, the intensity, average energy and variance, are respectively:
\begin{subequations}\label{eq:moments_STA_def}	
	\begin{eqnarray}\label{eq:sta_intensity_def}
	I_{\Xi}^{\alpha\beta}=\sum_{\substack{C\in \Xi \\ C'\in \Xi'}}\sum_{\substack{i\in C \\ j\in C'}}I_{ij},
	\end{eqnarray} 
	\begin{eqnarray}\label{eq:sta_energy_def}
	E_{\Xi}^{\alpha\beta}=\sum_{\substack{C\in \Xi \\ C'\in \Xi'}}\sum_{\substack{i\in C \\ j\in C'}}\dfrac{I_{ij}}{I_{\Xi}^{\alpha\beta}}(E_{j}-E_{i}),
	\end{eqnarray}
	\begin{eqnarray}\label{eq:sta_variance_def}
	\left(\Delta^{\alpha\beta}_{\Xi}\right)^{2} =\sum_{\substack{C\in \Xi \\ C'\in \Xi'}}\sum_{\substack{i\in C \\ j\in C'}}\dfrac{I_{ij}}{I_{\Xi}^{\alpha\beta}}\left( E_{j}-E_{i}-E_{\Xi}^{\alpha\beta}\right) ^{2},
	\end{eqnarray}
\end{subequations}
where  $ E_{i} $ is the energy of level $ i $ and the  line $ i\rightarrow j $ intensity is $ I_{ij}=N_{i}\sigma_{ij} $ with $ N_{i} $ the population of level $ i $ and $ \sigma_{ij} $ the line cross-section. The moments \eqref{eq:moments_STA_def} can be expressed as SC averages of occupation number polynomials that can be written in terms of generalized partition functions and evaluated numerically using recursion relations (\cite{BarShalom1989,wilson1999revised,gilleron2004stable,wilson2007further}), 	 known as the \textit{partition function algebra}.
The ionic distribution
\begin{eqnarray}
P_{Q}=\sum_{\substack{\Xi \ with: \\\sum_{\sigma}Q_{\sigma}=Q}}\sum_{C\in\Xi}P_{C},
\end{eqnarray}
where $ Q $ is the number of bound electrons and $ P_{C} $ is the population of the configuration $ C $, is also calculated via the partition function algebra. Ion distributions for iron and oxygen are given in \autoref{fig:pq} and the average ionization for various chemical elements as a function of the solar radius are given in 
\autoref{fig:elements_zf}.
It is evident that the ionization increases towards the solar core as a result of the higher temperature and pressure ionization.

The bound-bound spectra contains the contribution 
of all line profiles
from each orbital jump from all SCs. In this work we use Voigt profile, and the total bound-bound spectra is written as:
\begin{eqnarray}\label{eq:SPEC_STA_VOIGTS}
\sigma_{bb}(E) =\sum_{\alpha \rightarrow \beta}\sum_{\Xi}I^{\alpha  \beta}_{\Xi}
V\left( E-E^{\alpha \beta}_{\Xi},\sigma^{\alpha\beta}_{\Xi},\gamma_{\alpha\beta}\right).
\end{eqnarray}
The Voigt profile $ V $ is defined by the convolution of the Gaussian and Lorentzian profiles:
\begin{eqnarray}\label{eq:voigt_convolution}
V\left(E,\sigma,\gamma\right)=
\int_{-\infty}^{\infty} dE' \dfrac{1}{\sqrt{2\pi}\sigma}e^{ - E'^{2}/2\sigma^{2}} \\ \times \dfrac{1}{\pi}\dfrac{\gamma/2}{(E-E')^{2}+\left( \gamma/2\right) ^{2}}.
\end{eqnarray}
The total Gaussian width
\begin{eqnarray}
\sigma^{\alpha\beta}_{\Xi}=\sqrt{\left( \Delta^{\alpha  \beta}_{\Xi}\right) ^{2}+D_{\alpha\beta}^{2}}
\end{eqnarray} includes the Doppler width $ D_{\alpha\beta} $ and the statistical width $ \Delta^{\alpha  \beta}_{\Xi} $ resulting from the fluctuations in the occupation numbers of the various configurations in $ \Xi $ and from the UTA widths (\cite{bauche1985variance,bar1995effect,krief2015variance}).
 $ \gamma_{\alpha\beta} $ is the total Lorentz width due to the natural and Stark broadening effects. We use the well known semi-empirical formulas for electron impact broadening in the one-perturber approximation by \cite{dimitrijevic1980stark,dimitrijevic1987simple}.
 
 In the STA calculation, supershells are split recursively, giving rise to a larger number of SCs and a larger number of STAs in \autoref{eq:SPEC_STA_VOIGTS}. This procedure is repeated until the Rosseland mean opacity (or alternatively, any other criteria) has converged. For mid or high-Z hot dense plasmas, the convergence is very fast since the spectral lines are broadened and significantly overlap \cite{BarShalom1989}. We  note that the supershells are split according to the fluctuations in their occupation numbers $ \delta Q_{\sigma}=\sum_{s\in\sigma}\delta q_{s} $, which allows a very robust and fast convergence.
  
\subsection{Bound-Free}
The bound-bound STA formalism is straightforwardly generalized for bound-free transitions (\cite{bar1996photoelectric}) which is presented briefly in this subsection. The generalization is obtained by treating the continuum orbitals via the Fermi-Dirac statistics while not including them in the supershell structure. The total bound-free cross section contains contributions due to photoionization from all bound orbitals $ \alpha $:
\begin{eqnarray}
	\sigma_{bf}(E)=\sum_{\alpha }\Gamma ^{\alpha }(E),
\end{eqnarray}
where $ \Gamma ^{\alpha }(E) $ is given by the convolution:
\begin{eqnarray}
	\Gamma ^{\alpha }(E)=\int_{0}^{\infty }d\epsilon G^{\alpha }(E-\epsilon )M^{\alpha
		}\left( \epsilon \right),
\end{eqnarray}
over the free orbital energy $ \epsilon $.
$ 	G^{\alpha } $ is given by the sum
\begin{eqnarray}
		G^{\alpha }(E-\epsilon )=\sum_{\Xi }G_{\Xi }^{\alpha }(E-\epsilon ),
\end{eqnarray}
of all individual STA profiles:
\begin{eqnarray}
	G_{\Xi }^{\alpha }(E-\epsilon )=A_{\Xi }^{\alpha }V\left( \left[
	E-E_{\Xi }^{\alpha }\right] -\epsilon ,\sigma_{\Xi}^{\alpha},\gamma _{\alpha }\right),
\end{eqnarray}
centered at the energy $ E_{\Xi }^{\alpha } +\epsilon $ and include a Gaussian width
\begin{eqnarray}
\sigma_{\Xi}^{\alpha}=\sqrt{D_{\alpha }^{2}+\left(
	\Delta _{\Xi }^{\alpha }\right) ^{2}},
\end{eqnarray}
where $ D_{\alpha }$, $ \Delta _{\Xi }^{\alpha }  $ and $\gamma_{\alpha} $ are respectively the Doppler, STA and Lorentzian widths. The profile intensity is $ A_{\Xi }^{\alpha }=N_{\Xi }^{(1)}\left\langle q_{\alpha }\right\rangle _{\Xi
} $, where $ N_{\Xi }^{(1)} $ is the total SC population and $ \left\langle q_{\alpha }\right\rangle _{\Xi
}  $ is the SC average of the number of electrons in orbital $ \alpha $. These moments are easily calculated via the partition function algebra (\cite{oreg1997operator}) analogously to the bound-bound case.
$ M^{\alpha }\left( \epsilon \right) $ describes the total coupling of the orbital $ \alpha $ to the continuum via photon absorption:
\begin{eqnarray}\label{eq:m_bf_sum_kappa}
M^{\alpha }\left( \epsilon \right)=\sum_{lj }M_{\alpha,\epsilon lj}\left(2j+1 \right) \left( 1-n_{\epsilon }\right),
\end{eqnarray}
where the summation is over the total and spatial angular momentum quantum numbers of the free orbital $ (\epsilon lj) $
and the  coefficient $ \left(2j+1 \right)\left( 1-n_{\epsilon }\right) $ is the average number of available "holes" in the free orbital. 
$ M_{\alpha,\epsilon lj} $ is the total cross-section for the transition $ \alpha\rightarrow(\epsilon lj) $
and is calculated via the relativistic expressions of \cite{grant1970relativistic}. The sum in \eqref{eq:m_bf_sum_kappa} is restricted by the selection rules of the transition.

\section{solar opacity calculations}
The opacity at the solar core calculated with a variety of atomic models was compared by \cite{rose2001radiative,rose2004set}.
Solar opacities by the widely used OPAL 
(\cite{rogers1992radiative,rogers1995opal,iglesias1996updated,iglesias2015iron})
and OP 
(\cite{seaton1994opacities,badnell2005updated})
codes were compared in detail by
\cite{seaton2004comparison}.
Additional calculations and comparisons with the LEDCOP and ATOMIC codes can be found in
\cite{neuforge2001helioseismic,colgan2013light,colgan2016new}.
Average atom solar opacity calculations were presented by \cite{rozsnyai2001solar} and \cite{wang2004electronic}.
 A recent analysis of the solar opacity was carried by \cite{blancard2012solar} and \cite{mondet2015opacity}, using the OPAS detailed configuration accounting model.

\begin{figure}
	\centering
	\resizebox{0.5\textwidth}{!}{\includegraphics{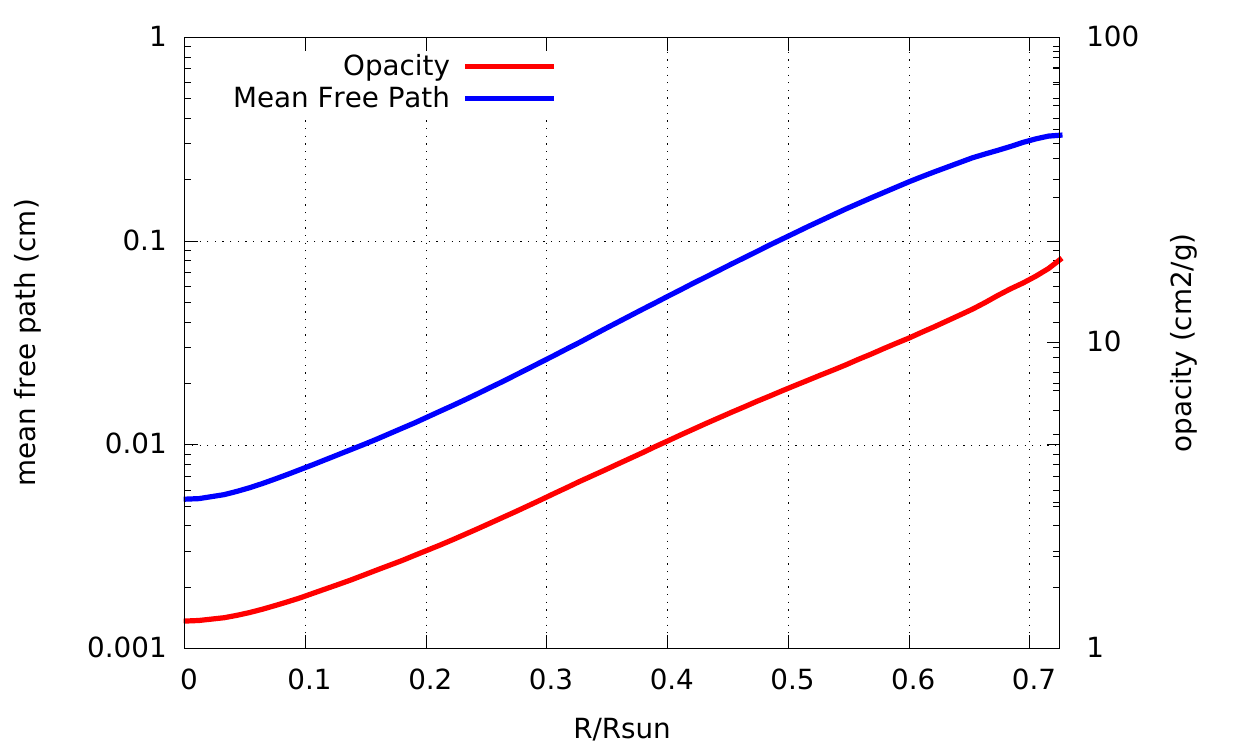}}
	\caption{The Rosseland mean opacity (red curve) and mean-free-path (blue curve) calculated by the STAR atomic code, for the solar mixture across the radiation zone.}
	\label{fig:krad_mfp}
\end{figure}

In this section we analyze the solar opacity  throughout the radiative zone, calculated by STAR. The resulting Rosseland opacity and mean free path $ l_{R}=1/\rho\kappa_{R} $ profiles, are shown in \autoref{fig:krad_mfp}. We note that even though the solar mixture is more opaque at larger radii, the mean free path is larger as well, due to the (approximate) exponential decrease in density (see \autoref{fig:rad_t_rho} and \autoref{fig:eff_densities}).

\subsection{Relative contribution of different elements along the radiation zone}
\begin{figure}[]
	\centering
	\resizebox{0.5\textwidth}{!}{\includegraphics{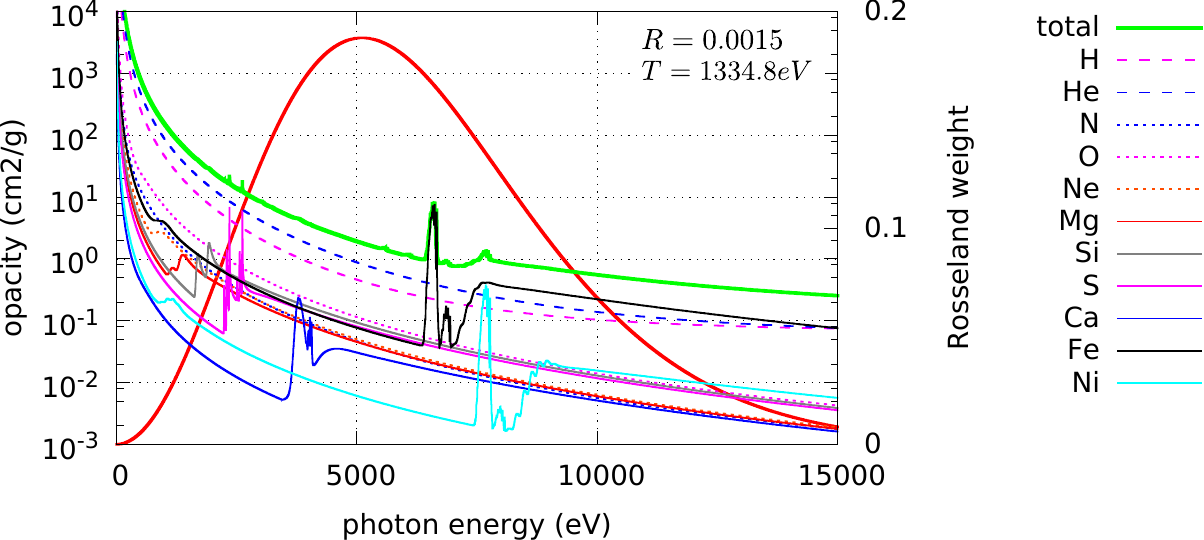}} 
	\resizebox{0.5\textwidth}{!}{\includegraphics{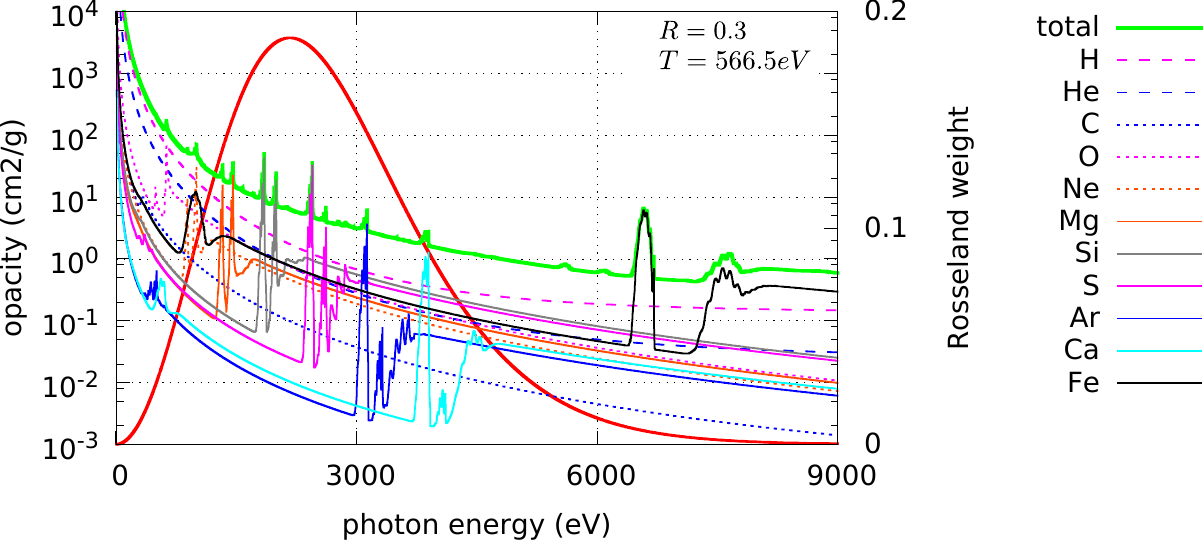}} 
	\resizebox{0.5\textwidth}{!}{\includegraphics{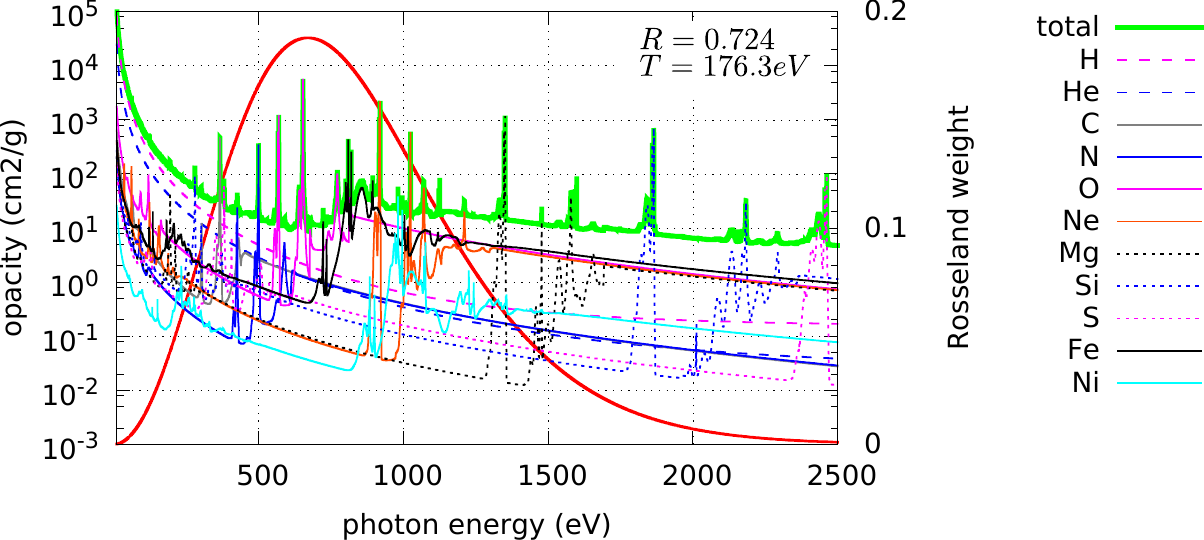}} 
				\caption{The total spectral opacity $ \kappa(E) $ (green solid curve), the elemental spectral contributions $ m_{i}\kappa_{i}(E) $  and the Rosseland weight (red solid curve), at three different solar radii.
					Only elements that contribute more than $ 1\% $ to the Rosseland mean are shown (see \autoref{fig:elements_ktot_fracs}).
					The solar radius and temperature are indicated in each figure.}
				\label{fig:spec_elements}
\end{figure}

\begin{figure}[]
	\centering
	\resizebox{0.5\textwidth}{!}{\includegraphics{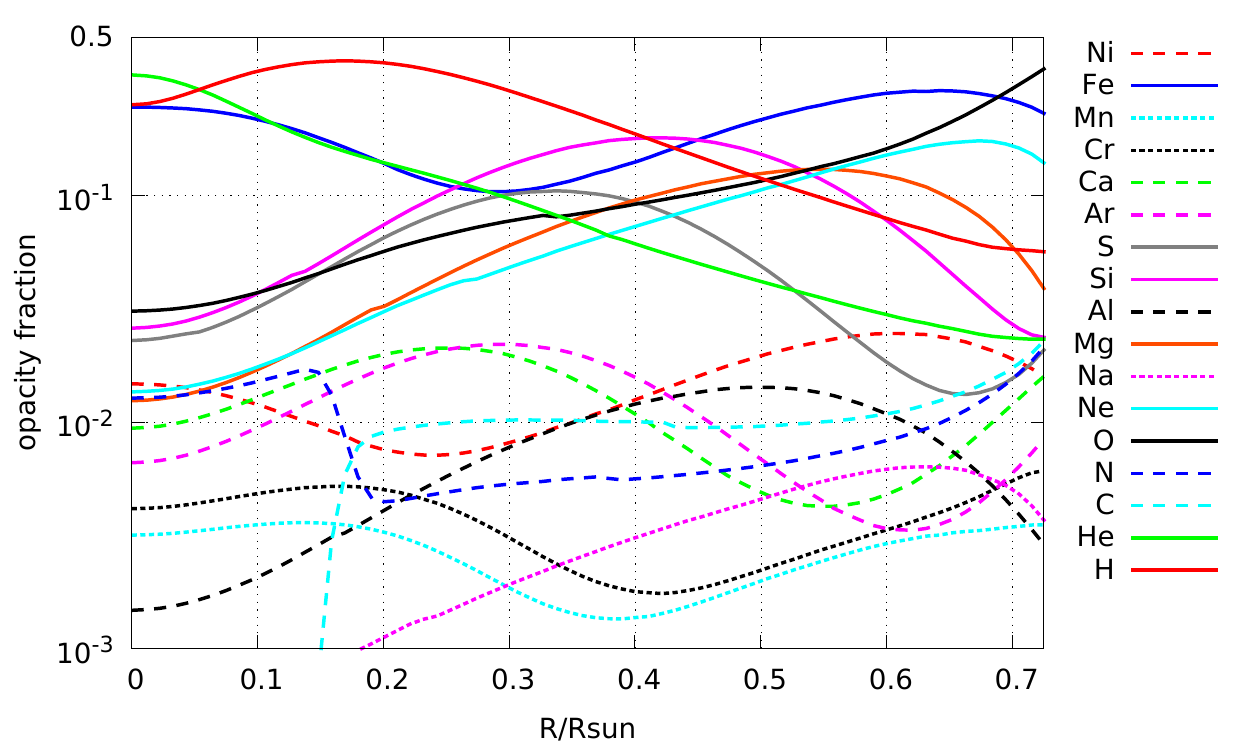}} 
	\caption{Relative contributions of the 17  elements with the largest contribution to the total solar Rosseland mean opacity, across the radiation zone.}
	\label{fig:elements_ktot_fracs}
\end{figure}

\begin{figure}[]
	\centering
	\resizebox{0.5\textwidth}{!}{\includegraphics{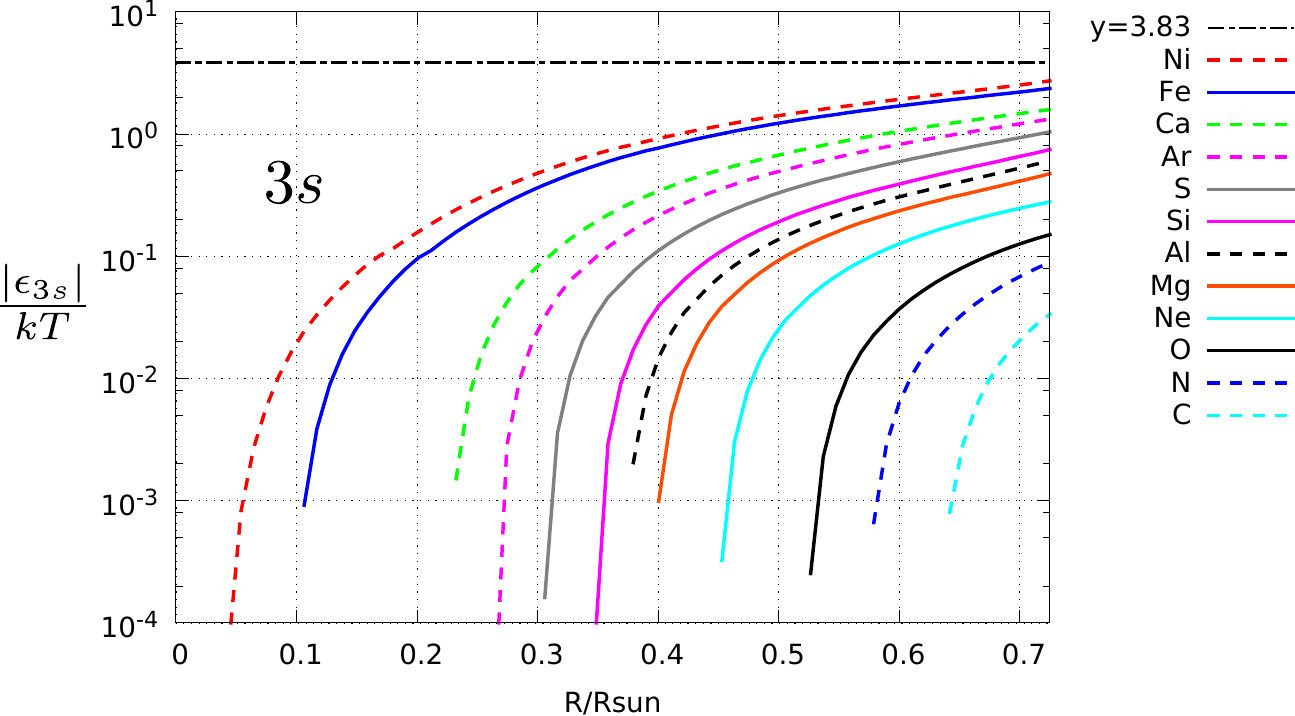}} 
	\resizebox{0.5\textwidth}{!}{\includegraphics{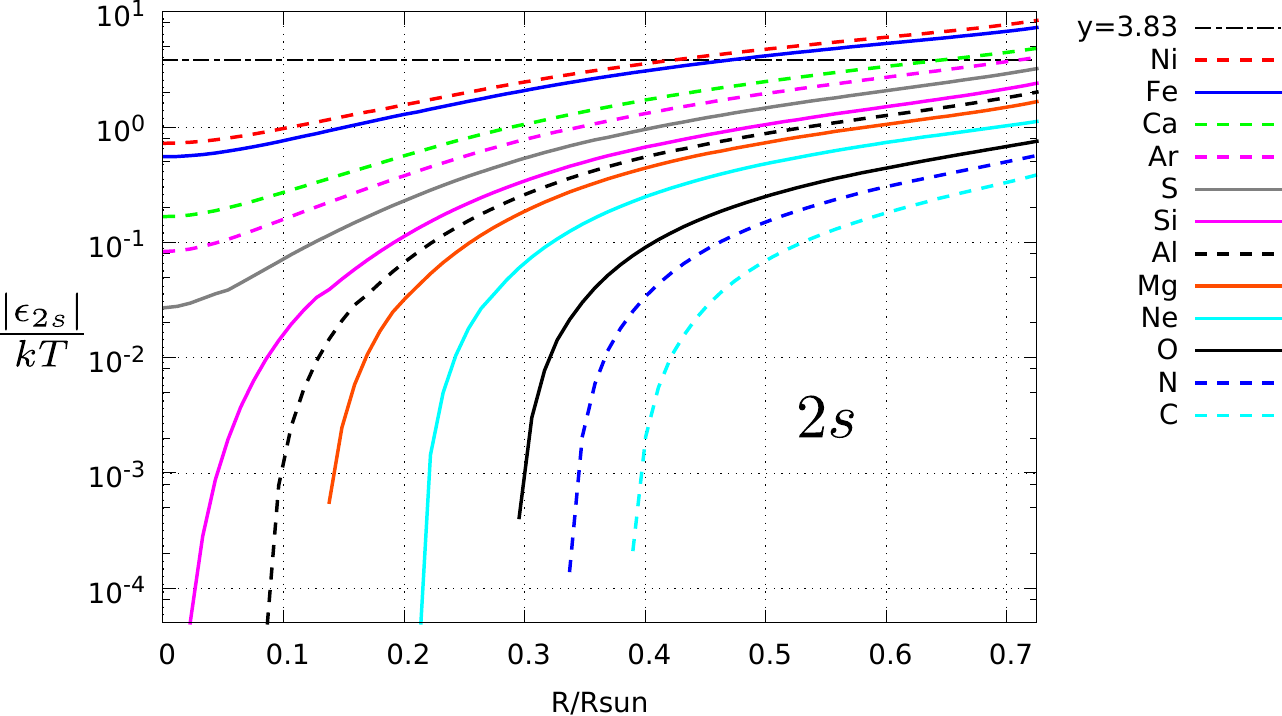}} 
	\resizebox{0.5\textwidth}{!}{\includegraphics{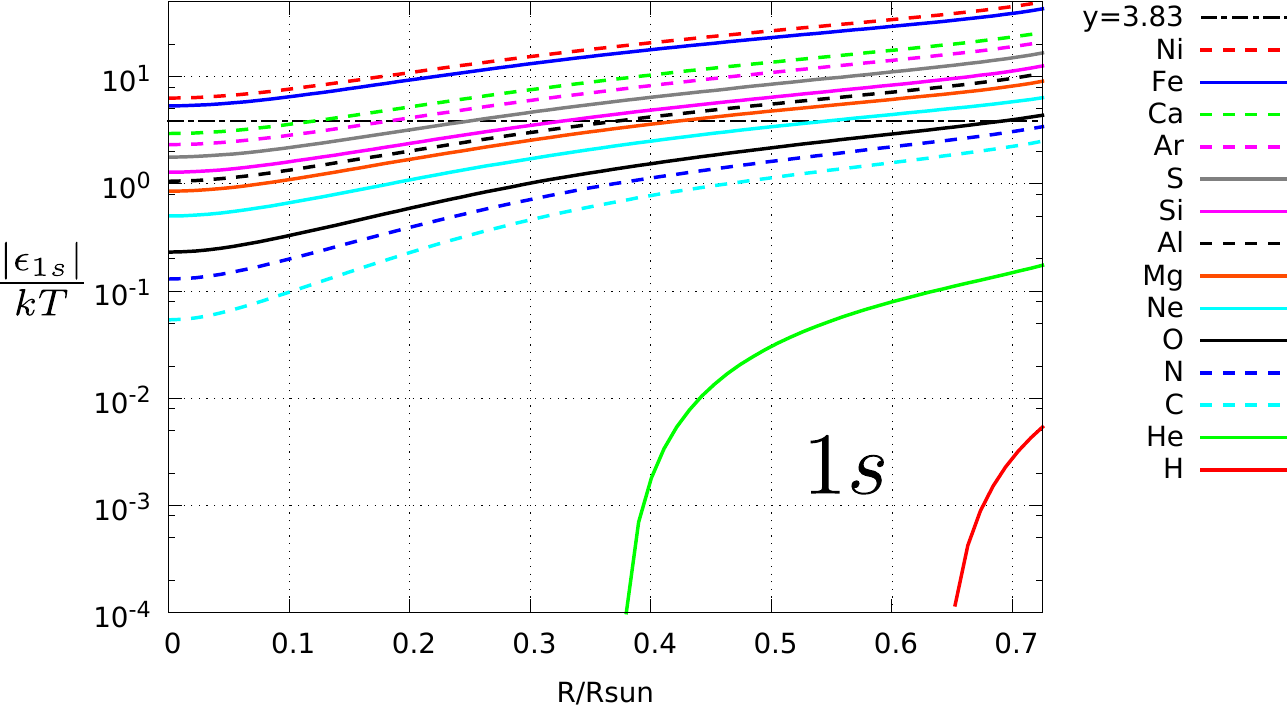}} 
	\caption{Orbital energies divided by $ k_{B}T $ for the $ 3s $ (upper figure), $ 2s $ (middle figure) and $ 1s $ (lower figure) orbitals and for various elements, across the radiation zone. The Rosseland peak is located on the $ y=3.83 $ dashed-dotted line.}
	\label{fig:es_over_kt}
\end{figure}

The total monochromatic opacity and the monochromatic contributions of various chemical elements, at three different solar radii are shown in \autoref{fig:spec_elements}, together with the Rosseland weight function \eqref{eq:ross_function}.
It is evident from the figure that, as is well known, despite their low abundance, metals contribute significantly to the opacity as some of them are not fully ionized and strong bound-bound transition lines and bound-free  edges appear nearby the Rosseland peak, whose location varies as $ \approx3.83k_{B}T $.
In \autoref{fig:elements_ktot_fracs} the contribution of various elements to the Rosseland mean opacity along the radiation zone is presented. 
These results are in agreement with Figure 10 of \cite{villante2010constraints} and Figure 2 of \cite{blancard2012solar}.
We note that the maxima and minima are caused by the metallic contributions as a direct result of the variations (mainly due to the varying temperature) in the positions of lines and edges  relative to the position of the Rosseland peak. For example, at the solar core or near the convection-radiation boundary, the iron K and L shell features are, respectively, nearby the Rosseland peak, while at $ R=0.3 $ they are both well outside the Rosseland distribution (as seen in \autoref{fig:spec_elements}). This results in a peak near the core, a minimum near $ R=0.3 $ and a peak near the radiation-convection interface, in the iron opacity contribution profile, as shown in \autoref{fig:elements_ktot_fracs}.
A complete plot of orbital energies relative to $ k_{B}T $ for the $ 3s $, $ 2s $ and $ 1s $ orbitals and for various elements across the radiation zone is shown in \autoref{fig:es_over_kt}. The relevance of each element due to each of its orbitals is inferred from the proximity to the Rosseland peak. 
It is also evident that orbital energies become less negative towards the core since the screening of the nucleus is increasing due to the increase in density. As a result, some orbitals can be pressure ionized into the continuum at a finite radii.
Indeed, as seen in \autoref{fig:es_over_kt}, the $ 1s $ orbital of hydrogen and helium, the $ 2s $ orbital of C, N, O, Ne, Mg, Al and Si and the $ 3s $ orbital of all elements in the solar mixture, pressure ionize at a finite radii.

\subsection{Relative contributions of different atomic processes along the radiation zone}
\begin{figure}[H]
	\centering
	\resizebox{\textwidth}{!}{
		\begin{minipage}{\textwidth}
			\subfloat[H]{
				\includegraphics[width=60mm]{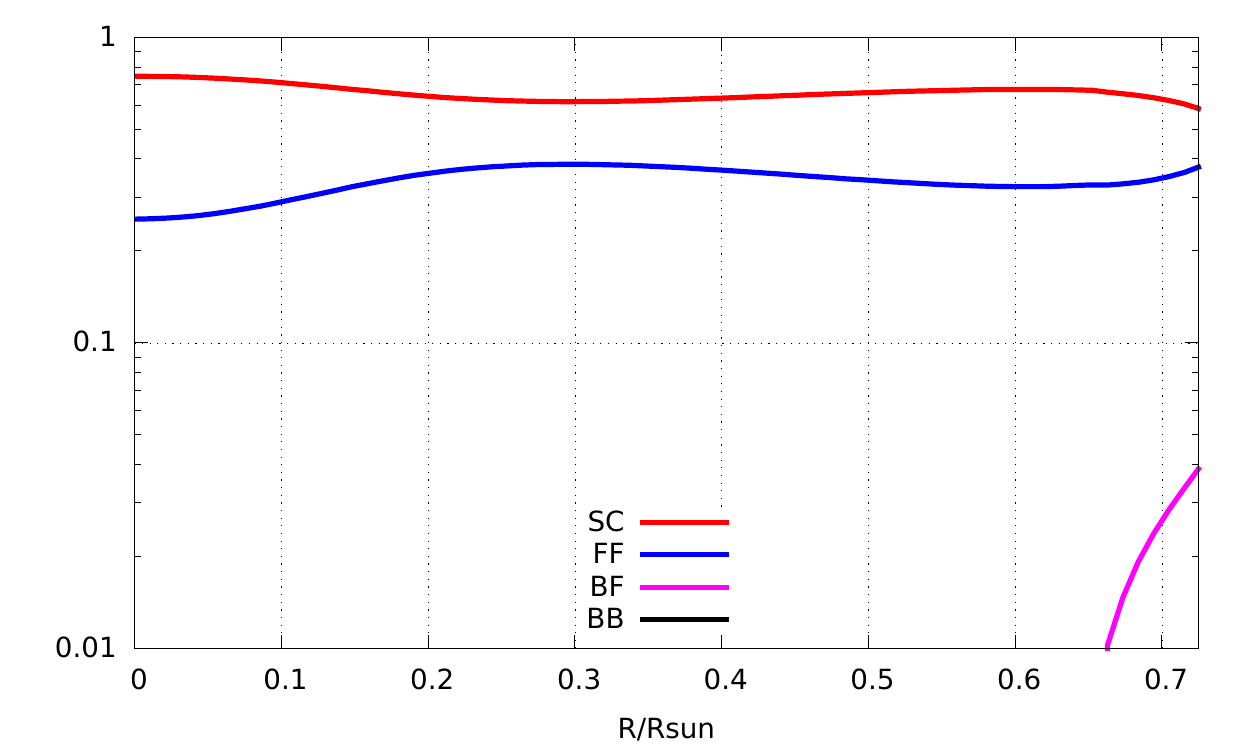}
				
			}
			\subfloat[He]{
				\includegraphics[width=60mm]{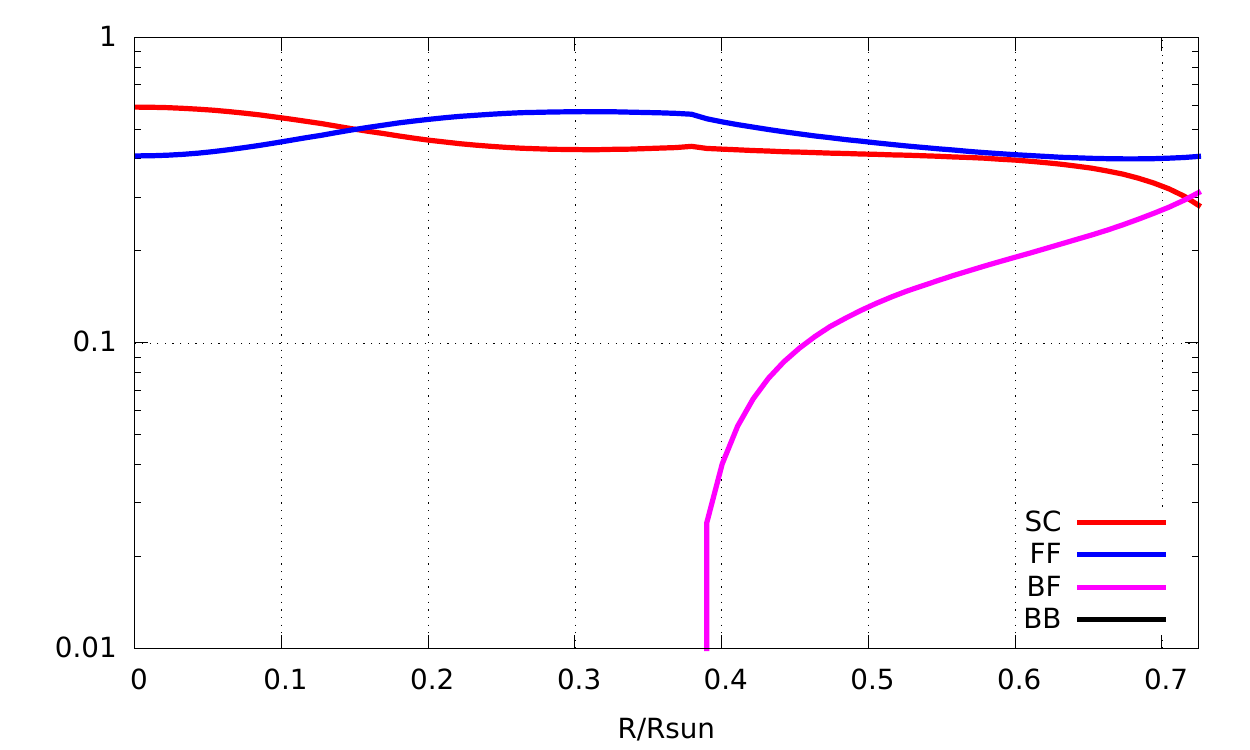}
			}
			\subfloat[C]{
				\includegraphics[width=60mm]{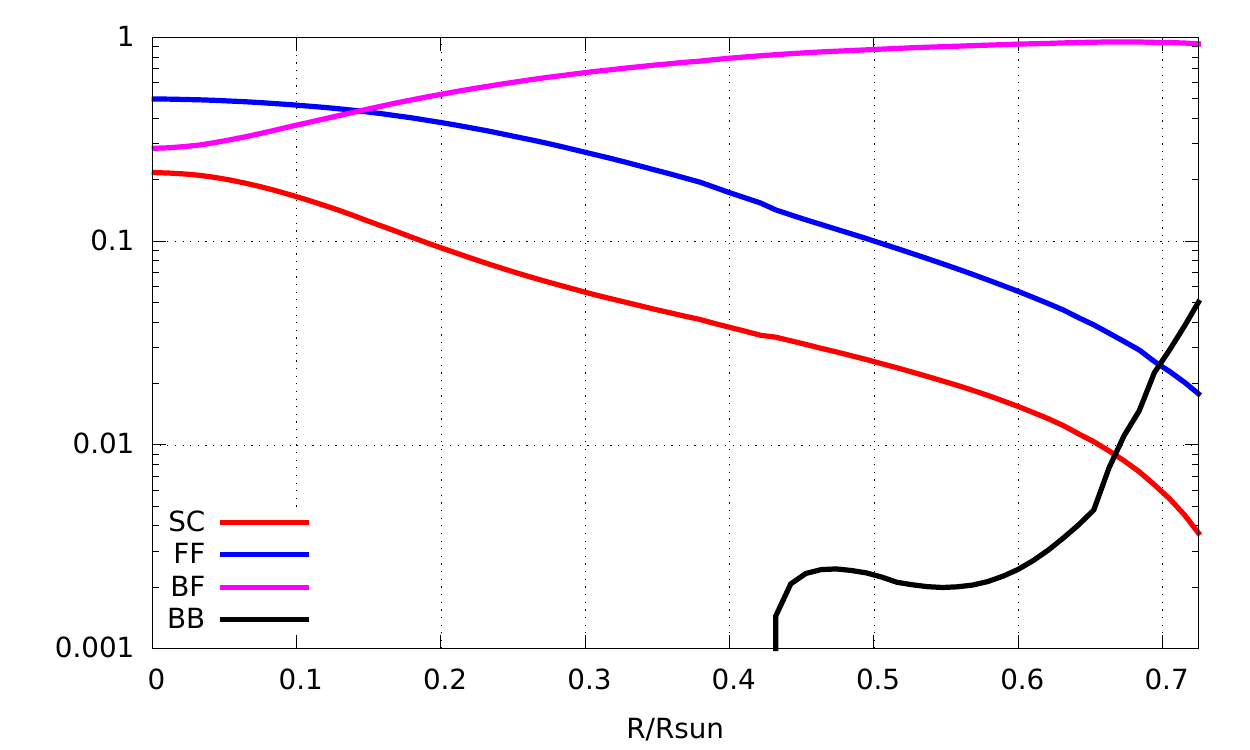}
			}
			\newline
			\subfloat[N]{
				\includegraphics[width=60mm]{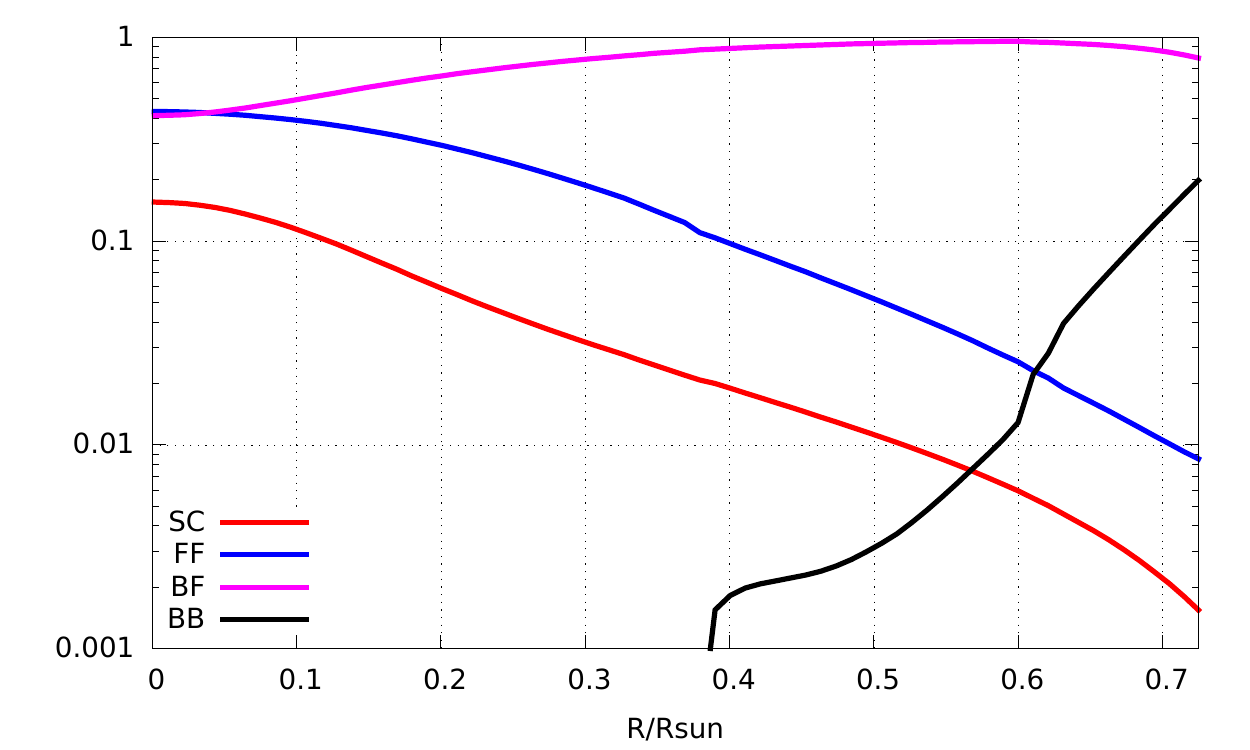}
			}
			\subfloat[O]{
				\includegraphics[width=60mm]{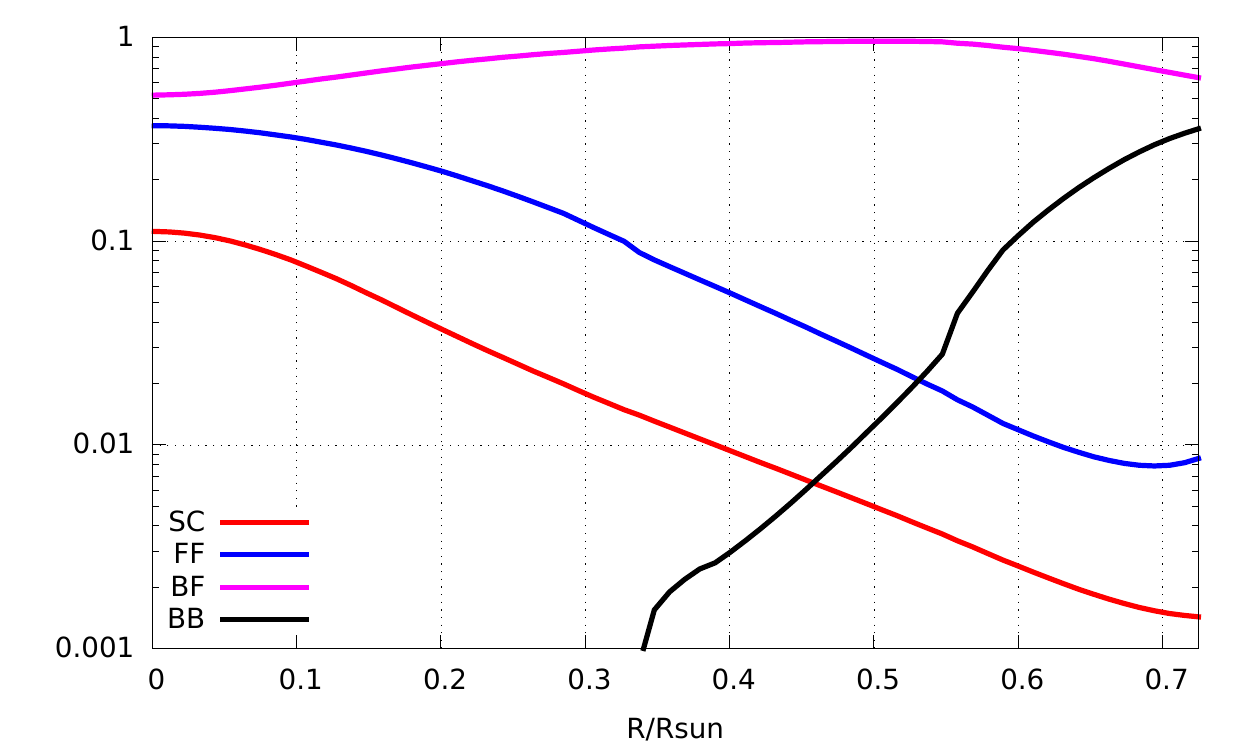}
			}
			\subfloat[Ne]{
				\includegraphics[width=60mm]{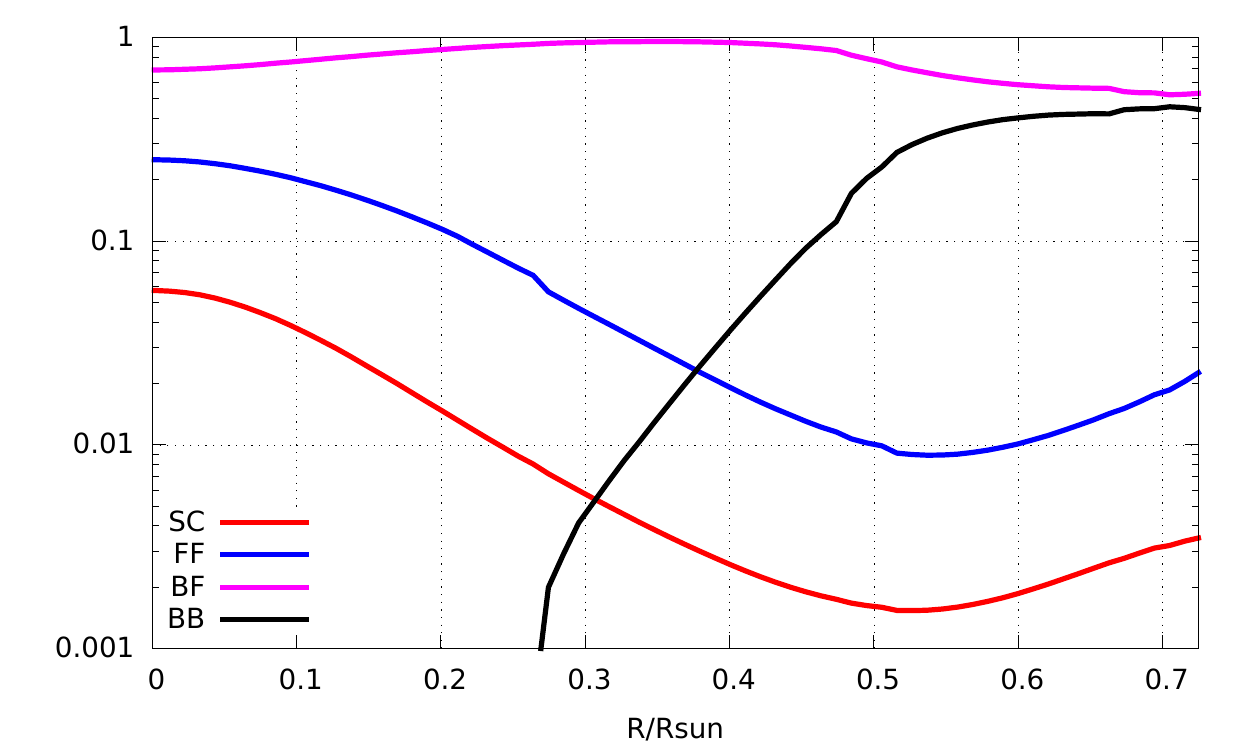}
			}
			\newline
			\subfloat[Na]{
				\includegraphics[width=60mm]{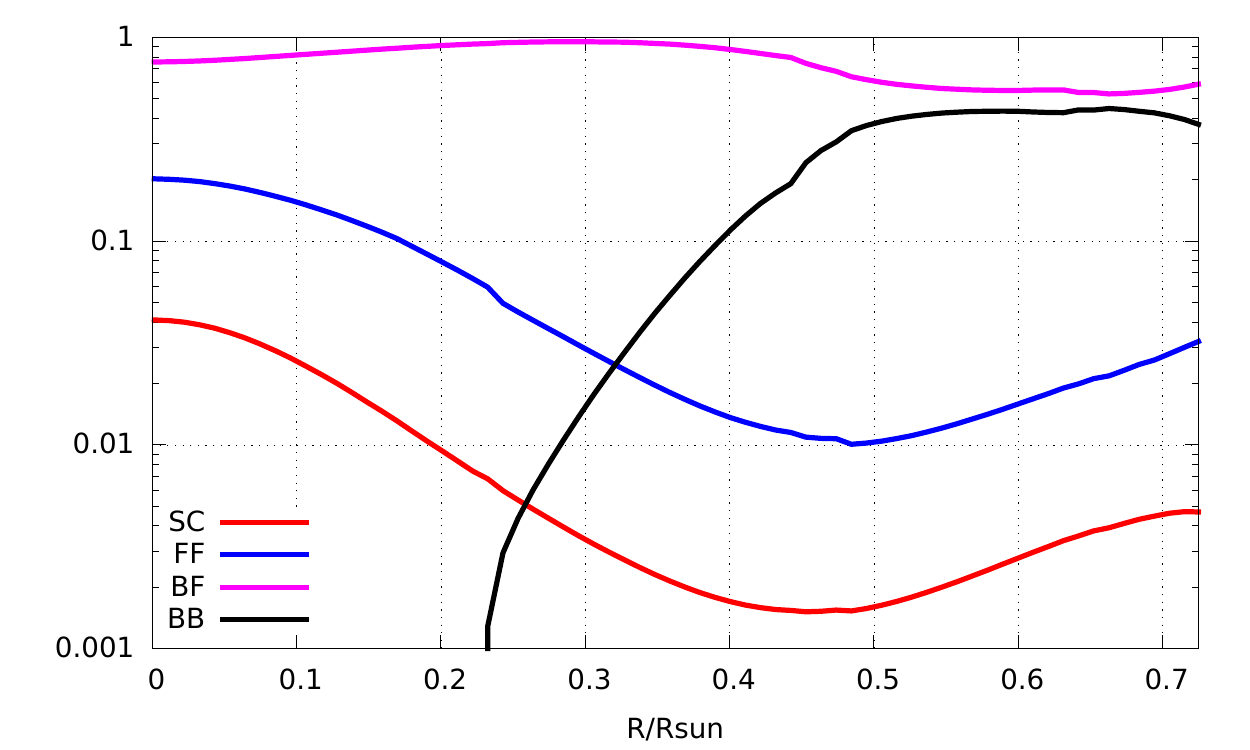}
			}
			\subfloat[Mg]{
				\includegraphics[width=60mm]{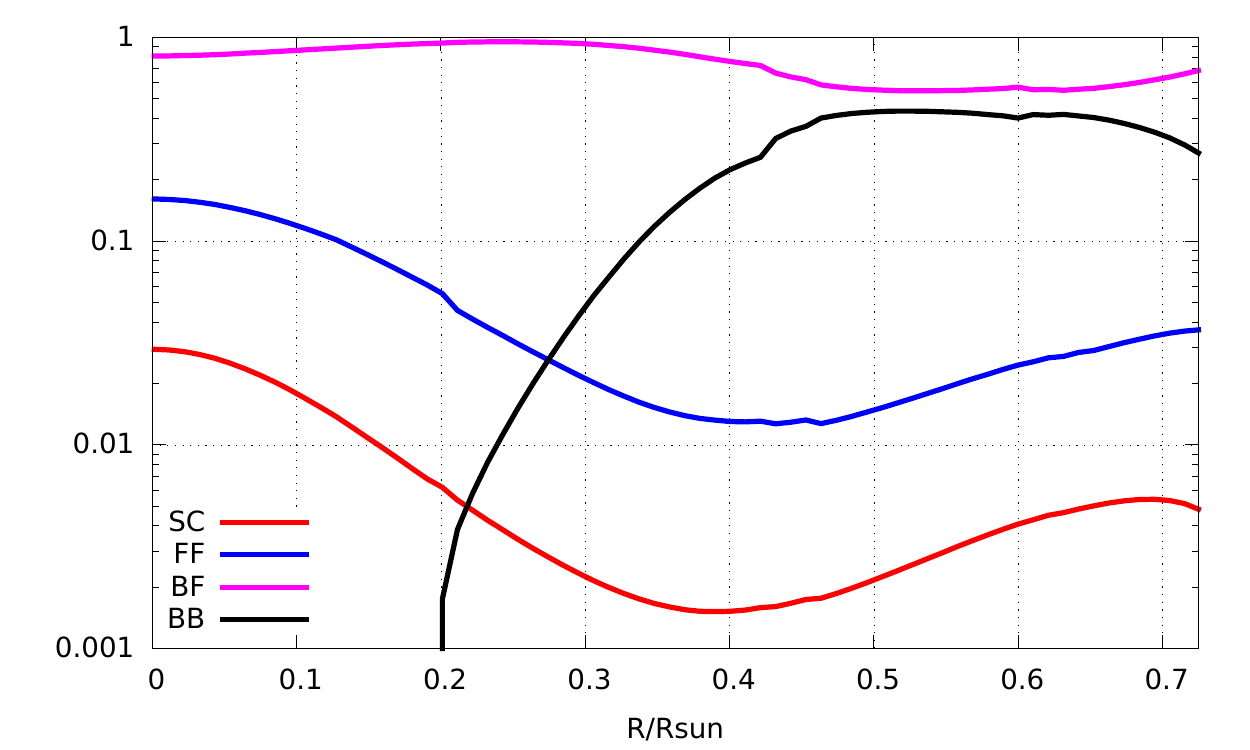}
			}
			\subfloat[Al]{
				\includegraphics[width=60mm]{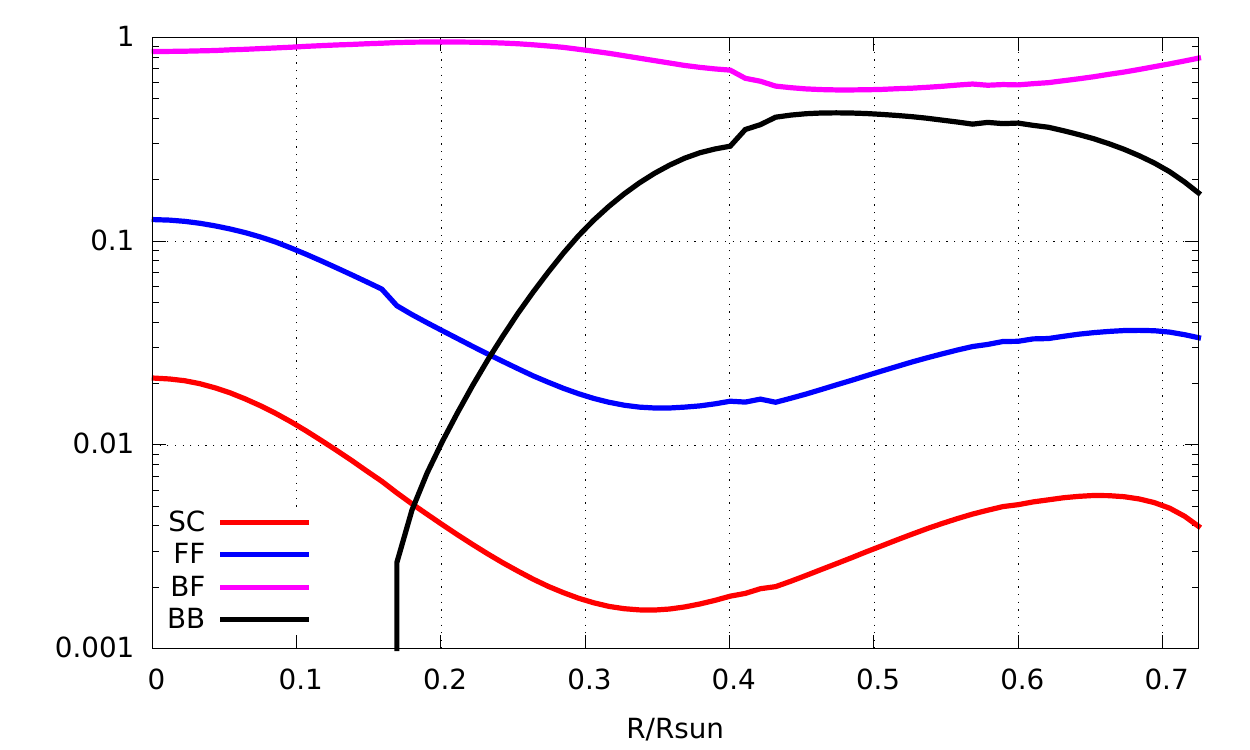}
			}
			\newline
			\subfloat[Si]{
				\includegraphics[width=60mm]{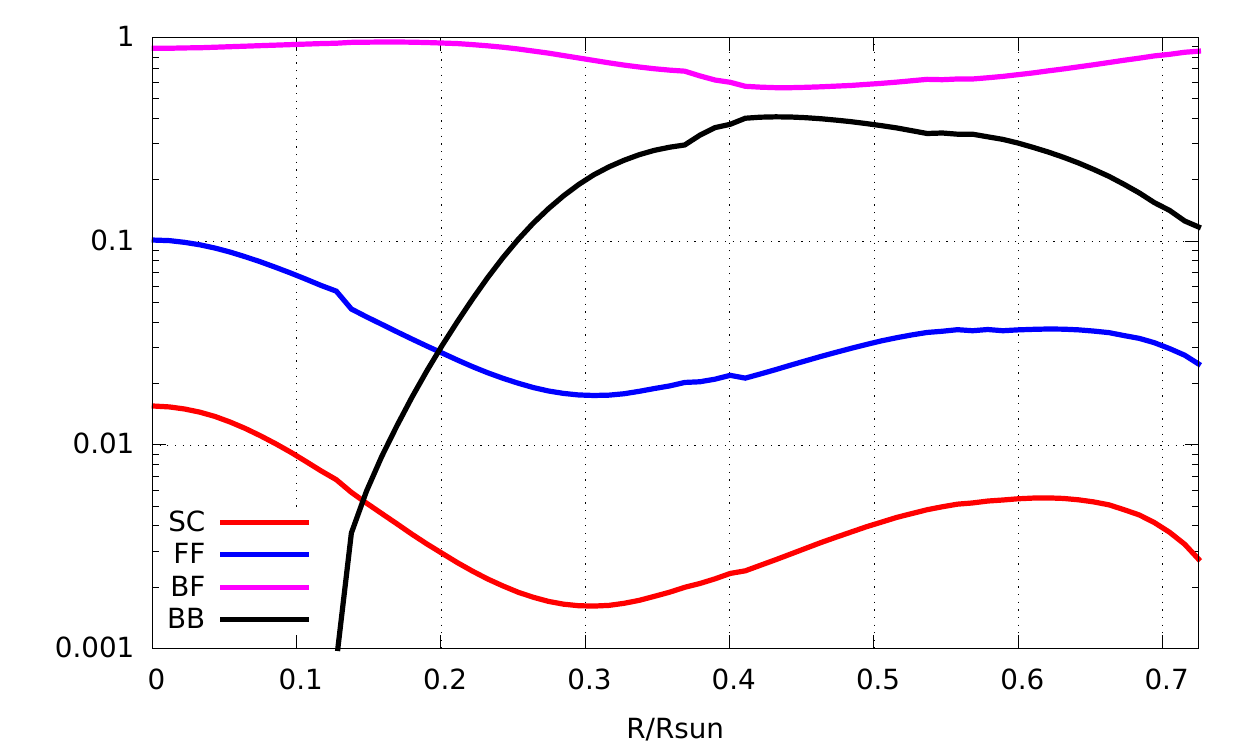}
			}
			\subfloat[S]{
				\includegraphics[width=60mm]{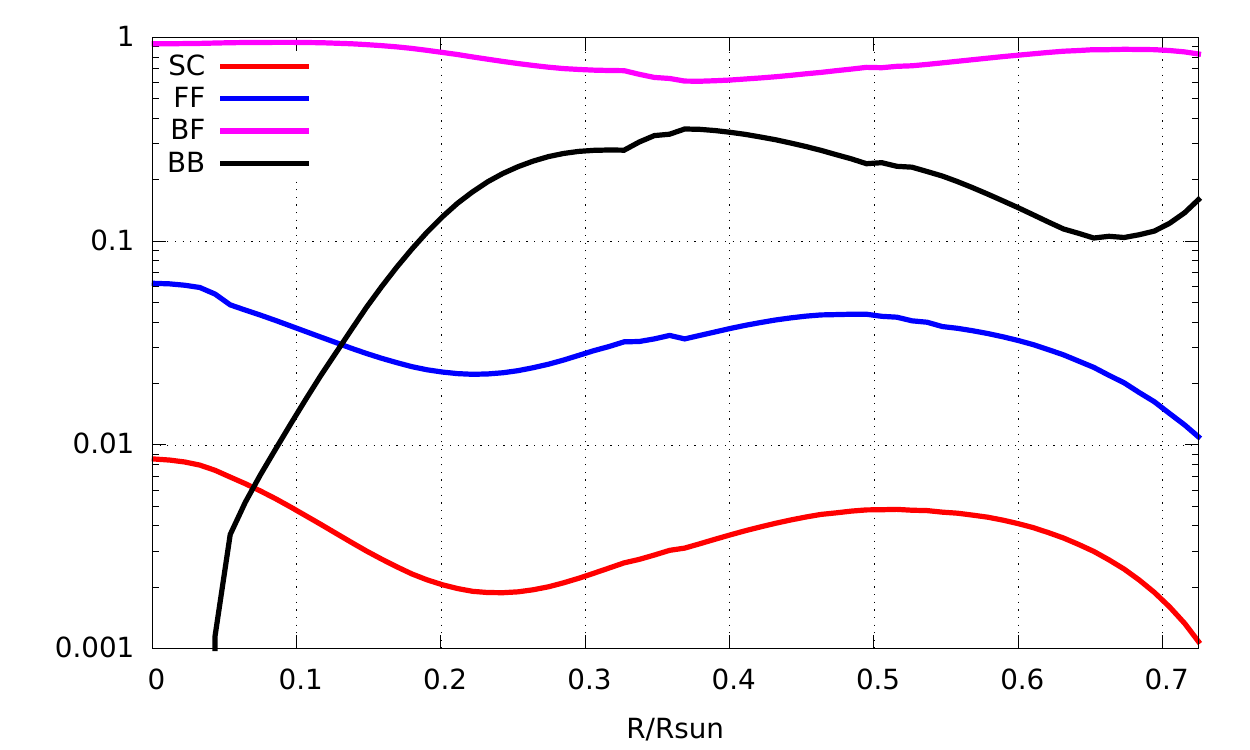}
			}
			\subfloat[Ar]{
				\includegraphics[width=60mm]{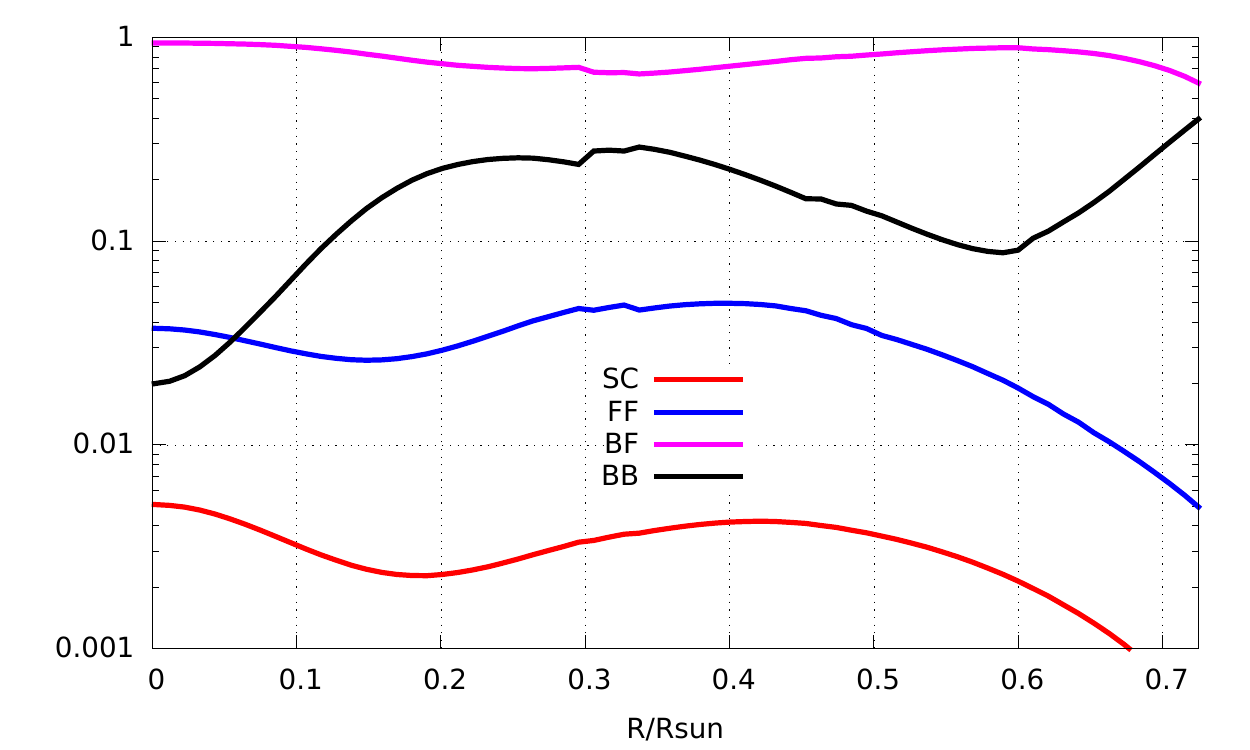}
			}
			\newline
			\subfloat[Ca]{
				\includegraphics[width=60mm]{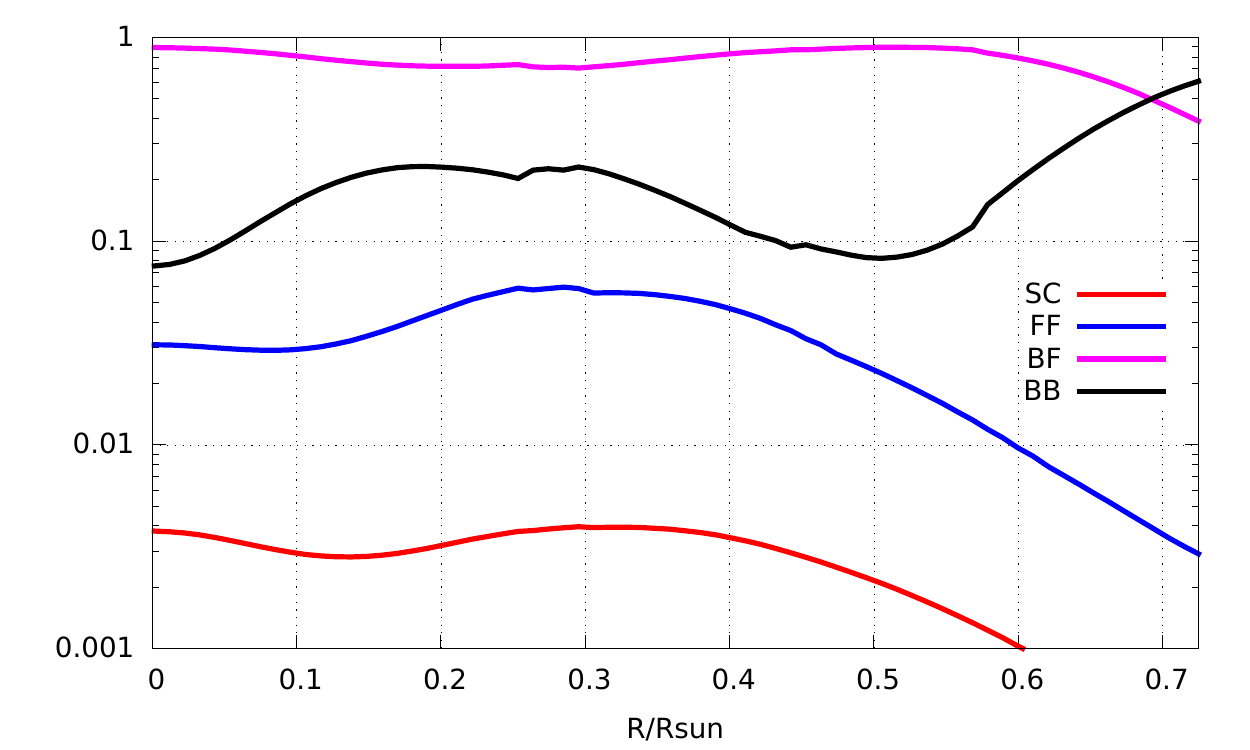}
			}
			\subfloat[Fe]{
				\includegraphics[width=60mm]{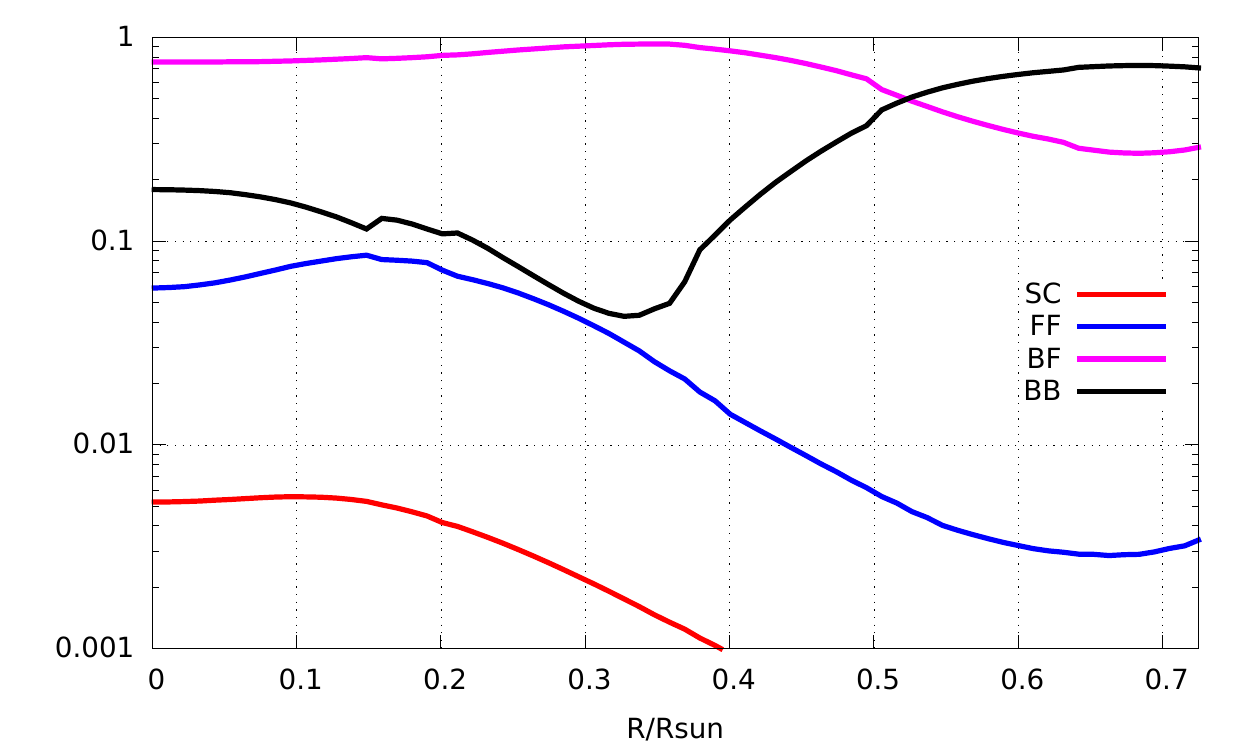}
			}
			\subfloat[Ni]{
				\includegraphics[width=60mm]{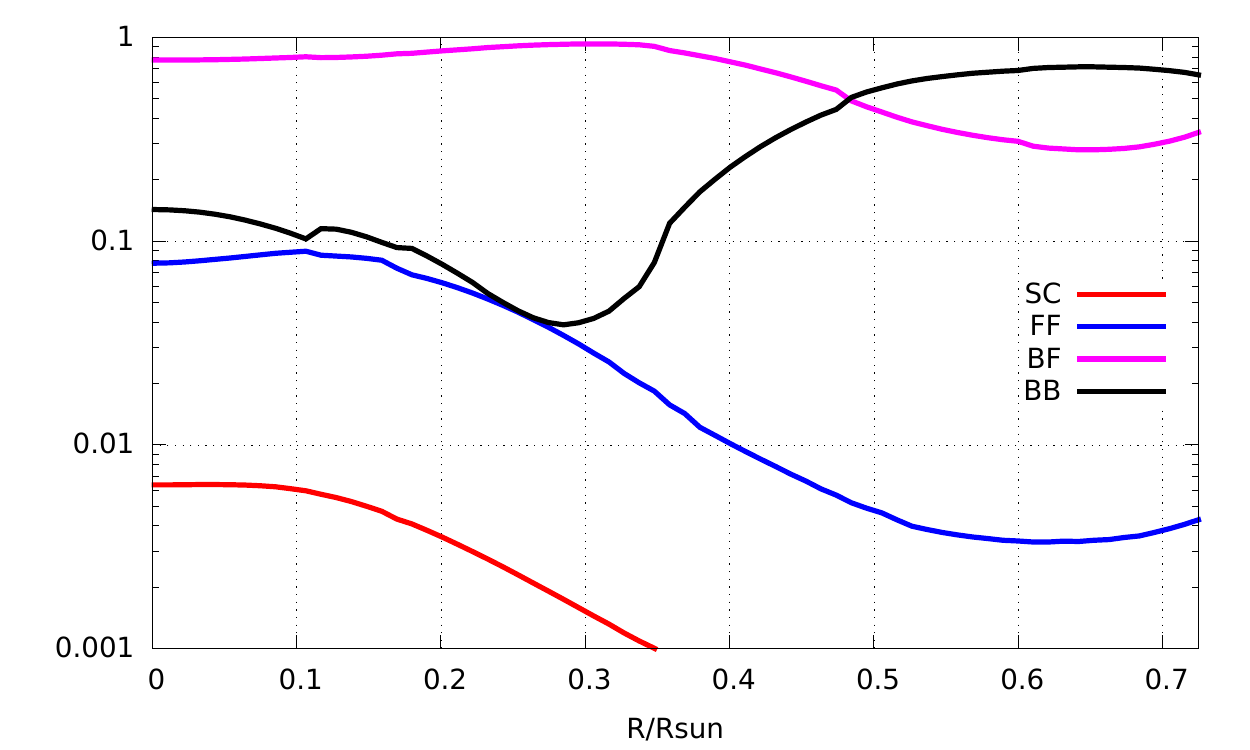}
			}
			\newline	
			\caption{The relative contributions of scattering (red), free-free (blue), bound-free (magenta) and bound-bound (black) processes to the opacity of several elements (as specified under each plot), across the radiation zone.}
			\label{fig:fracs_dk_elements}
		\end{minipage}
	}
\end{figure}

The contribution of the bound-bound and bound-free processes to the Rosseland opacity depends on the positions of photoabsorption lines and edges relative to the position of the Rosseland peak and on the population of the various bound states. Relative contributions of the four different atomic processes: scattering, free-free, bound-free and bound-bound, to the elemental Rosseland opacity of various elements throughout the radiation zone, are presented in \autoref{fig:fracs_dk_elements}.
The results are in agreement with Figure 2 of \cite{blancard2012solar}.
The behavior of the bound-bound and bound-free contributions in \autoref{fig:fracs_dk_elements} can be qualitatively understood from \autoref{fig:es_over_kt} and \autoref{fig:spec_elements}. For hydrogen and helium, all orbitals are pressure ionized up to some finite radii ($ R\approx 0.65 $ and $ R\approx 0.38 $, respectively) and the bound-free contribution  is zero up these threshold radii. In addition, orbitals with principle quantum number $ n>1 $ are pressure ionized throughout the radiation zone so that there is no bound-bound contribution for hydrogen and helium. On the other hand, for all other elements the K-shell exists throughout the radiation zone and the bound-free contribution is always non-zero. However, several elements have a finite threshold radius for the existence of the L-shell orbitals and therefore a finite threshold radius for the bound-bound contribution (see Fig. \ref{fig:fracs_dk_elements}c to Fig.\ref{fig:fracs_dk_elements}k). This threshold radius is lower for heavier elements, for which the $ L $-shell orbitals are closer to the heavier nucleus and a larger density is required for pressure ionization. For the heaviest elements (Fig. \ref{fig:fracs_dk_elements}l to Fig.\ref{fig:fracs_dk_elements}o), the L-shell exists throughout the radiation zone and therefore, they always have a non-zero bound-bound contribution. 
As we have mentioned, as the temperature decreases outward, bound-bound lines move with respect to the Rosseland peak (\autoref{fig:es_over_kt}). K-shell lines of certain light elements, only approach the Rosseland peak from below, explaining the monotonic behavior of their bound-bound contributions (Fig. \ref{fig:fracs_dk_elements}c to Fig.\ref{fig:fracs_dk_elements}f), while heavier elements K-shell lines cross the Rosseland peak and have a local maximum in their bound-bound contribution (Fig. \ref{fig:fracs_dk_elements}g to Fig.\ref{fig:fracs_dk_elements}j).
For heavier elements, after their K-shell lines pass the Rosseland peak, at large enough radii, their L-shell lines also begin to approach it, giving rise to local maximum followed by a local minimum in the bound-bound contribution (Fig. \ref{fig:fracs_dk_elements}k to Fig.\ref{fig:fracs_dk_elements}m). Finally, we note that the iron and nickel K-shell lines at $ R=0 $ are already past the Rosseland peak, while their L-shell lines cross it near the radiation-convection boundary, giving rise to a local minimum followed by a local maximum (Fig. \ref{fig:fracs_dk_elements}n and Fig.\ref{fig:fracs_dk_elements}o).

\begin{figure}
	\resizebox{0.5\textwidth}{!}{\includegraphics{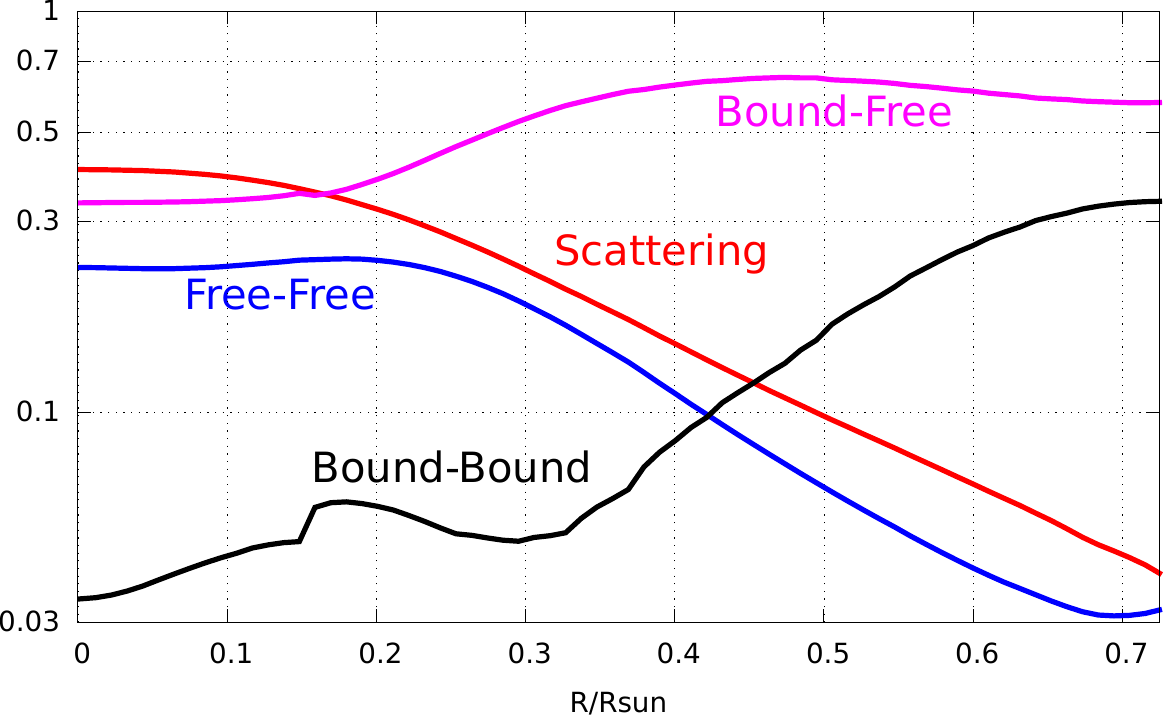}}
	\caption{Relative contributions of the different atomic processes to the total SSM mixture opacity, across the radiation zone.}
	\label{fig:fracs_tot}
\end{figure}

The total contribution of the different atomic processes (due to all elements combined) to the SSM mixture Rosseland opacity is shown in \autoref{fig:fracs_tot}.
\newpage
\newpage

\subsection{Comparison with the Opacity-Project}
\begin{figure}
	\centering
	\resizebox{0.5\textwidth}{!}{\includegraphics{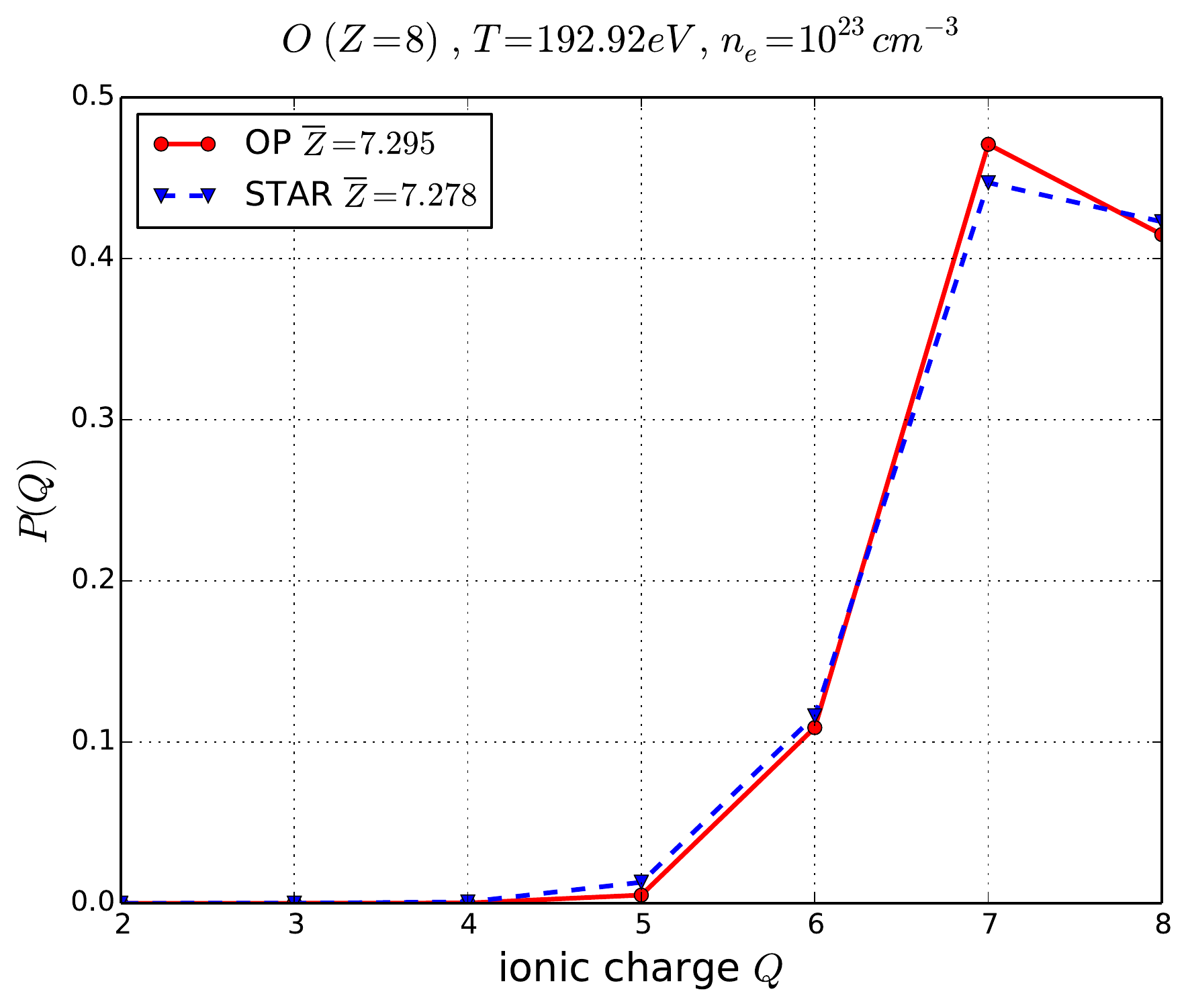}} 
	\caption{Ion charge distributions for oxygen at $ T=192.92eV $ and $ n_{e}=10^{23}cm^{-3} $ calculated by OP (red solid line) and STAR (blue dashed line). The mean ionizations are given in the legend.}
	\label{fig:pq_op_o}
\end{figure}

\begin{figure}
	\centering
	\resizebox{0.5\textwidth}{!}{\includegraphics{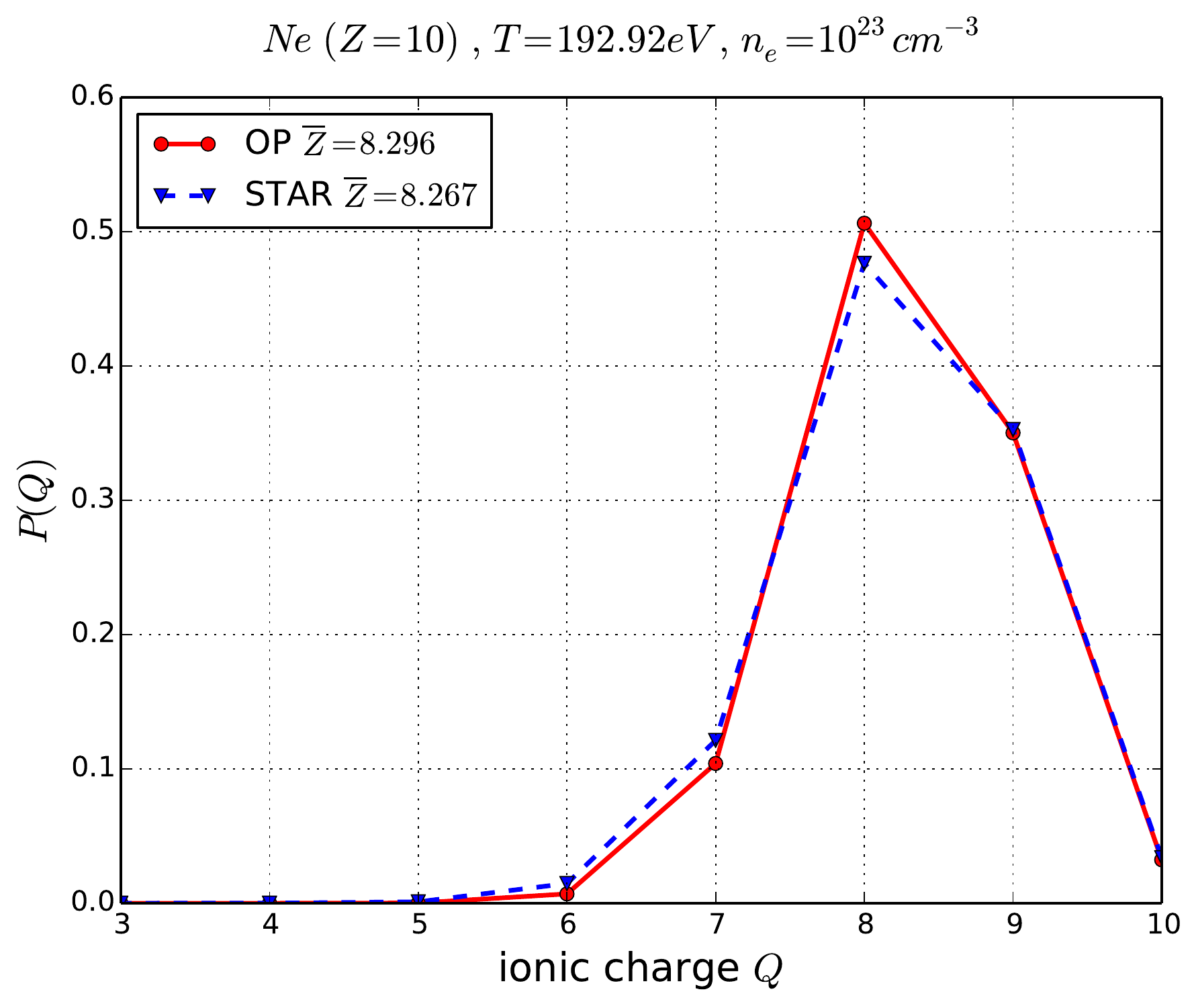}} 
	\caption{Ion charge distributions for neon at $ T=192.92eV $ and $ n_{e}=10^{23}cm^{-3} $ calculated by OP (red solid line) and STAR (blue dashed line). The mean ionizations are given in the legend.}
	\label{fig:pq_op_ne}
\end{figure}

\begin{figure}
	\centering
	\resizebox{0.5\textwidth}{!}{\includegraphics{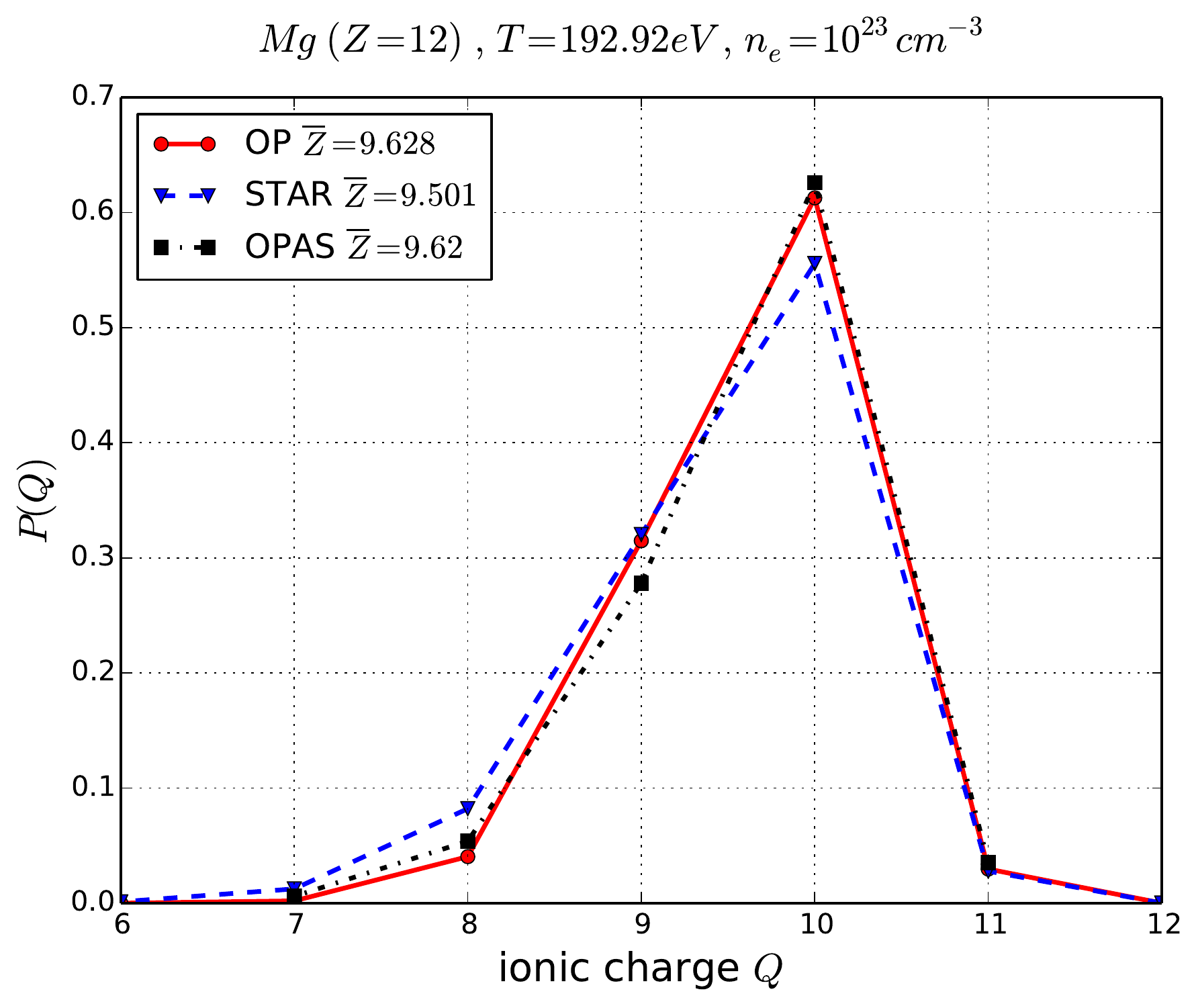}} 
	\caption{Ion charge distributions for magnesium at $ T=192.92eV $ and $ n_{e}=10^{23}cm^{-3} $ calculated by OP (red solid line), STAR (blue dashed line) and OPAS (black dashed-dotted line). The mean ionizations are given in the legend.}
	\label{fig:pq_op_mg}
\end{figure}

\begin{figure}
	\centering
	\resizebox{0.5\textwidth}{!}{\includegraphics{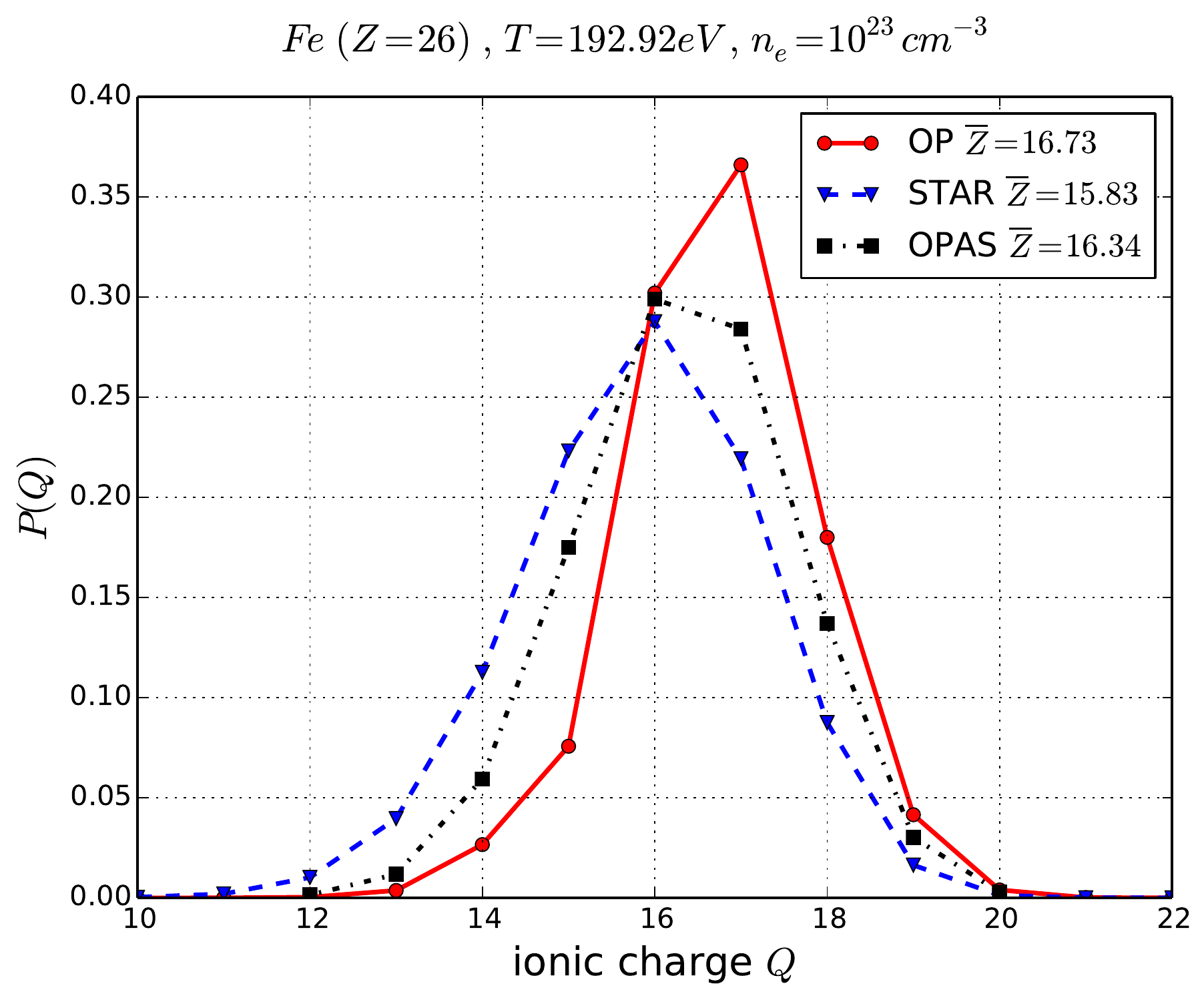}} 
	\caption{Ion charge distributions for iron at $ T=192.92eV $ and $ n_{e}=10^{23}cm^{-3} $ calculated by OP (red solid line), STAR (blue dashed line) and OPAS (black dashed-dotted line). The mean ionizations are given in the legend.}
	\label{fig:pq_op_fe}
\end{figure}

We compare STAR and Opacity-Project (OP) monochromatic opacities and ion charge state distributions for O, Ne, Mg and Fe, which have a major opacity contribution near the radiation-convection interface.
Comparisons are made at $ T= 192.92eV $ and $ n_{e} =10^{23}$ available  in the TOPbase database (\cite{cunto1994topbase}), which is used to generate OP data. The same comparisons were made by \cite{blancard2012solar} for Fe and Mg with OPAS and by \cite{colgan2013light} for Fe and O using the LEDCOP and ATOMIC opacity codes. The ion distributions are given in figures \ref{fig:pq_op_o}-\ref{fig:pq_op_fe}. The OPAS charge distributions given in \cite{blancard2012solar} for Mg and Fe are also shown in figures \ref{fig:pq_op_mg}-\ref{fig:pq_op_fe}. For O and Ne, mean ionizations $ \overline{Z} $ of OP and STAR are in excellent agreement, though the charge distributions slightly differ. We note that for Fe, larger differences appear and for the lower charge states ($ Q<16 $), STAR populations are much larger. In these lower ionic states, ground and excited configurations must have a non-empty M-shell. This tendency is similar to that shown by OPAS, though the STAR populations are somewhat larger than OPAS at the lower charge states.

\begin{figure}[]
	\centering
	\resizebox{0.5\textwidth}{!}{\includegraphics{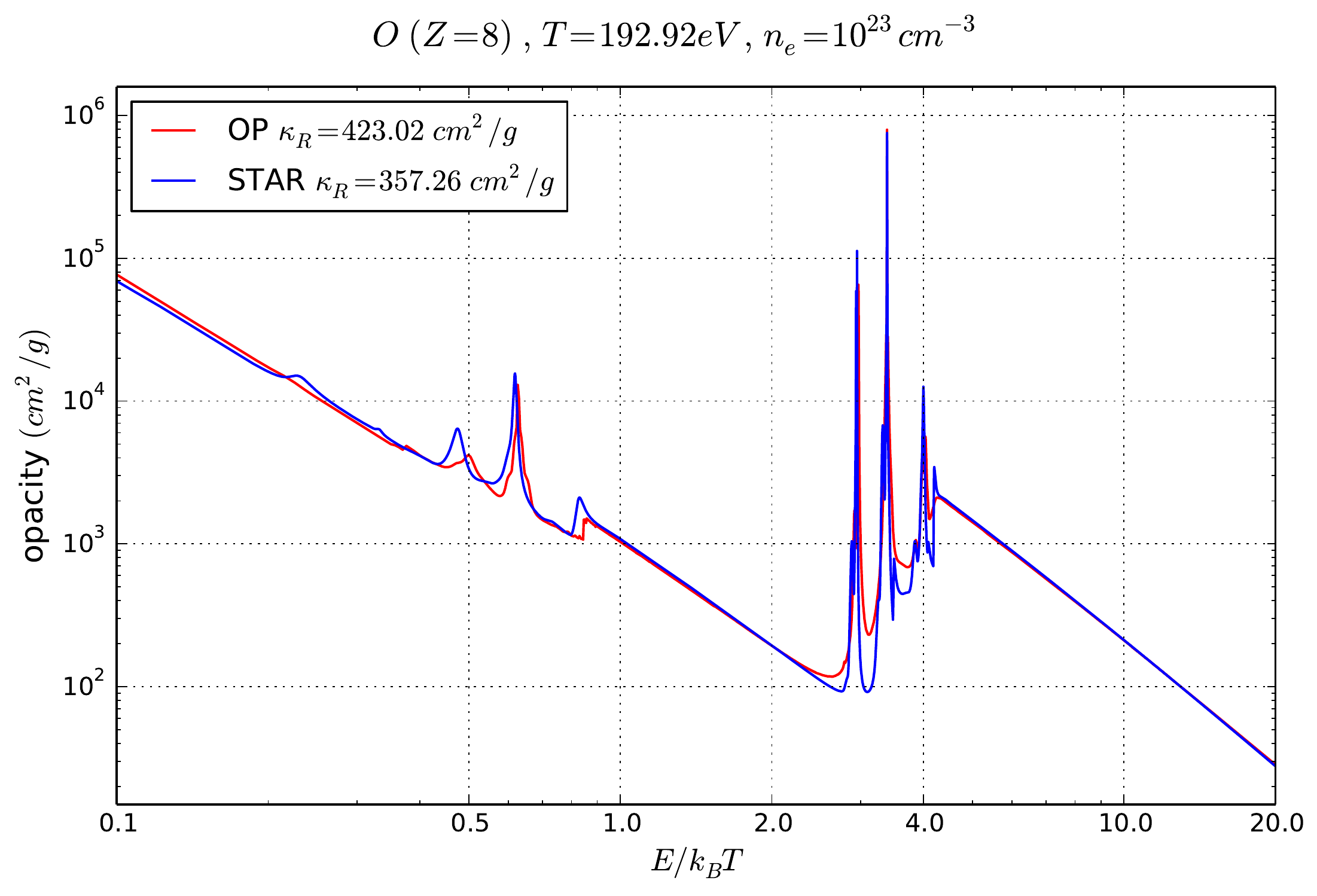}} 
	\caption{Comparison of STAR (blue line) and OP (red line) monochromatic opacity calculations for oxygen at $ T=192.92eV $ and $ n_{e}=10^{23}cm^{-3} $. The Rosseland mean opacities are given in the legend.}
	\label{fig:op_specs_o}
\end{figure}

\begin{figure}[]
	\centering
	\resizebox{0.5\textwidth}{!}{\includegraphics{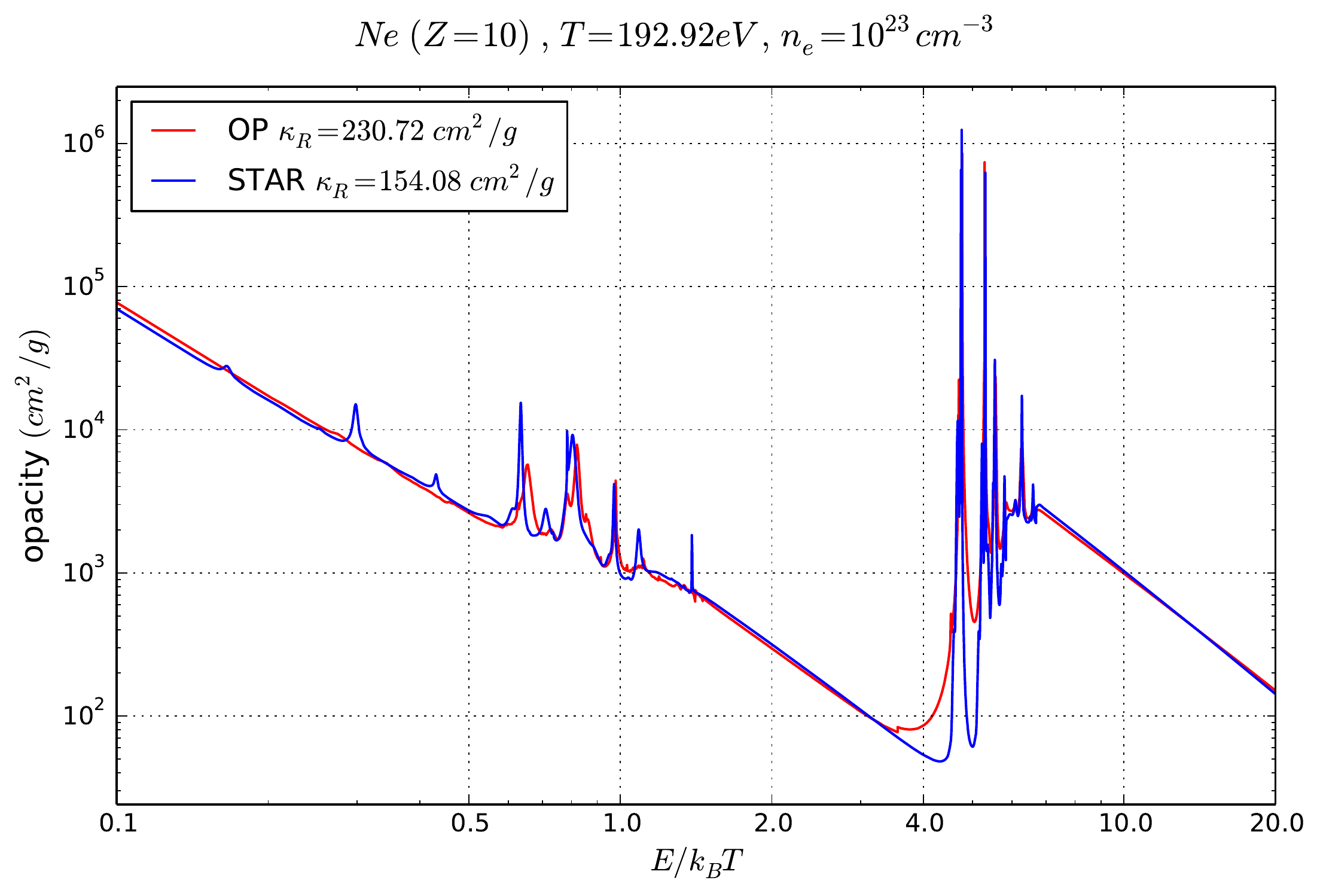}} 
	\caption{Same as \autoref{fig:op_specs_o}, for neon at $ T=192.92eV $ and $ n_{e}=10^{23}cm^{-3} $.}
	\label{fig:op_specs_ne}
\end{figure}

\begin{figure}[]
	\centering
	\resizebox{0.5\textwidth}{!}{\includegraphics{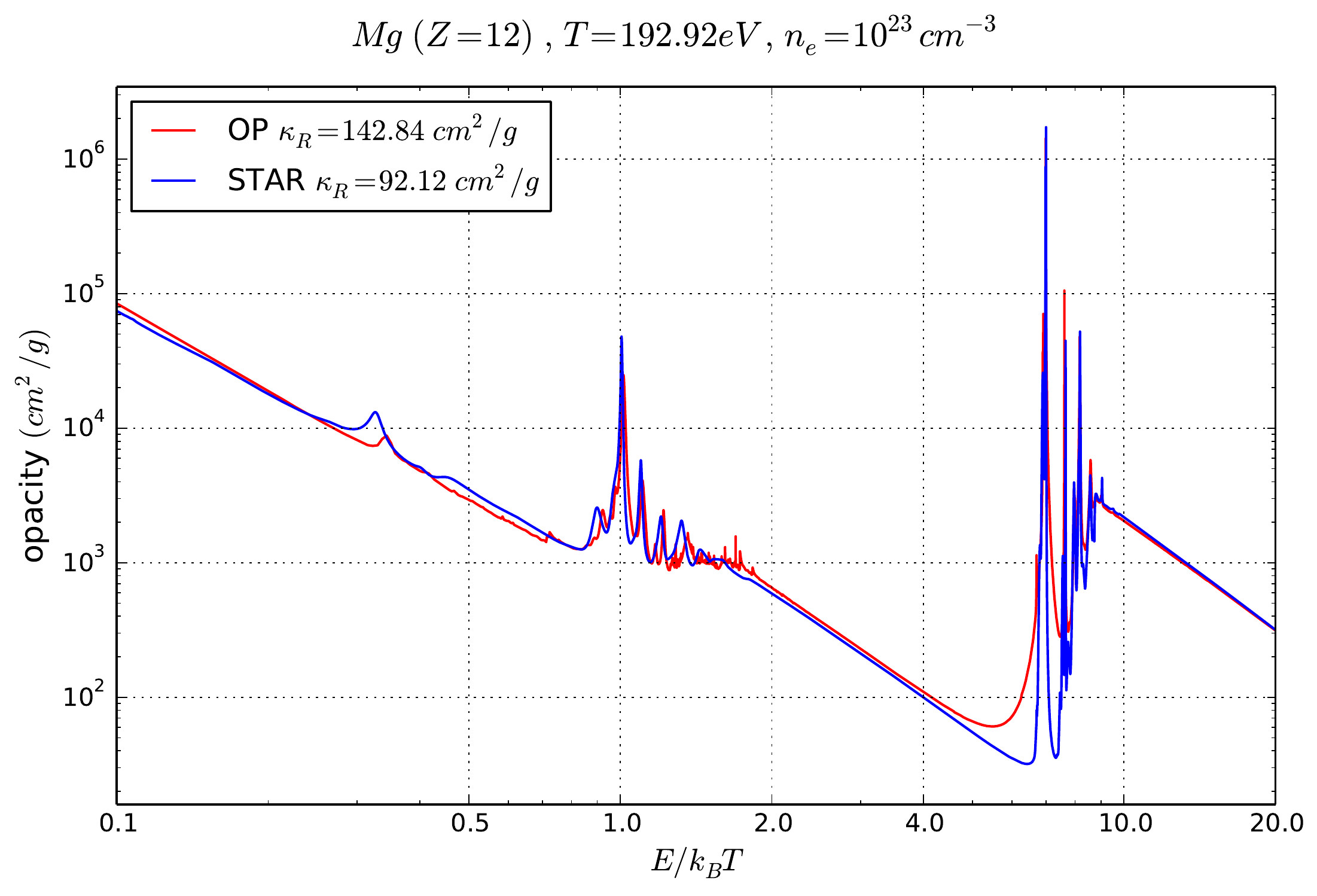}} 
	\caption{Same as \autoref{fig:op_specs_o}, for magnesium at $ T=192.92eV $ and $ n_{e}=10^{23}cm^{-3} $.}
	\label{fig:op_specs_mg}
\end{figure}

\begin{figure}[]
	\centering
	\resizebox{0.5\textwidth}{!}{\includegraphics{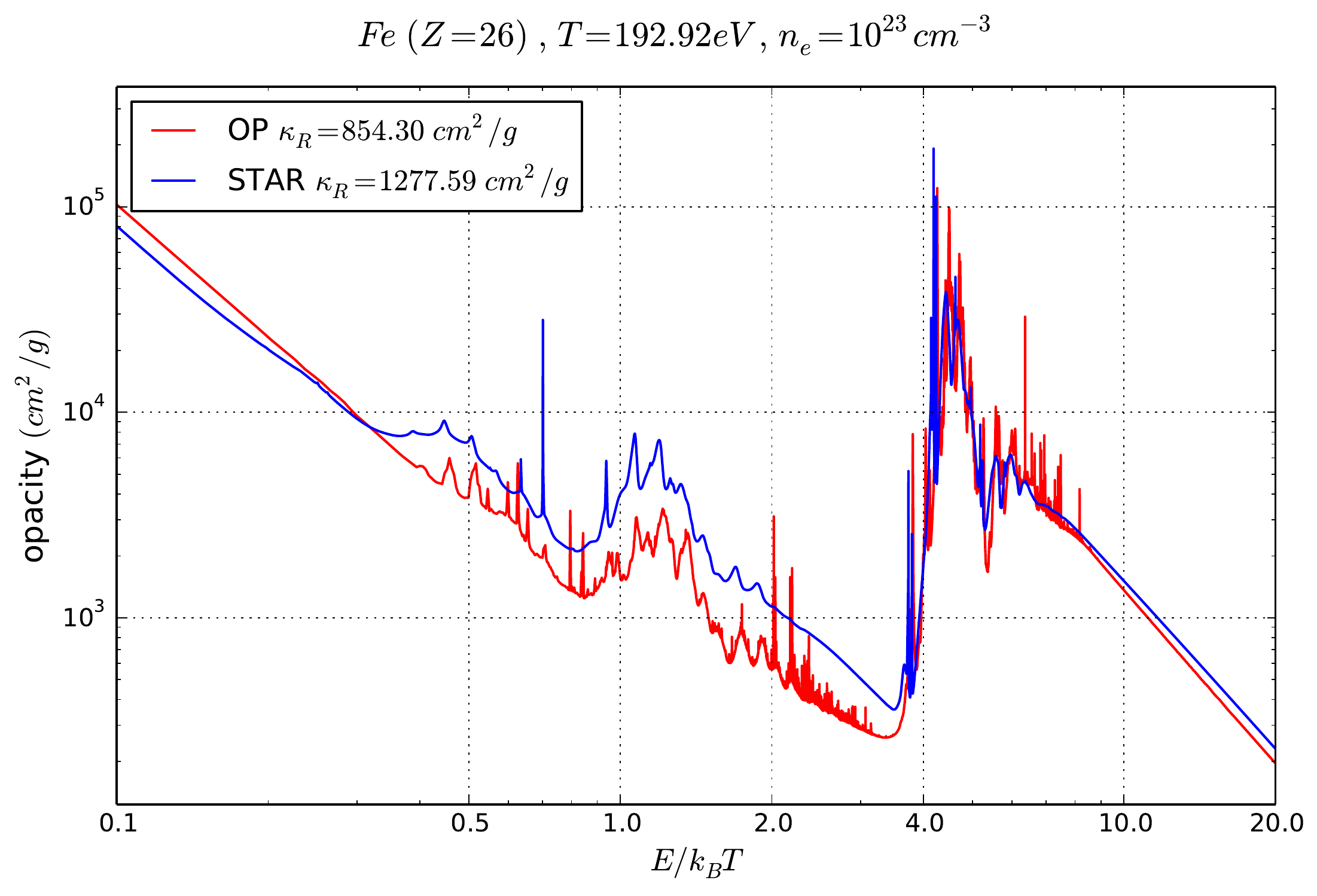}} 
	\caption{Same as \autoref{fig:op_specs_o}, for iron at $ T=192.92eV $ and $ n_{e}=10^{23}cm^{-3} $.}
	\label{fig:op_specs_fe}
\end{figure}

A comparison between OP and STAR O, Ne, Mg and Fe monochromatic opacities are given in figures \ref{fig:op_specs_o}, \ref{fig:op_specs_ne}, \ref{fig:op_specs_mg} and \ref{fig:op_specs_fe}, respectively. The Rosseland means are also given in the figures. 
It is evident that OP K-shell lines that are shown for O, Ne and Mg, are significantly broader than the STAR lines. The broad OP K-shell wings which lie nearby the Rosseland peak, contribute significantly to the Rosseland mean, as seen in Figures
\ref{fig:op_specs_o}, \ref{fig:op_specs_ne}, \ref{fig:op_specs_mg}. Such differences were also observed by \cite{blancard2012solar} for Mg and by \cite{colgan2013light} for O.
Both STAR and OP line width calculations include electron impact broadening  in the one-perturber approximation.
However, OP uses collisional rates obtained from comprehensive quantum calculations that include unitarization
(\cite{seaton1987atomic,burke2011r}), while STAR uses the widely-used semi-empirical formulas by \cite{dimitrijevic1987simple}.
A comparison of the iron monochromatic opacity is shown in \autoref{fig:op_specs_fe}.
It is evident that the STAR opacity is larger than OP near the M-shell bound-bound and bound-free regions, which contribute significantly to the Rosseland mean.
Similar differences were also found by \cite{blancard2012solar,colgan2016new}.
This larger M-shell contribution is explained by the larger STAR  population of the lower ionic states (\autoref{fig:pq_op_fe}), for which ground and excited configurations have a non empty M-shell. The STA method takes into account the huge number of excited configurations in these lower ion states by the robust superconfiguration approach. 

\begin{figure}
	\centering
	\resizebox{0.5\textwidth}{!}{\includegraphics{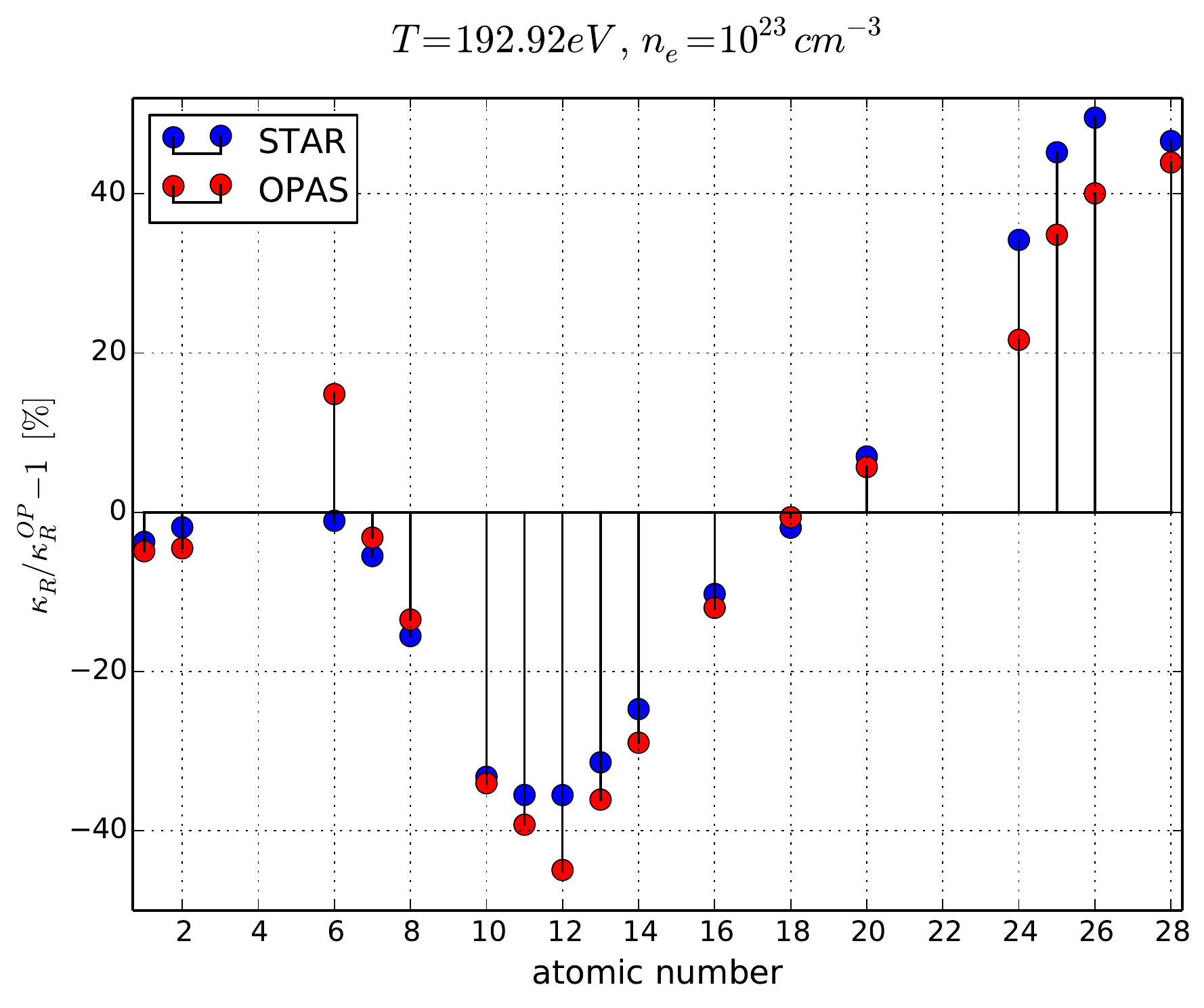}} 
	\caption{Relative difference $ \kappa_{R}/\kappa^{OP}_{R}-1 $  between OP and STAR (blue dots) and between OP and OPAS (red dots, taken from \cite{blancard2012solar}), for all elements available in the TOPbase database at $ T=192.92eV $ and $ n_{e}=10^{23}cm^{-3} $.}
	\label{fig:op_opas_diff}
\end{figure}

A comparison of STAR and OPAS with OP Rosseland mean opacities at $ T=192.92eV $ and $ n_{e}=10^{23}cm^{-3} $ for the $ 17 $  elements available in the TOPbase database are given in \autoref{fig:op_opas_diff} (The OPAS data was taken from \cite{blancard2012solar}). 
For Ne, Na, Mg, Al, and Si, the OP opacity is significantly higher than STAR due to the larger widths of the K-shell lines. On the other hand, for Cr, Mn, Fe and Ni,  STAR opacity is significantly larger than OP due to higher population of non-empty M-shell configurations.
It is also evident that the STAR differences are in a relatively good agreement with those of OPAS, except for carbon for which the STAR agrees better with OP.



\begin{figure}
	\centering
	\resizebox{0.5\textwidth}{!}{\includegraphics{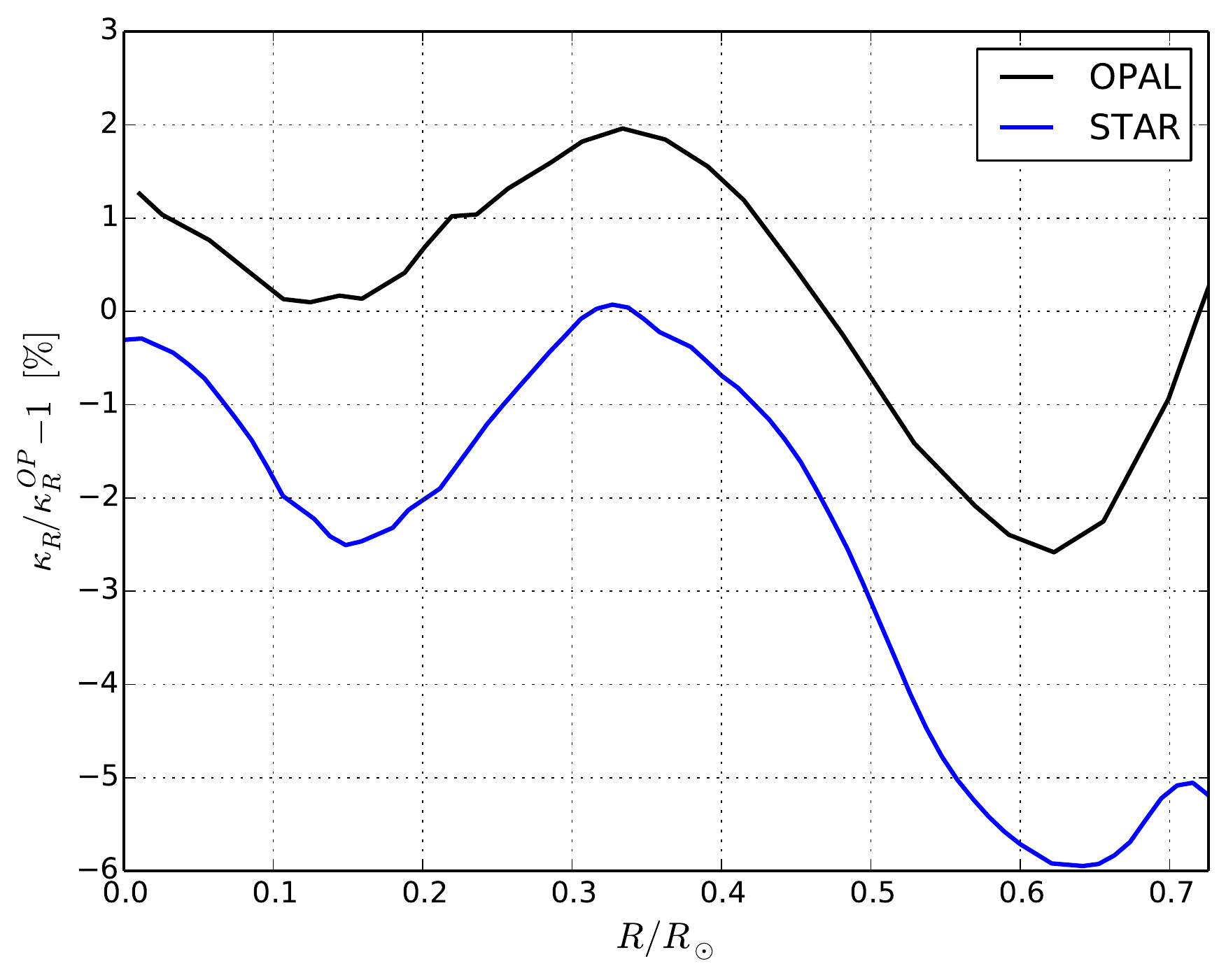}} 
	\caption{Relative difference $ \kappa_{R}/\kappa^{OP}_{R}-1 $ between OP and OPAL (black curve) and between OP and STAR (blue curve), across the radiation zone.}
	\label{fig:err_op_opal}
\end{figure}

A comparison of the solar opacity profile throughout the radiation zone, between calculations via STAR and OPAL (given by \cite{villante2014chemical}), to the opacity profile
	calculated by OP (\cite{seaton1994opacities,badnell2005updated}), is presented in \autoref{fig:err_op_opal}. It can be seen that the STAR and OP calculations agree within $ 6\% $  throughout the radiation zone. 
The solar mixture opacity is quite insensitive to the individual element opacities, so that, as was shown by \cite{blancard2012solar}, despite the large differences between several elemental opacities (\autoref{fig:op_opas_diff}), the total mixture STAR opacity is in good agreement with OP. This is possible since for Ne, Na, Mg, Al and Si, STAR opacities are significantly lower than OP, while for Cr, Mn, Fe and Ni, STAR opacities are significantly higher than OP.

\subsection{The effect of heavy metals}
Metals heavier than nickel have a very low abundance in the solar mixture, and therefore  are usually neglected in SSMs. However, at solar conditions, heavy metals may have a large number of bound electrons and therefore their spectra usually contain a huge number of strongly absorbing lines, which may in principle, affect the solar opacity. 
Due to the huge number of lines, a tractable calculation of hot high-Z elements spectra is only possible using the STA or Average-Atom methods. Solar opacities including heavier elements from the
\cite{grevesse1993cosmic} composition
 were calculated and analyzed by \cite{iglesias1995effects} using the STA program of \cite{BarShalom1989}. It was shown that even though the heavy elements are very strong photon absorbers, their very low abundance leads to a negligible effect on the Rosseland mean opacity.
 
\begin{figure}[]
	\centering
	\resizebox{0.5\textwidth}{!}{\includegraphics{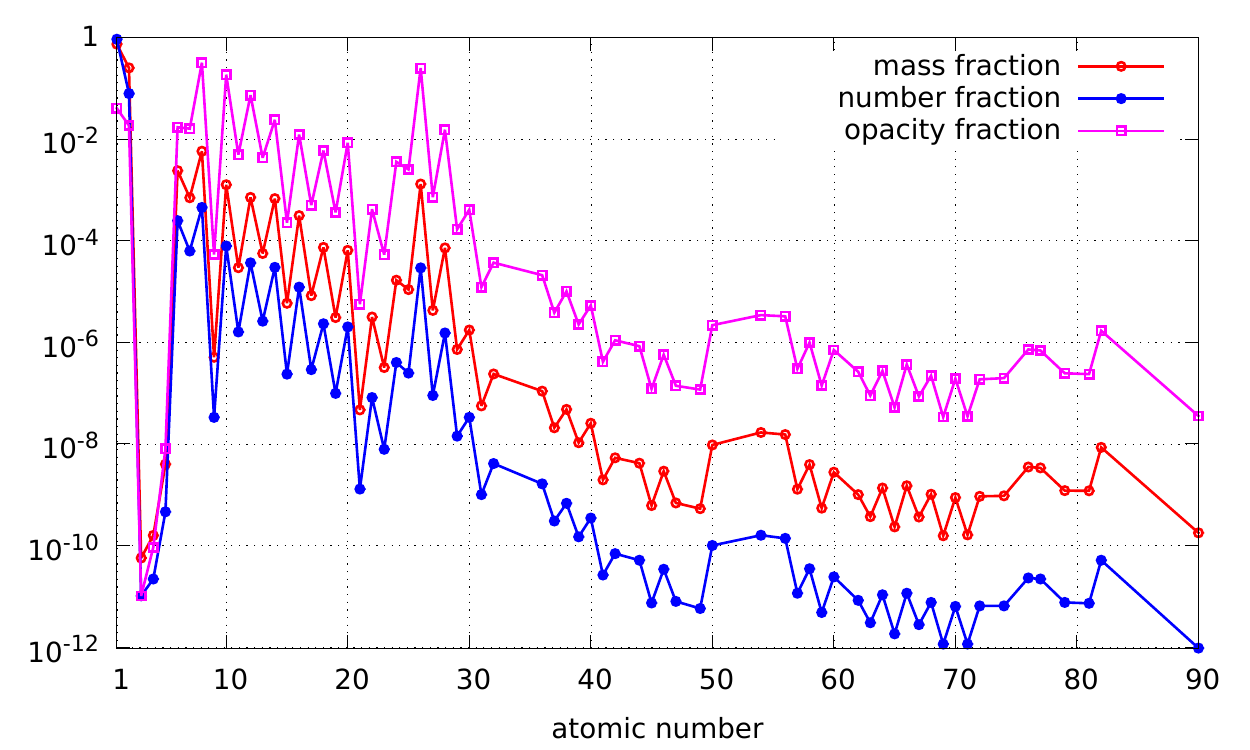}} 
	\caption{Visualization of \autoref{tab:highz_tab}: mass fractions (red), number fractions (blue) and Rosseland opacity fractions (magenta) for the elements in the full AGSS09 admixture. The mixture opacity was calculated nearby the radiation-convection boundary at $ T=176.3eV, \ \rho=0.1623g/cm^{3} $.}
	\label{fig:mr}
\end{figure}

\begin{figure}[]
	\centering
	\resizebox{0.5\textwidth}{!}{\includegraphics{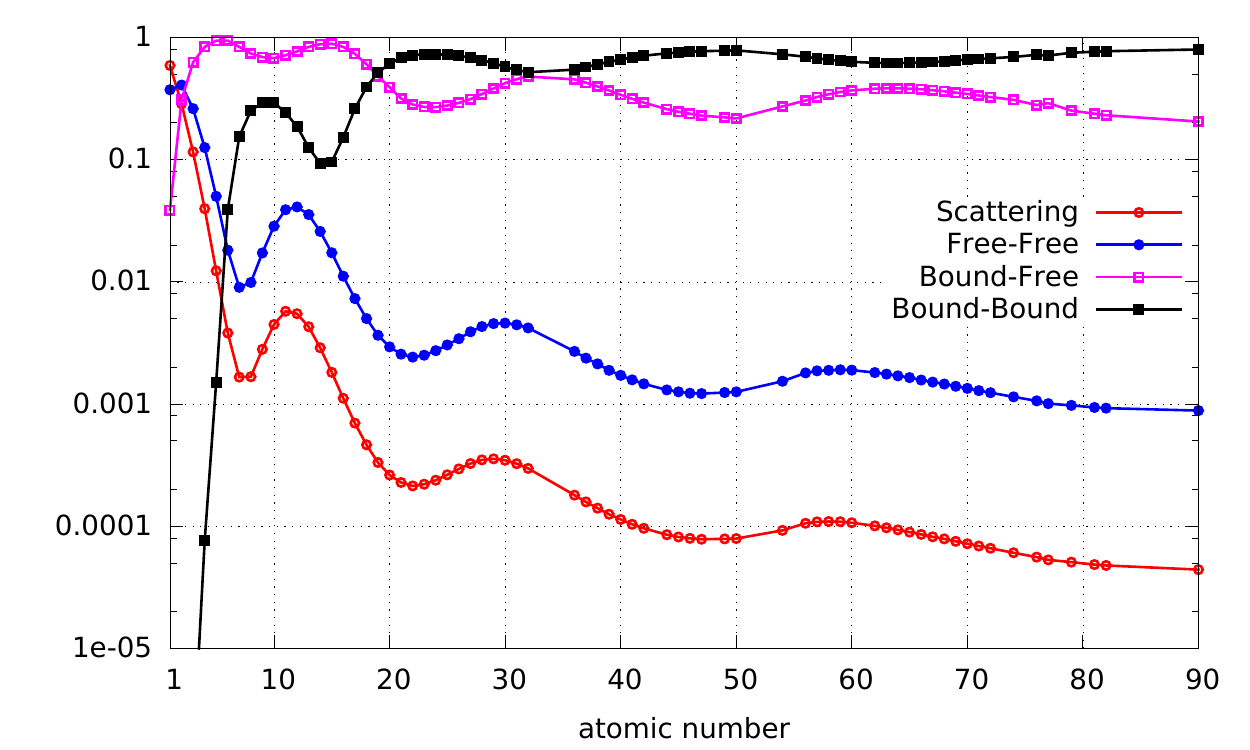}} 
	\caption{Relative contributions of the different atomic processes for each of the elements in the full AGSS09 admixture, to the opacity
		calculated at $ T=176.3eV, \rho=0.162g/cm^{3} $.}
	\label{fig:fracs_trans_highz}
\end{figure}

\begin{figure}[]
	\centering
	\resizebox{0.5\textwidth}{!}{\includegraphics{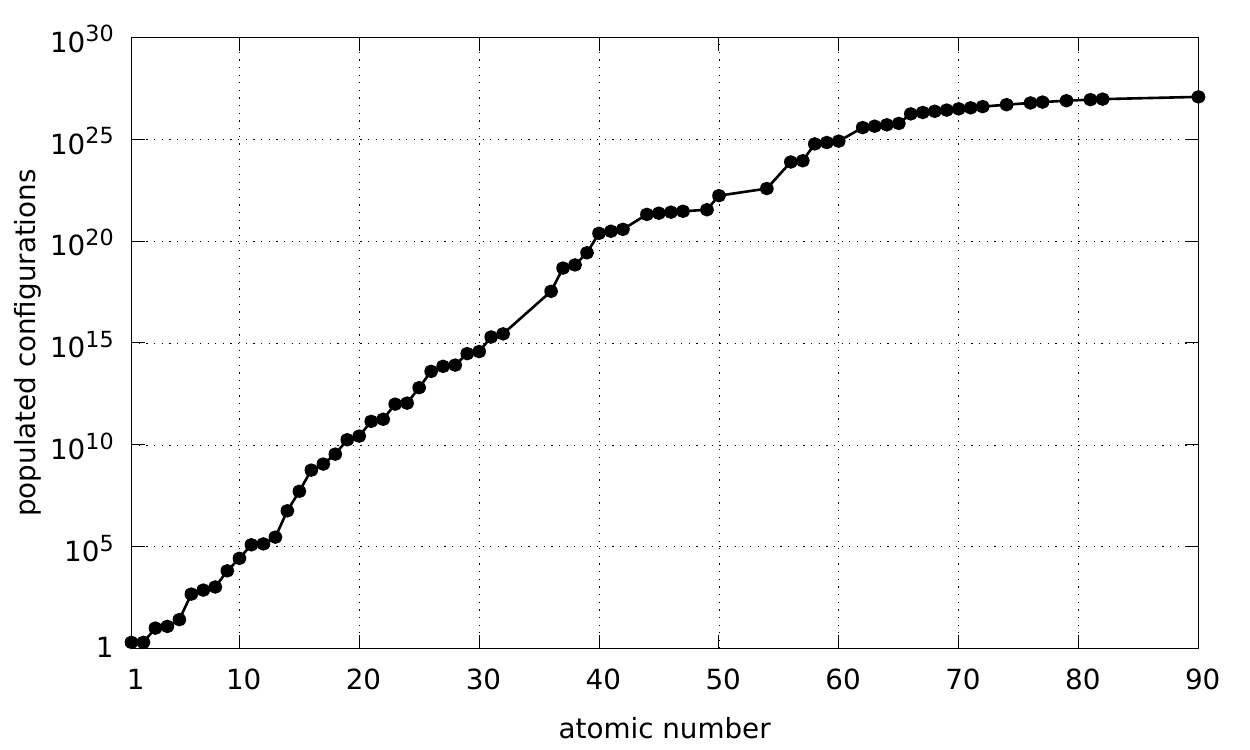}} 
	\caption{The number of populated electronic configurations $ \mathcal{N}_{C} $ for the elements in the full AGSS09 admixture nearby the radiation-convection boundary, $ T=176.3eV$ and $\rho=0.1623g/cm^{3} $.}
	\label{fig:numc}
\end{figure}

 Here we consider the Rosseland mean opacity of the solar mixture including \textit{all} elements in the recent AGSS09 (\cite{asplund2009chemical}) photospheric composition.
 The calculation is done at $ T=176.3eV$ and $\rho=0.16g/cm^{3} $ that exist nearby the radiation-convection interface. In \autoref{tab:highz_tab} the mass and number fractions of all elements in the AGSS09 mixture are given, together with the elemental contribution to the mixture Rosseland mean opacity. The table is also visualized in \autoref{fig:mr}. It is evident that the heavy metals opacity fractions are larger by more than four (two) orders of magnitude than their number (mass) fractions, indicating that heavy metals are indeed good absorbers. However, as was also shown by \cite{iglesias1995effects}, it is seen that due to their extremely small abundance, the effect of the heavy metals on the total Rosseland mean opacity is completely negligible. The contribution of the different atomic processes for the various elements are given in \autoref{fig:fracs_trans_highz}. We note that the local maxima in the bound-bound contribution represent the $ K, \ L $ and $ M $ shell lines crossing of the Rosseland peak (here, the crossing occurs due to the dependence of line energies on the atomic number). The estimated number of populated configurations (\autoref{eq:num_config_binomial}) of the different elements is given in \autoref{fig:numc}. It is evident that according to the estimate \eqref{eq:num_config_binomial}, for mid and high-Z metals, a complete DCA (and of course, DLA) calculation is absolutely intractable, and STA or Average-Atom (\cite{Rozsnyai1972,shalitin1984level}) methods must be used.

\begin{table}[h]
	\centering
		\caption{Specification of the AGSS09 photospheric admixture mass and number fractions. Elemental contribution to the Rosseland opacity, calculated by STAR at $ T=176.3eV, \rho=0.1623g/cm^{3} $ that are found nearby the radiation-convection interface, are also listed. 
			$ a(-b) $ represents $ a\times 10^{-b} $
		}
		\label{tab:highz_tab}
	\begin{tabular}{@{}lllll@{}}
		\toprule
		$ Z $ & element &  \begin{tabular}[c]{@{}l@{}}mass\\fraction\end{tabular}  & \begin{tabular}[c]{@{}l@{}}number\\fraction\end{tabular} & \begin{tabular}[c]{@{}l@{}}opacity\\fraction\end{tabular}\\ \midrule
		$1$ & H & $        7.36 \ (-01)$ & $        9.21 \ (-01)$ & $        4.02 \ (-02)$ \\ 
		$2$ & He & $        2.51 \ (-01)$ & $        7.84 \ (-02)$ & $        1.83 \ (-02)$ \\ 
		$3$ & Li & $        5.73 \ (-11)$ & $        1.03 \ (-11)$ & $        1.02 \ (-11)$ \\ 
		$4$ & Be & $        1.59 \ (-10)$ & $        2.21 \ (-11)$ & $        9.23 \ (-11)$ \\ 
		$5$ & B & $        3.99 \ (-09)$ & $        4.61 \ (-10)$ & $        8.07 \ (-09)$ \\ 
		$6$ & C & $        2.38 \ (-03)$ & $        2.48 \ (-04)$ & $        1.70 \ (-02)$ \\ 
		$7$ & N & $        6.97 \ (-04)$ & $        6.22 \ (-05)$ & $        1.59 \ (-02)$ \\ 
		$8$ & O & $        5.77 \ (-03)$ & $        4.51 \ (-04)$ & $        3.18 \ (-01)$ \\ 
		$9$ & F & $        5.08 \ (-07)$ & $        3.34 \ (-08)$ & $        5.33 \ (-05)$ \\ 
		$10$ & Ne & $        1.26 \ (-03)$ & $        7.84 \ (-05)$ & $        1.84 \ (-01)$ \\ 
		$11$ & Na & $        2.94 \ (-05)$ & $        1.60 \ (-06)$ & $        4.89 \ (-03)$ \\ 
		$12$ & Mg & $        7.12 \ (-04)$ & $        3.67 \ (-05)$ & $        7.31 \ (-02)$ \\ 
		$13$ & Al & $        5.60 \ (-05)$ & $        2.59 \ (-06)$ & $        4.26 \ (-03)$ \\ 
		$14$ & Si & $        6.69 \ (-04)$ & $        2.98 \ (-05)$ & $        2.38 \ (-02)$ \\ 
		$15$ & P & $        5.86 \ (-06)$ & $        2.37 \ (-07)$ & $        2.25 \ (-04)$ \\ 
		$16$ & S & $        3.11 \ (-04)$ & $        1.21 \ (-05)$ & $        1.22 \ (-02)$ \\ 
		$17$ & Cl & $        8.25 \ (-06)$ & $        2.91 \ (-07)$ & $        4.99 \ (-04)$ \\ 
		$18$ & Ar & $        7.39 \ (-05)$ & $        2.31 \ (-06)$ & $        5.93 \ (-03)$ \\ 
		$19$ & K & $        3.08 \ (-06)$ & $        9.87 \ (-08)$ & $        3.58 \ (-04)$ \\ 
		$20$ & Ca & $        6.45 \ (-05)$ & $        2.01 \ (-06)$ & $        8.44 \ (-03)$ \\ 
		$21$ & Sc & $        4.67 \ (-08)$ & $        1.30 \ (-09)$ & $        5.63 \ (-06)$ \\ 
		$22$ & Ti & $        3.14 \ (-06)$ & $        8.21 \ (-08)$ & $        4.17 \ (-04)$ \\ 
		$23$ & V & $        3.19 \ (-07)$ & $        7.84 \ (-09)$ & $        5.31 \ (-05)$ \\ 
		$24$ & Cr & $        1.67 \ (-05)$ & $        4.02 \ (-07)$ & $        3.55 \ (-03)$ \\ 
		$25$ & Mn & $        1.09 \ (-05)$ & $        2.48 \ (-07)$ & $        2.52 \ (-03)$ \\ 
		$26$ & Fe & $        1.30 \ (-03)$ & $        2.91 \ (-05)$ & $        2.49 \ (-01)$ \\ 
		$27$ & Co & $        4.24 \ (-06)$ & $        9.00 \ (-08)$ & $        7.17 \ (-04)$ \\ 
		$28$ & Ni & $        7.17 \ (-05)$ & $        1.53 \ (-06)$ & $        1.52 \ (-02)$ \\ 
		$29$ & Cu & $        7.24 \ (-07)$ & $        1.43 \ (-08)$ & $        1.67 \ (-04)$ \\ 
		$30$ & Zn & $        1.75 \ (-06)$ & $        3.34 \ (-08)$ & $        4.06 \ (-04)$ \\ 
		$31$ & Ga & $        5.63 \ (-08)$ & $        1.01 \ (-09)$ & $        1.22 \ (-05)$ \\ 
		$32$ & Ge & $        2.39 \ (-07)$ & $        4.11 \ (-09)$ & $        3.70 \ (-05)$ \\ 
		$36$ & Kr & $        1.10 \ (-07)$ & $        1.64 \ (-09)$ & $        2.10 \ (-05)$ \\ 
		$37$ & Rb & $        2.08 \ (-08)$ & $        3.05 \ (-10)$ & $        3.84 \ (-06)$ \\ 
		$38$ & Sr & $        4.78 \ (-08)$ & $        6.83 \ (-10)$ & $        1.01 \ (-05)$ \\ 
		$39$ & Y & $        1.06 \ (-08)$ & $        1.49 \ (-10)$ & $        2.24 \ (-06)$ \\ 
		$40$ & Zr & $        2.55 \ (-08)$ & $        3.50 \ (-10)$ & $        5.32 \ (-06)$ \\ 
		$41$ & Nb & $        1.97 \ (-09)$ & $        2.66 \ (-11)$ & $        4.14 \ (-07)$ \\ 
		$42$ & Mo & $        5.36 \ (-09)$ & $        6.98 \ (-11)$ & $        1.10 \ (-06)$ \\ 
		$44$ & Ru & $        4.18 \ (-09)$ & $        5.18 \ (-11)$ & $        8.36 \ (-07)$ \\ 
		$45$ & Rh & $        6.16 \ (-10)$ & $        7.48 \ (-12)$ & $        1.22 \ (-07)$ \\ 
		$46$ & Pd & $        2.91 \ (-09)$ & $        3.42 \ (-11)$ & $        5.70 \ (-07)$ \\ 
		$47$ & Ag & $        6.91 \ (-10)$ & $        8.02 \ (-12)$ & $        1.40 \ (-07)$ \\ 
		$49$ & In & $        5.33 \ (-10)$ & $        5.81 \ (-12)$ & $        1.17 \ (-07)$ \\ 
		$50$ & Sn & $        9.58 \ (-09)$ & $        1.01 \ (-10)$ & $        2.17 \ (-06)$ \\ 
		$54$ & Xe & $        1.68 \ (-08)$ & $        1.60 \ (-10)$ & $        3.43 \ (-06)$ \\ 
		$56$ & Ba & $        1.53 \ (-08)$ & $        1.39 \ (-10)$ & $        3.24 \ (-06)$ \\ 
		$57$ & La & $        1.29 \ (-09)$ & $        1.16 \ (-11)$ & $        2.99 \ (-07)$ \\ 
		$58$ & Ce & $        3.92 \ (-09)$ & $        3.50 \ (-11)$ & $        9.90 \ (-07)$ \\ 
		$59$ & Pr & $        5.44 \ (-10)$ & $        4.83 \ (-12)$ & $        1.40 \ (-07)$ \\ 
		$60$ & Nd & $        2.79 \ (-09)$ & $        2.42 \ (-11)$ & $        7.07 \ (-07)$ \\ 
		$62$ & Sm & $        1.01 \ (-09)$ & $        8.40 \ (-12)$ & $        2.63 \ (-07)$ \\ 
		$63$ & Eu & $        3.70 \ (-10)$ & $        3.05 \ (-12)$ & $        8.95 \ (-08)$ \\ 
		$64$ & Gd & $        1.36 \ (-09)$ & $        1.08 \ (-11)$ & $        2.75 \ (-07)$ \\ 
		$65$ & Tb & $        2.33 \ (-10)$ & $        1.84 \ (-12)$ & $        5.22 \ (-08)$ \\ 
		$66$ & Dy & $        1.51 \ (-09)$ & $        1.16 \ (-11)$ & $        3.69 \ (-07)$ \\ 
		$67$ & Ho & $        3.67 \ (-10)$ & $        2.78 \ (-12)$ & $        8.60 \ (-08)$ \\ 
		$68$ & Er & $        1.02 \ (-09)$ & $        7.66 \ (-12)$ & $        2.21 \ (-07)$ \\ 
		$69$ & Tm & $        1.57 \ (-10)$ & $        1.16 \ (-12)$ & $        3.38 \ (-08)$ \\ 
		$70$ & Yb & $        8.81 \ (-10)$ & $        6.37 \ (-12)$ & $        1.93 \ (-07)$ \\ 
		$71$ & Lu & $        1.62 \ (-10)$ & $        1.16 \ (-12)$ & $        3.50 \ (-08)$ \\ 
		$72$ & Hf & $        9.30 \ (-10)$ & $        6.52 \ (-12)$ & $        1.86 \ (-07)$ \\ 
		$74$ & W & $        9.58 \ (-10)$ & $        6.52 \ (-12)$ & $        1.98 \ (-07)$ \\ 
		$76$ & Os & $        3.52 \ (-09)$ & $        2.31 \ (-11)$ & $        7.21 \ (-07)$ \\ 
		$77$ & Ir & $        3.39 \ (-09)$ & $        2.21 \ (-11)$ & $        6.90 \ (-07)$ \\ 
		$79$ & Au & $        1.21 \ (-09)$ & $        7.66 \ (-12)$ & $        2.46 \ (-07)$ \\ 
		$81$ & Tl & $        1.19 \ (-09)$ & $        7.31 \ (-12)$ & $        2.37 \ (-07)$ \\ 
		$82$ & Pb & $        8.58 \ (-09)$ & $        5.18 \ (-11)$ & $        1.69 \ (-06)$ \\ 
		$90$ & Th & $        1.79 \ (-10)$ & $        9.64 \ (-13)$ & $        3.54 \ (-08)$ \\ 
		\bottomrule
	\end{tabular}
\end{table}


\section{Conclusion}
A new opacity code, STAR, implementing the STA method of \cite{BarShalom1989}, was presented.
It was used to calculate and analyze opacities throughout the radiation zone for a solar model implementing the recent AGSS09 composition.
The relative contribution of various elements and atomic processes to the Rosseland mean opacity were compared in detail.
STAR and OP monochromatic opacities and charge state distributions were compared for several elements near the radiation-convection interface.
For the heavier elements, such as iron, STAR Rosseland opacity is significantly larger than OP, which is explained by the larger population of lower ionization states.
For the lighter elements, such as magnesium, OP Rosseland opacity is significantly larger than STAR, due to a much larger line broadening of the K-shell lines shown by OP. Very similar results were reported by \cite{blancard2012solar} using the OPAS code.
STAR Rosseland opacities for the solar mixture were calculated throughout the radiation zone and a good agreement with OP
and OPAL was achieved. Finally, an explicit STAR calculation of the full AGSS09 photospheric mixture, including all heavy metals was performed. It was shown that due to their extremely low abundance, and despite being very good photon absorbers, the heavy elements do not affect the Rosseland opacity, in agreement with \cite{iglesias1995effects}.
\\
\\
The authors thank Aldo Serenelli and Francesco Villante for providing various solar profiles and for
useful suggestions and comments.
		
\acknowledgments{

}
	
\bibliographystyle{apj}

\end{document}